\documentclass[
    aps,
    pra,
    longbibliography,
    superscriptaddress,amsmath,amssymb,amsfonts, 
    tightenlines,
    twocolumn,
    ]{revtex4-1}
    
\usepackage[utf8]{inputenc}  
\usepackage[T1]{fontenc}     
\usepackage[english]{babel}  
\usepackage{hyperref}  
\usepackage{graphicx} 
\usepackage[babel]{microtype}  
\usepackage{amsmath,amssymb,amsthm,bm,amsfonts,mathrsfs,bbm} 
\usepackage{setspace}
\usepackage{times}
\usepackage{braket}
\usepackage[bottom]{footmisc}
\usepackage{blochsphere}
\usepackage{pgfplots}
 
\pgfplotsset{compat=1.17}
\usetikzlibrary{calc,3d,shapes, pgfplots.external, intersections}
\usepackage{algorithm}
\usepackage{physics}
\usepackage[noend]{algorithmic}
\usepackage{eqparbox}

\usepackage{physics}
\usepackage{tikz}
\usepackage{mathtools}
\usetikzlibrary{arrows.meta}
\usepackage{csquotes}

\usepackage{xspace}  
\usepackage{pgfplots}
\usepackage{verbatim}
\usepackage{amsmath}

\newtheorem{theorem}{Theorem}

\newcommand\id{\leavevmode\hbox{\small1\kern-3.3pt\normalsize1}}

\newcommand{\x}{{\bf x}}

\definecolor{jens}{rgb}{0.1,0.4,0.6}

\newcommand{\newtext}[1]{{\textcolor{black}{#1}}}

\usepackage{pdfpages}
\usepackage{pgffor}
\makeatletter
\AtBeginDocument{\let\LS@rot\@undefined}
\makeatother

\begin{document}

\title{Towards provably efficient quantum algorithms for large-scale machine-learning models}

\author{Junyu Liu}
\affiliation{Pritzker School of Molecular Engineering, The University of Chicago, Chicago, IL 60637, USA}
\affiliation{Department of Computer Science, The University of Chicago, Chicago, IL 60637, USA}
\affiliation{Chicago Quantum Exchange, Chicago, IL 60637, USA}
\affiliation{Kadanoff Center for Theoretical Physics, The University of Chicago, Chicago, IL 60637, USA}
\affiliation{qBraid Co., Chicago, IL 60615, USA}
\affiliation{SeQure, Chicago, IL 60615, USA}

\author{Minzhao Liu}

\affiliation{Department of Physics, The University of Chicago, Chicago, IL 60637, USA}
\affiliation{Computational Science Division, Argonne National Laboratory, Lemont, IL 60439, USA}

\author{Jin-Peng Liu}

\affiliation{Simons Institute for the Theory of Computing, University of California, Berkeley, CA 94720, USA}
\affiliation{Department of Mathematics, University of California, Berkeley, CA 94720, USA}
\affiliation{Center for Theoretical Physics, Massachusetts Institute of Technology, Cambridge, MA 02139, USA}

\author{Ziyu Ye}
\affiliation{Department of Computer Science, The University of Chicago, Chicago, IL 60637, USA}

\author{Yunfei Wang}

\affiliation{Martin A. Fisher School of Physics, Brandeis University, Waltham, MA 02453, USA}

\author{Yuri Alexeev}
\affiliation{Computational Science Division, Argonne National Laboratory, Lemont, IL 60439, USA}
\affiliation{Department of Computer Science, The University of Chicago, Chicago, IL 60637, USA}
\affiliation{Chicago Quantum Exchange, Chicago, IL 60637, USA}
\author{Jens Eisert}
\affiliation{Dahlem Center for Complex Quantum Systems,
Free University Berlin, Berlin, 14195, Germany}

\author{Liang Jiang}

\affiliation{Pritzker School of Molecular Engineering, The University of Chicago, Chicago, IL 60637, USA}
\affiliation{Chicago Quantum Exchange, Chicago, IL 60637, USA}

\maketitle

{\bf Large machine learning models are revolutionary technologies of artificial intelligence whose bottlenecks include huge computational expenses, power, and time used both in the pre-training and fine-tuning process. In this work, we show that fault-tolerant quantum computing could possibly provide provably efficient resolutions for generic (stochastic) gradient descent algorithms, scaling as $\mathcal{O}(T^2 \times \text{polylog}(n))$, where $n$ is the size of the models and $T$ is the number of iterations in the training, as long as the models are both sufficiently dissipative and sparse, with small learning rates. Based on earlier efficient quantum algorithms for dissipative differential equations, we find and prove that similar algorithms work for (stochastic) gradient descent, the primary algorithm for machine learning. In practice, we benchmark instances of large machine learning models from 7 million to 103 million parameters. We find that, in the context of sparse training, a quantum enhancement is possible at the early stage of learning after model pruning, motivating a sparse parameter download and re-upload scheme. Our work shows solidly that fault-tolerant quantum algorithms could potentially contribute to most state-of-the-art, large-scale machine-learning problems.  
}

It is widely believed that large-scale machine learning might be one of the most revolutionary technologies benefiting society \cite{lecun2015deep}, including already important breakthroughs in digital arts \cite{dalle2}, conversation like \texttt{GPT-3} \cite{brown2020language,chatgpt}, and mathematical problem solving \cite{lewkowycz2022solving}. However, training such models with considerable parameters is costly and has high carbon emissions. For instance, twelve million dollars and over five-hundred tons of CO$_2$ equivalent emissions have been produced to train \texttt{GPT-3} \cite{patterson2021carbon}. Thus, on the one hand, it is important to make large-scale machine-learning models (like large language models, LLM) sustainable and efficient. 

On the other hand, machine learning might possibly be one of the flag applications of quantum technology. Running machine learning algorithms on quantum devices, implementing readings of so-called 
\emph{quantum machine learning}, is widely seen as a potentially very fruitful application of quantum algorithms \cite{biamonte2017quantum}. \newtext{Specifically, many quantum approaches are proposed to enhance the capability of classical machine learning and hopefully find some useful applications, like \cite{rebentrost2014quantum,zhao2019quantum}}. Despite rapid development and significant progress, current quantum machine learning algorithms feature substantial limitations both in theory and practice. First, practical applications of quantum machine learning algorithms for near-term devices are often lacking theoretical grounds that guarantee or at least plausibly suggest to outperform their classical counterparts. Second, for fault-tolerant settings of quantum machine learning problems \cite{peruzzo2014variational,mcclean2016theory,mcardle2020quantum,cerezo2021variational,farhi2014quantum,ebadi2022quantum,farhi2018classification,havlivcek2019supervised,liu2022noise}, 
rigorous super-polynomial quantum speedups can actually be 
proven \cite{PACLearning,TemmeML,DensityModelling} for highly structured problems. 
That said, these prescriptions are arguably still far from real state-of-the-art applications of classical machine learning. Some of them are primarily using quantum states as training data instead of classical data, which can be---highly encouraging as these approaches are---argued to be not the currently most important classical machine learning application \cite{lloyd2014quantum,TemmeML,huang2021power,huang2020predicting,huang2022quantum}. 
Efforts need to be made to extend our understanding of quantum machine learning, in the sense that we have to understand how they could have theoretical guarantees and how they could solve 
timely and natural problems, at least in principle, of classical machine learning. For instance, they should relate to scalable and sustainable 
natural problems in large-scale machine-learning.

\begin{figure} [ht]
   \begin{center}
   \includegraphics[width=8.5cm]{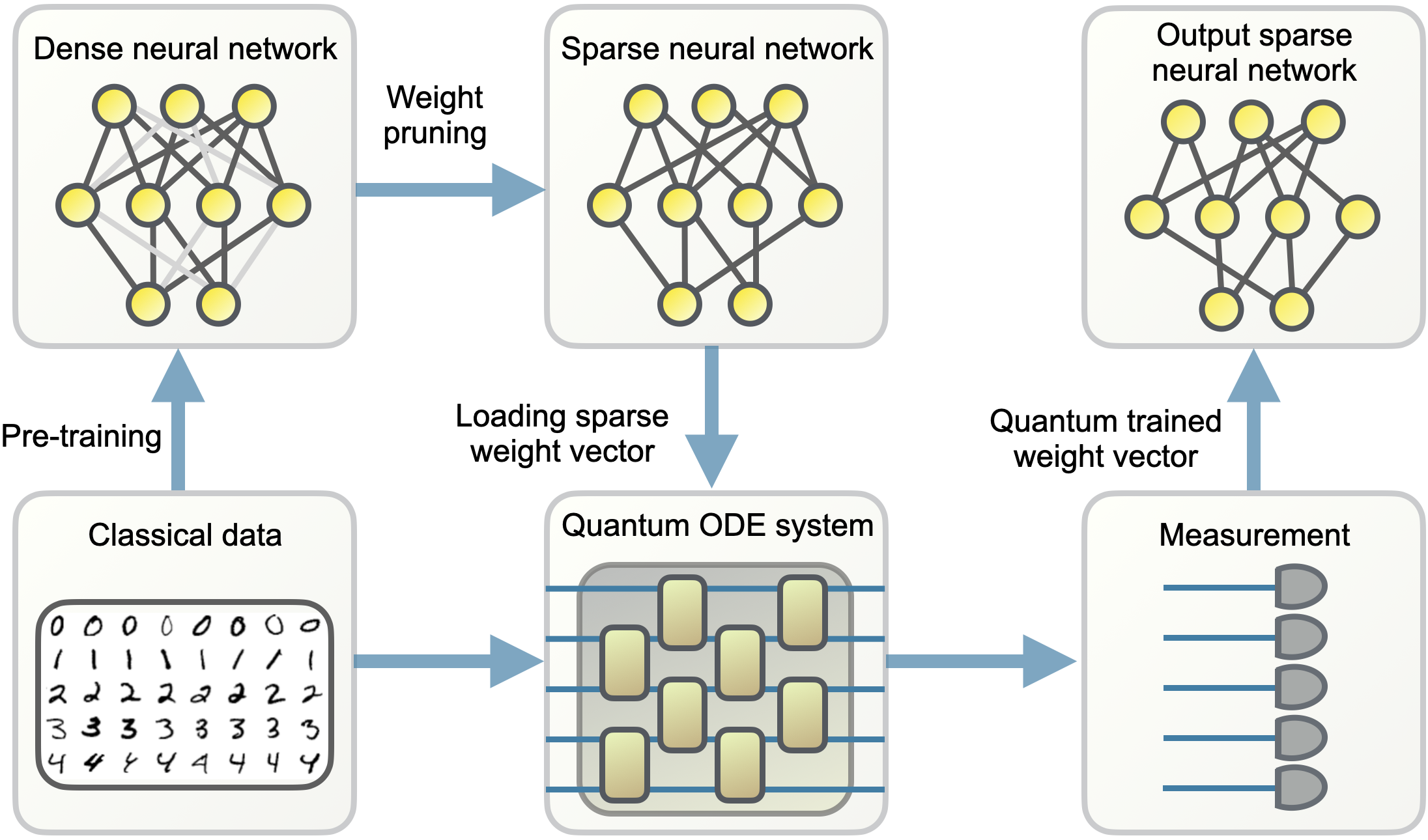}
   \end{center}
   \caption{A possible learning process in large-scale models, which might use sparse training, whose early stage in learning might admit possible quantum enhancement. 
    A dense neural network is pre-trained classically. The neural network weights are then pruned and only a small fraction is preserved. A quantum ordinary difference equation system that corresponds to the sparse training dynamics is created using the sparse network and the training data. To allow quantum enhancement, the system must be sparse and dissipative. Measurement on the solution state is performed to obtain the final trained parameters, used to construct a trained classical sparse neural network.}
   { \label{fig:illulearning}
}
   \end{figure}

In this work, we take significant steps in this direction by designing end-to-end quantum machine learning algorithms that are expected to be timely for the current machine learning community and that are to an extent equipped with guarantees. Based on a typical large-scale (classical) machine-learning process (see Fig.~\ref{fig:illulearning} for an illustration), we find that after a significant number of neural network training parameters have been pruned (sparse training) \cite{hoefler2021sparsity,lee2018snip,wang2020Picking,tanaka2020pruning} and the classical training parameters compiled to a quantum computer, 
we suggest to find a quantum enhancement at the early state of training before the error grows exponentially. At its heart, the quantum algorithm part of the work includes suitable modifications of the quantum algorithm \cite{liu2021efficient} for solving differential equations to running (stochastic) gradient descent algorithms---presumably the primary classical machine learning algorithm---into a quantum processor after linearization. The expectation of a possible quantum enhancement is rooted in an application of a variant of the so-called \emph{Harrow-Hassidim-Lloyd} (HHL) algorithm \cite{harrow2009quantum}, an efficient quantum algorithm for sparse matrix inversion that solves the problem within $\mathcal{O}(\log n)$ time for suitably conditioned
$n\times n$ sparse matrices. We find that our algorithm can solve 
large-scale model-dimension-$n$ machine learning problems in $\mathcal{O}(\text{polylog}(n)\times T)$ or $\mathcal{O}(\text{polylog}(n)\times T^2)$ time, where $T$ is the number of iterations. The scaling in $n$ outperforms
the scaling of any classical algorithms we know of. However, for a given machine learning problem with required performances, there is no guarantee that our hybrid quantum-classical algorithm will necessarily outperform all other conceivable classical algorithms 
for related, but different tasks
(for instance, for algorithms that are not gradient-based). 
Thus, our result gives, to the best of our knowledge, rise to a potential substantial quantum speedup or enhancement of particular classical algorithms, 
instead of a quantum advantage over the entire
problem class. 

\newtext{From a quantum algorithms perspective, stochastic gradient descent processes are solved here using quantum
\emph{ordinary differential equation}
(ODE) solvers derived from the findings of Ref.~\cite{liu2021efficient}, based on linearizing non-linear equations using so-called quantum Carleman linearization. We find that the corresponding differential equation solvers can, in principle, also be used in the discrete setting and for stochastic gradient descent in machine learning. However, in the discrete setting, the theoretical details are significantly different from those applicable in the small learning rate limit. In this work, we systematically establish a novel discrete Carleman linearization in the supplemental material, including reformulations of the Carleman linearization theory, a tensor network diagrammatic notation for the discretization error, analytic derivations of higher-order corrections, and explicit examples for lower order expansions. Further details about the novelty of our algorithms beyond the findings of Ref.~\cite{liu2021efficient} are summarized in the supplemental material. }

It is important to stress that the above algorithm has a number of requirements 
that do admit a quantum enhancement. First, both the machine learning model and the weight vectors have to be sufficiently \emph{sparse}, which will ensure a fast interface between classical and quantum processors (this requirement could be relaxed in the presence of \emph{quantum random access memory} (QRAM) \cite{giovannetti2008quantum}, a fast uploader towards quantum data, but we stress that this is \emph{not} required and there are no hidden resources in our scheme). Second, the model has to be sufficiently \emph{dissipative}. For dissipative systems, the linearization error is well controlled, ensuring that the HHL algorithm can obtain reliable results even with non-linear machine learning models. We find dissipation happens generically in the early training process of large-scale machine learning. 

We corroborate the intuition developed here by a number of theorems, as well as extensive numerical experiments. The formal definition of dissipation, sparsity, and quantum speedups are rigorously proven in the supplementary material. Informal readings of the main theorems are presented in Section \ref{sec:theorems}, while solid numerical evidence up to $103$ million training parameters are provided in Section \ref{sec:experiments}. Finally, a conclusion providing also an outlook will be provided in Section \ref{sec:concoutlook}.

\section{Theorems}\label{sec:theorems}

In this section, we will lay out the informally formulated main theorems that are established in this work. Details can be found in the supplementary material.

\begin{theorem}[Informal]\label{thm:posdef}
For a sparse machine learning model with model size $n$, running $T$ iterations, with the algorithm being 
fully dissipative with small learning rates (whose formal definition is given in the \newtext{supplemental material}), 
there is a quantum algorithm that runs in 
\begin{align}
{\mathcal{O}}\left( {T \times {\mathop{\rm poly}\nolimits} \left( {\log n,\frac{1}{\epsilon }} \right)} \right)
\end{align}
time with precision $\epsilon>0$. The sparsity condition also ensures the efficiency of uploading and downloading quantum states towards classical processors. 
\end{theorem}

\begin{theorem}[Informal]\label{thm:almostdissipative}
For a sparse machine learning model with model size $n$, running in $T$ iterations, and the algorithm being almost dissipative with small learning rates (whose formal definition is given in the \newtext{supplemental material}), then there is a quantum algorithm runs in 
\begin{align}
{\mathcal{O}}\left( {T^2 \times {\mathop{\rm poly}\nolimits} \left( {\log n,\frac{1}{\epsilon }} \right)} \right)
\end{align}
time with precision $\epsilon>0$. The sparsity condition also ensures the efficiency of uploading and downloading quantum states towards classical processors. 
\end{theorem}
In these expressions, $m:=\log_2(n)$ takes the role of a system size of the quantum system. First, we describe the problem we are trying to solve. A machine learning model is defined partially by a function $\mathcal{L}_{\mathcal{A}}$, called the \emph{loss function}, as a function of weight vector (variational angle) $\theta \in \mathbb{R}^n = (\theta_\mu)$, 
%
%
%
and the \emph{input training set} $\mathcal{A}$. The weight vector has $n$ components if an $n$-dimensional model. 
The task is to minimize the function $\mathcal{L}_\mathcal{A}$ by adjusting $\theta$ making use of $T$ iterations. 

The presumably most widely utilized algorithm in machine learning is called (stochastic) gradient descent. Starting from the initial weight vector $\theta(t=0)$, we implement the following ordinary differential equation from $t=0$ to $t=T$ with small, positive learning rate $\eta$,
\begin{align}
\theta_\mu(t+1)=\theta_\mu(t)-\left.\eta \frac{d \mathcal{L}_{\mathcal{A}}}{d 
\theta_\mu}\right|_{\theta(t)} .
\end{align}
Variants of the gradient descent algorithms also include adding random noise $\xi_\mu (t)$ in 
each step, so-called stochastic gradient descent algorithms. One can show that in many cases, at the end of training, $\theta_\mu (t=T)$ can make the loss function $\mathcal{L}_{\mathcal{A}} (\theta (t=T))$ sufficiently small. 

The quantum algorithm with the promised efficiency in Theorem \ref{thm:posdef} and 
Theorem \ref{thm:almostdissipative} \newtext{is} described in the following.
\begin{itemize}
\item Our starting point of the algorithm is given by a initial weight vector, $\theta (0)$, the maximal number of iterations $T$, and the machine learning architecture $\mathcal{L}_\mathcal{A}$, with model size $n$.
\item In a first step, we use so-called \emph{quantum Carleman linearization} introduced 
in Ref.~\cite{liu2021efficient}, to linearize the model $\mathcal{L}_\mathcal{A}$ with 
the matrix $M$ (see the supplemental material for more details).
\item Then, we need to upload the sparse weight vector $\theta(0)$ as a state vector in quantum devices, using tools of Ref.~\cite{gleinig2021efficient} or alternatively
more sophisticated and at the same time challenging architectures like 
\emph{quantum random access memory} (QRAM) \cite{giovannetti2008quantum}.

\item Then, in a further step, we use a variant of the HHL solver that has been introduced 
in Ref.~\cite{harrow2009quantum} and the supplementary material, 
to solve the state vector at the end $t=T$. \newtext{The pipeline is runnable under the condition of sparsity and dissipation, which is satisfied by our models. Sparsity includes the sparsity of model themselves, and the sparsity of weight vectors (ensured by the assumptions of sparse training), while dissipation is a natural property of the early steps of training, extensively discussed in Section \ref{sec:experiments}.} 
\item Finally, we exploit  tomographic methods described in, for instance, Refs.~\cite{guctua2020fast,huang2020predicting} and the supplemental material, 
to obtain the classical model parameters $\theta(T)$. 
\end{itemize}

\newtext{Finally, we wish to mention that this potential enhancement from quantum computing is concerning the size of the model, not necessarily the precision. For instance, we use tomographic method to download the sparse quantum state at the end of quantum ODE solver with the precision scaling as $1/\epsilon^2$. This scaling might be optimal in the quantum setting \cite{guctua2020fast}, but may not be ideal compared to purely classical algorithms (although the precise relationship between the error and the performances of classical machine learning models is generically not clear to date). We leave those interesting issues for future works.}

\section{Linearizing classical neural networks}
\newtext{In this section, we provide a short and heuristic description of how to solve stochastic gradient descent using HHL algorithms. For a given (stochastic) gradient descent process, the recursion relation is given by}
\begin{align}\label{eq:model}
   \delta u&=u(t+1)-u(t)=\sum\limits_{\ell =0}^{q}{{{F}_{\ell }}{{u}^{\otimes \ell }}(t)} \nonumber\\
 & ={{F}_{q}}{{u}^{\otimes q}}(t)+\ldots +{{F}_{2}}{{u}^{\otimes 2}}(t)+{{F}_{1}}u(t)+{{F}_{0}} ~,
\end{align}
\newtext{with the initial condition $u(0)={{u}_{\text{in}}}$. Here, $u(t)=(\theta_\mu)(t)$ is a set of weight vectors at the iteration $t$, $\delta o\equiv o(t+1)- o(t)$ represents the discrete difference between two time steps $(t+1)$ and $(t)$ for a variable $o$ (see Ref.~\cite{liu2021efficient} for the continuous version), and ${{u}^{\otimes \ell}}$ is the $\ell$-th order tensor product. Thus, 
Eq.~(\ref{eq:model}) 
characterizes the dynamics of $q$-th order non-linearity in classical neural networks. Now, we introduce a linear process designed
to approximate the non-linear model (\ref{eq:model}), called 
\emph{quantum Carleman linearization},
as} 
\begin{align}\label{eq:qcl}
\delta \left[ \begin{matrix}
   u  \\
   {{u}^{\otimes 2}}  \\
   {{u}^{\otimes 3}}  \\
   \vdots   \\
   {{u}^{\otimes (N-1)}}  \\
   {{u}^{\otimes N}}  \\
   \vdots   \\
\end{matrix} \right]=A\left[ \begin{matrix}
   u  \\
   {{u}^{\otimes 2}}  \\
   {{u}^{\otimes 3}}  \\
   \vdots   \\
   {{u}^{\otimes (N-1)}}  \\
   {{u}^{\otimes N}}  \\
   \vdots   \\
\end{matrix} \right]+\left[ \begin{matrix}
   {{F}_{0}}  \\
   0  \\
   0  \\
   \vdots   \\
   0  \\
   0  \\
   \vdots   \\
\end{matrix} \right]~.
\end{align}
\newtext{In this linear process, the vector space (whose vectors could be denoted by $\hat{y}$) is given by the weight vectors and all possible tensor products thereof, while $A$ is a large matrix with matrix elements given by the $F_\ell$, the so-called 
\emph{quantum Carleman matrix} (QCM). In principle, this linear relation is infinitely dimensional, so we are replacing the original non-linear recursions to an infinite set of linear relations. If we wish to solve this infinite process by a digital system, we need to make truncation. In this work, we show that for dissipative systems (whose QCMs have enough negative eigenvalues, roughly corresponding to large enough positive eigenvalues for the Hessian of the loss functions in classical neural networks; the positive eigenvalues are called 
\emph{dissipative modes} in the supplementary material, while negative eigenvalues are called \emph{divergent modes}), the truncation error can be well-controlled.}

\newtext{For sparse, dissipative systems, Eq.~(\ref{eq:qcl}) can be treated as a matrix inversion problem, thus solved by the HHL algorithm using quantum computers,}
\begin{align}
&  \begin{bmatrix}
    I &  &  &  &  \\
    -(I\!+\!A) & I &  &  &  \\
     & \ddots & \ddots &  &  \\
     &  & -(I\!+\!A) & I &  \\
     &  &  & -(I\!+\!A) & I \\
  \end{bmatrix}
  \begin{bmatrix}
    \hat{y}(0) \\
    \hat{y}(1) \\
    \vdots \\
   \hat{y}(T-1) \\
    \hat{y}(T) \\
  \end{bmatrix} =
  \begin{bmatrix}
    \hat{y}_{\text{in}} \\
    b \\
    \vdots \\
    b \\
    b \\
  \end{bmatrix}.
\end{align}
\newtext{Here, we are considering $T+1$ iterations in total from $t=0$ to $t=T$, and the vector space has been further extended $T+1$ times. $I$ is the identity matrix, and $\hat{y}_{\text{in}}$ is the initial weight vector corresponding to $u_{\text{in}}$, written as a tensor product. This quantum ODE solver is our primary strategy towards solving stochastic gradient descent equations using quantum computers.}

\textcolor{black}{Although  similar to its continuous version \cite{liu2021efficient}, the distinct differences between our algorithms and 
those of Ref.~\cite{liu2021efficient} are extensively discussed in the supplementary material. More precisely, the discrete contributions will lead to higher order terms in the learning rate $\eta$. In fact, in the continuous case $A$ is linearly depending on various $F$, while in the discrete case, $A$ also has contributions scaling as $\eta^2 F^2 + \eta ^3 F^3 \cdots$ (see the supplementary material for detailed examples). In the limit where $\eta \to 0$, the discrete contributions become identical to the continuous ones. It is also worth clarifying that those discrete contributions go beyond those considered in Ref.~\cite{liu2021efficient}. In fact, although one has to discretize the differential equations eventually in the ODE setup of Ref.~\cite{liu2021efficient}, the time derivative is computed before discretization for higher order Carleman linearization. For instance, $d(u^{\otimes 2}) (t)$ is treated as $du\otimes u+ u\otimes du$ both at the time $t$, while in the discrete case one has to consider contributions both from the $(t+1)$-st step and the $t$-th step. This contribution is the primary difference between the continuous and the discrete ODEs.
Finally, in our problem setup, we always assume that an explicit form of the gradient descent equation is accessible, such that one can construct the Carleman linearization and make it available to quantum devices. This may not always be true for generic complicated classical neural networks whose complexity of analytic decomposition might grow with the size of the network. We leave a more thorough treatment to future research.}

\section{Numerical analysis}\label{sec:experiments}
In this part of our work, we focus on providing numerical evidence of a potential quantum enhancement for large-scale machine-learning models. Commercial large-scale LLMs like \texttt{GPT-3} can have $\mathcal{O}(100)$ billion parameters and even more, 
which is challenging as a starting point due to its tremendous computational costs. Instead, here we provide examples of classification and computer vision machine learning models, which are relatively small compared to language models used in industry. Our computational resources allow us to achieve the scale up to $\mathcal{O}(100)$ million, which is both practically minded and reachable. We expect that LLMs and other models will feature a similar behavior to those examples since our algorithm works in general as a replacement for stochastic gradient descent. 

Thus, in order to provide evidence of the functioning 
of our quantum algorithm in the context of 
practically minded machine learning, we 
perform numerical experiments on a state-of-the-art machine vision architectures, namely the so-called \emph{ResNet}, to tentatively outline schemes with a potential quantum enhancement. First, we study a model with $7$ million trainable parameters trained to distinguish images of $100$ classes \cite{cifar100}. We first pre-train the neural network, use the largest 10\% of learned parameters for initialization, and use the quantum ODE system to obtain a sparse output model. We record the Hessian spectra during sparse training, allowing us to track the evolution of an error bound related quantity, given by 
\begin{equation}
\frac{1}{N_c}
\int_{-\infty}^{-0.4}\rho(a)\vert(1+a)^t\vert da+\frac{1}{N_c}\int_{0.4}^{\infty}\rho(a)\vert(1+a)^t\vert da,
\end{equation}
where $\rho$ is the eigenvalue density, $a$ is the negative of Hessian eigenvalues, and $N_c$ is the renormalization constant implicitly defined by
\begin{equation}
    \frac{1}{N_c}
\int_{-\infty}^{-0.4}\rho(a)da+\frac{1}{N_c}\int_{0.4}^{\infty}\rho(a)da=1.
\end{equation}
This error proxy discards small magnitude Hessian eigenvalues because they are close to 0, extremely abundant, and renders the error proxy stationary.

\begin{figure} [ht]
   \begin{center}
   \includegraphics[width=8.5cm]{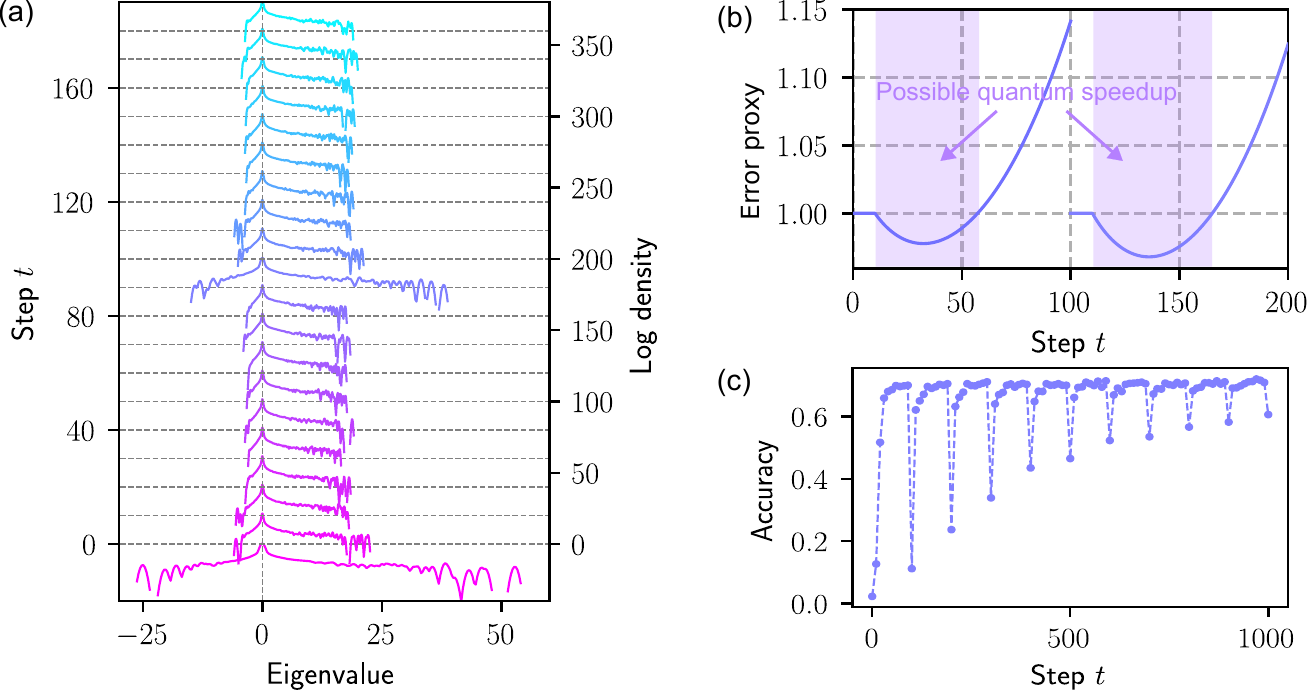}
   \end{center}
   \caption{(a) - (c) Numerical results on ResNet as a function of step. Each step corresponds to a step of stochastic gradient descent based on the derivatives of the loss computed from 2048 randomly selected training samples. (a) ResNet Hessian spectra during training. (b) Estimated error proxy during training. (c) Training accuracy evolution for ResNet.}
   { \label{main_figures}}
   \end{figure}

\newtext{This numerical prescription is created according to criteria towards positivity of Hessian eigenvalues (dissipative modes). }More dissipative systems have more positive Hessian eigenvalues, more negative $a$, and a better behaved error proxy.  Specifically, the dissipative nature of the training dynamics initially leads to a reduction in this error proxy, which then gets overtaken by divergent modes and leads to an exponentially increasing error bound as shown in Fig.~\ref{main_figures} (b). This motivates us to download the quantum trained model parameters sparsely and re-upload to the quantum computer to continue training every $100$ steps. The effect of this procedure is that the exponentially increasing error restarts at $0$ after re-uploading, with the side effect of Hessian broadening and accuracy reduction as shown in 
Fig.~\ref{main_figures}.

\begin{figure} [ht]
   \begin{center}
   \includegraphics[width=6cm]{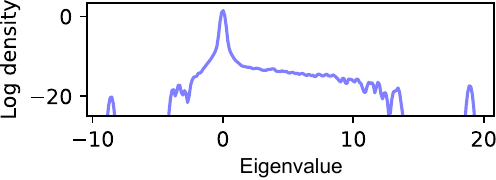}
   \end{center}
   \caption{Hessian of the pruned $103$ million parameter model immediately after pruning without any additional training.}
   { \label{100MHessian}}
   \end{figure}

There is another strategy assuming the existence of QRAM. To combat the effect of Hessian broadening on the error proxy, we train the model classically for 10 steps after download before re-uploading of the new dense parameters, during which no training error is accrued. Although classical training has a  cost linear in $n$, it is a small fraction of the entire training process. The accuracy dips immediately after download improves as training progresses, so our quantum training scheme is capable of producing useful sparse models.
Finally, we examine the Hessian of a $103$ million parameter ResNet. We start with a pre-trained model and prune $90\%$ of the parameters. Due to the immense computational cost of computing Hessian for a large machine learning model (a relatively large-scale model for computational vision based on our computational resources), we only benchmark the Hessian spectra to provide evidences of dissipation and potential quantum enhancements. Fig.~\ref{100MHessian} shows the initial Hessian, which clearly shows the dominance of dissipative modes over divergent modes similar to the $7$ million parameter model. Since the Hessian improves with training for the $7$ million parameter model, we believe this is evidence that the $103$ million parameter model will have similarly manageable error growth.

\section{Conclusion and outlook}\label{sec:concoutlook}

In our work, we have provided quantum algorithm strategies that are presumably helpful for solving the (stochastic) gradient descent dynamics for large-scale classical machine learning models, like LLMs such as \texttt{GPT-3}. We identify certain precisely stated dissipative and sparse regimes of the model where quantum devices could meaningfully contribute, providing an end-to-end HHL-type quantum algorithm application that could outperform known classical algorithms. The observation that an efficient classical algorithm for efficiently solving all 
instances of non-linear dissipative differential equations would imply an efficient classical algorithm for any problem that can be solved efficiently by a quantum computer \newtext{(is BQP hard)} \cite{liu2021efficient}
can be seen as \newtext{an argument that our algorithm is implausible} to be de-quantized by classical proposals along the lines of Ref.~\cite{Tang}. Frankly, the core thesis of this work is that a main application of quantum computers may be in the \emph{training of classical neural networks.}

Indeed, we claim that our algorithm might significantly increase the scalability and sustainability of classical large-scale machine-learning models and provide evidence for our claims numerically up to $103$ million training parameters. Our work provides solid theoretical guarantees and intersections with state-of-the-art classical machine learning research. 
It sharply deviates from the mindset of variational quantum algorithms, and instead aims at augmenting classical machine learning by a key quantum step that constitutes 
a bottleneck for the classical training. In a way, it can be seen as 
adding flesh to the expectation 
that quantum formulations of neural networks may lead to new computational tools \cite{Welling}. \textcolor{black}{Specifically, our model requires the sparsity to be kept as a constant (or 
feature a polynomial scaling)
in the size of the model
to maintain 
a potential enhancement, which is consistent with the so-called 
\emph{lottery ticket hypothesis} \cite{frankle2018lottery}. The setup 
is expected to be favorable in large-scale machine learning numerical experiments, although the sparsity ratio will generically decay.}

Our work is expected to open up several potential directions in the field of 
quantum machine learning where one can reasonably hope for algorithmic improvements.
In the supplemental material, we hint at a number of potentially particularly fruitful directions for future research.
%
In short, they include the development of an alternative, time-dependent version during gradient descent trajectories, the identification of better formal criteria for dissipation, 
work on connections to diffusion models in classical machine learning and LLMs \cite{ho2020denoising}, theoretical improvements on the truncated HHL algorithms, and 
the identification of mechanisms of possible quantum speedups beyond notions of 
dissipation. 
We hope that this work can provide some stimulus for this type of research.\\

\subsection*{Acknowledgments}
We thank Senrui Chen, Vedran Dunjko, Aram Harrow, Robert Huang, Daliang Li, John Preskill, Jonah Sherman, Umesh Vazirani, Carl Vondrick, Han Zheng, Sisi Zhou and Peter Zoller for many valuable discussions. This research used the resources of the Argonne Leadership Computing Facility, which is a U.S. Department of Energy (DOE) Office of Science User Facility supported under Contract DE-AC02-06CH11357.
J.~L.~is supported in part by International Business Machines (IBM) Quantum through the Chicago Quantum Exchange, and the Pritzker School of Molecular Engineering at the University of Chicago through AFOSR MURI (FA9550-21-1-0209).
M.~L.~acknowledges support from DOE Q-NEXT.
J.-P.~L.~acknowledges the support by the NSF (grant CCF-1813814, PHY-1818914), an NSF QISE-NET triplet award (DMR-1747426), an NSF QLCI program (OMA-2016245), a Simons Foundation award 
(No.~825053), and the Simons Quantum Postdoctoral Fellowship. 
Y.~A.~acknowledges support from the Office of Science, U.S. Department of Energy, under contract DE-AC02-06CH11357 at Argonne National Laboratory and Defense Advanced Research Projects Agency (DARPA) under Contract No.~HR001120C0068.
J.~E.~acknowledges funding of the 
ERC (DebuQC), the 
BMBF (Hybrid), the Quantum Flagship (Millenion, PasQuans2), the BMWK (PlanQK, EniQma), the Munich Quantum Valley (K-8), the QuantERA (HQCC), the 
DFG (The Berlin Mathematics Research Center MATH+ (EXC-2046/1, project ID: 390685689), CRC 183), and the Einstein Research Foundation (Einstein Research Unit on Quantum Devices).
L.~J.~acknowledges support from the the ARO(W911NF-23-1-0077), ARO MURI (W911NF-21-1-0325), AFOSR MURI (FA9550-19-1-0399, FA9550-21-1-0209), AFRL (FA8649-21-P-0781), DoE Q-NEXT, NSF (OMA-1936118, ERC-1941583, OMA-2137642), NTT Research, and the Packard Foundation (2020-71479).

\subsection{Data availability statement}
The full code and data for this work are available. 

\subsection{Inclusion and ethics statement} 
All subjects gave their informed consent for inclusion before they participated in the study. The study has been conducted in accordance with the Declaration of Helsinki. This work obviously does not include any data obtained by experiments with living beings.

\bibliographystyle{naturemag}

\newpage

 
\pagebreak
\clearpage
\foreach \x in {1,...,\the\pdflastximagepages}
{
	\clearpage
	

\section*{Supplementary material}

# Supplemental material:
# Towards provably efficient quantum algorithms for large-scale machine learning models

## I. MOTIVATION

### A. Quantum machine learning

Quantum computing is considered one of the most compelling alternatives of von Neumann architectures in the post-Moore era [1]. Making use of quantum features such as superposition and entanglement along the computation, quantum computing is expected to outperform its classical counterparts for many specific tasks, including factoring [2], database search [3], the simulation of complex quantum systems [4], sampling [5–7], and tasks of linear algebra [8]. Those algorithms might be useful in multiple areas, while applications in quantum machine learning might combine several of them at the same time. *Quantum machine learning* [9] utilizes quantum devices to run suitably modified machine learning algorithms. It has been argued that it may have the potential to lead to dramatically faster algorithms compared to what classical machine learning algorithms can provide in certain scenarios. However, substantial limitations exist for currently known quantum machine learning models.

- There are many studies about hybrid quantum-classical implementations of quantum machine learning, so-called *variational quantum algorithms* [10]. Relevant algorithms of such a type have shown potential applications in quantum computational chemistry [10–13], in combinatorial optimization [14, 15], and data-driven learning tasks [16, 17]. However, there is increasing evidence that such variational algorithms run on near-term quantum devices [11] are challenged by noise beyond logarithmic depth [18–21] (even though in some instances, noise can even be of some help [22]). Those algorithms might still be implementable on fault-tolerant quantum computing devices, but it is to date unclear whether and to what extent they are generically expected to have a quantum advantage compared to classical ones.

- Full-scale quantum machine learning algorithms have at the same time been established in the fault-tolerant setting, including steps of principle component analysis [23] (which has been largely de-quantized in the meantime [24]), quantum kernel methods [25, 26], or notions of quantum machine learning of quantum physical states and systems [27, 28]. Those algorithms are promising when fault-tolerant quantum devices will eventually be available, and some may even be equipped with theoretical guarantees. For example, it has been shown that quantum computers fare better in notions of distribution learning [29, 30] and applications of kernel methods to achieve robust quantum speed-ups in supervised machine learning [25]. However, to be fair, those algorithms are still quite far away from the realm of applicability of state-of-the-art classical machine learning applications. To start with, most data in common applications arising in everyday life are classical, not quantum. So if quantum computing requires quantum data to have a speedup, the applicability in our current world, especially from commercial perspectives, seems narrow. In other instances, the data have to be highly structured. It is to date not clear if those theoretical results are helpful for the general classical machine learning community at the moment. For instance, Ref. [25] presents an algorithm showing a robust quantum advantage in notions of machine learning, but it requires constructing a problem in the discrete logarithmic scenario, and there is no immediate state-of-the-art classical machine learning usage for such structured distributions as far as the authors know. Similar in spirit, theoretical contributions about other quantum machine learning works [31]—despite them contributing substantially to our conceptual and mathematical understanding—often provide information-theoretic bounds using statistical learning theory which might miss intricacies of practical real-world applications [32].

The algorithm designed in this work manages to overcome several of the above issues. Compared to existing literature, our work can be seen as contributing the following improvements.

- Instead of decoupling from applications of real-world classical machine learning, our algorithm is relatively practical and expected to be beneficial to the state-of-the-art classical machine learning community, touching upon one of the most important tasks of the current interest: large-scale machine learning models, including *large language models* (LLM) [33–35]. In order to validate our theory, our work also contains a practically minded simulations of around 100 million classical parameters, which is, as far as we know, a record for the quantum computing literature at the moment.

- The performance of our algorithm is, at least in several aspects, amenable to rigorous guarantees. Our algorithm is based on a provably efficient quantum algorithm, the HHL [8] algorithm, which promises to feature exponentially faster performance in sparse matrix inversions over known classical counterparts. Based on the HHL algorithm and Carleman linearization introduced for solving *dissipative ordinary differential equations* (ODEs) [36], we constructively show that there is an algorithm with poly-logarithmic performance in the number of model parameters and linear or quadratic dependence in the number of iterations when the training is sparse, dissipative, and close to a local minimum of the cost function. Moreover, in the more realistic scenarios, we also have solid theoretical bounds on the performance, which are supported by our numerical simulations.

- In the sparse training setup, our work could be applicable to classical data as well and is not relying on QRAM [37], while in the case of dense training, QRAM is required to ensure fast uploading, and tomographic methods could be a natural pruning approach for further classical machine learning steps. Under the assumptions of sparse training [38–41], we could create efficient interfaces between classical and quantum processors. Moreover, our algorithm only requires that interface to be established *once* for totally $T$ iterations, which maximally reduces the burden of transmissions between classical and quantum worlds. Since our inputs and outputs are sparse in the quantum device, we do not necessarily need access to QRAM, which might be challenging to build in the large-scale and fault-tolerant regime [42, 43]. On the other hand, in the dense training task, QRAM could be used to ensure efficient uploading from the classical processors to quantum devices, and in the example of Section II D, the tomographic algorithm could read $r$ components with largest norms with high probability, thus could serve a pruning method for dense or sparse training in the classical processor in the follow-up steps naturally.

- Our algorithm can be seen as a *quantum-classical hybrid algorithm*, but this is meant in a very different way than what is commonly considered in variational quantum algorithms. The quantum algorithm introduced here tackles a key step in otherwise classical machine learning algorithms.

One of the critical concepts in our work is sparse training, which makes the quantum implementation feasible and practical. We will summarize the sparse training perspectives in Section I B.

### B. Making large-scale machine learning sustainable

Large-scale deep learning models have been shown to be an efficient method for enhancing performance across various tasks [33–35]. Those models are widely used to interact with humans [44], to solve mathematical problems [45], in the digital arts [46]. After all, they might after all be revolutionary for the entire society. Large-scale models, especially LLMs, are beyond reasonable doubt, some of most impressive and cutting-edge technologies in the world. However, training such models with considerable numbers of parameters is costly and can have high carbon emissions. For instance, twelve millions dollars and over five-hundred tons of $CO_2$ equivalent emissions are being produced to train `GPT-3` [47]. It is hard to imagine that, in order to build such a machine, we have to use the equivalent of the annual electricity energy consumption of a small city. If one were to vote for the specific areas where quantum computing devices could contribute, LLMs might be the most promising playgrounds.

In our work, we make a substantial step towards such a topic, where quantum and classical computing might both help in the regime of *sparse training*, namely, training an otherwise classical network with most parameters being zero or very close to zero. Recently, sparse training has emerged as a promising direction to reduce the training cost and improve the inference efficiency [38]. One may say that the future of deep learning is sparse: By inducing sparsity in neural networks, we are able to train models at scale with unprecedented efficiency, paving the way for the next-generation artificial intelligence architectures. Sparse training is also a field where multiple research efforts have intersections, including sparse fine-tuning [48, 49], sparse-scratch [39–41], different pruning techniques and the techniques built on the *lottery ticket hypothesis* [50, 51]. One of the intuitive reasons for the advantages of sparse training is apparent: The task itself only fundamentally requires a few data and the current artificial neural networks are highly over-parametrized. In principle, the parameters required for given tasks should be fixed, hence in large-scale tasks, such parameters should be set to zero (see Ref. [52] for a comprehensive discussion).

Appealingly, sparsity is also one of the requirements of the HHL algorithm to obtain an exponential improvement over classical algorithms [52]. Moreover, the sparsity is also required for establishing interfaces between classical and quantum devices: The uploading step of dense quantum states necessitates realizations of QRAM, whereas downloading dense quantum states necessitates readings of tomography, which requiring significant resources for generic dense quantum states but efficient for sparse states. See more discussions around Section II D and Section V A.

Fig. 1 illustrates a typical process that elicidates how modern large-scale machine learning models work. The early stage involves pre-training whose data inputs and data processing are typically dense (which might be still sparse after the technology improves). In those processes, since they are more data-dependent, it will be harder to think about how quantum technologies could possibly help. The second process is the actual training where quantum technologies might contribute more substantially, where also another key factor of our quantum algorithm comes into play: dissipation.

When applying the ODE algorithm related to Ref. [36], dissipation is an essential condition. In fact, the algorithm presented in Ref. [36] employs linearization techniques to convert a system non-linear ODEs (in our case, the stochastic gradient descent equations in machine learning) into a sparse linear system problem to which the HHL algorithm applies. If the system is dissipative, the linearization error is dissipative as well and thus exponentially decaying, making it possible to linearize the system in a controllable fashion. In practical machine learning algorithms that are able to generalize efficiently (the generalization regime [52]), we find that the actual training process undergoes two phases: First, the neural network is actually learning non-trivial and significant examples. The system becomes dissipative, and equivalently, sensitive during the phase. Later, when the system accumulates enough data, the model becomes less dissipative.

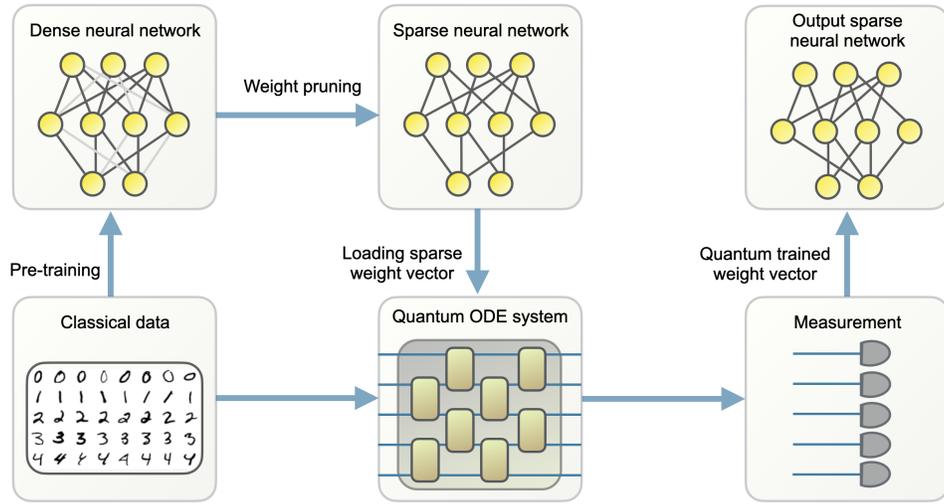

FIG. 1. A possible learning processes in large-scale models. A dense neural network is pre-trained classically. The neural network weights are then pruned and only a small fraction is preserved. A quantum ordinary difference equation system that corresponds to the sparse training dynamics is created using the sparse weight parameters and the training data. To allow for a quantum enhancement, the system must be sparse and dissipative. Measurement on the solution state is performed to obtain the final trained weight parameters, which is used to construct a trained classical sparse neural network.

In our quantum algorithm, we quantify the level of dissipation by the positivity of Hessian matrices during gradient descent (more precisely, *quantum Carleman matrices* (QCM), and the above phenomena might be intuitively explained from well-established information bottleneck theory [52–54]. When the system is actually learning knowledge, the system is more sensitive, and the trace of the Fisher information matrix tends to be higher (see Ref. [52]). When the learning is effectively ending, the Fisher information is approaching zero. Because the Hessian matrix is a second-order approximation of the Fisher matrix, the above observation also holds for Hessian matrices. Thus, we identify here important and specific circumstances where quantum algorithms could meaningfully contribute: This is when the system is both sparse and dissipative. Then we are likely to be able to employ quantum algorithms to accelerate large-scale machine learning models which are super-polynomially more efficient than known classical algorithms for this task. This regime is likely to occur in the core learning process of sparse training.

### C. Theorems

In this subsection, we will primarily state our theorems of the main text and present the respective proofs in full.

**Theorem I.1** (First main theorem). *For a sparse machine learning model with model size n, running T iterations, with the algorithm being fully dissipative (see Eq. (II.1)), there is a quantum algorithm that runs in*

$$O\left(sqT \times \text{poly}\left(m, \frac{2}{\log(1/\|\bar{u}\|)} \log \frac{3\sum_{\ell=2}^{q}(\ell-1)\|\bar{F}_\ell\|}{\|A\|}, \frac{1}{\epsilon}\right)\right) \quad (I.1)$$

*time with error $\epsilon > 0$, with $s$, $\|\bar{u}\|$, $\|\bar{F}_\ell\|$, $q$, $\|A\|$ are model-related parameters, and $m := \log_2(n)$ is the number of qubits. The algorithm also assumes that the learning rate is small, such that the condition in V.30 is satisfied.*

*Moreover, if we assume that the output weight vectors are $r$-sparse, the tomographic cost is $O(m^2 r^3/\epsilon^2)$, ignoring log factors to bring the quantum states to the classical devices. The algorithm ignores the state preparation cost to prepare the initial weight vector in the quantum device, which is also computationally efficient for sparse training.*

**Theorem I.2** (Second main theorem). *For a machine learning model with model size n, running T iterations, with the algorithm*

*being almost dissipative (see Eq. (II.2)), there is a quantum algorithm that runs in*

$$O\left(s^2q^2T^2 \times \text{poly}\left(m, \frac{2}{\log(1/\|\bar{u}\|)} \log \frac{3\sum_{\ell=2}^{q}(\ell-1)\|F_\ell\|}{\|A\|}, \frac{1}{\epsilon}\right)\right) \quad (I.2)$$

*time with error $\epsilon > 0$, with $s$, $\|\bar{u}\|$, $\|\bar{F}_\ell\|$, $q$, $\|A\|$ are model-related parameters, under the assumption of Theorem II.4, and $m = \log_2(n)$ is the number of qubits. The algorithm also assumes that the learning rate is small, such that the condition in V.30 is satisfied.*

*Moreover, again, if we assume that the output weight vectors are r-sparse, the tomographic cost is $O(m^2r^3/\epsilon^2)$ ignoring log factors to bring the quantum states to the classical devices. The algorithm ignores the state preparation cost to prepare the initial weight vector in the quantum device, which is also computationally efficient for sparse training.*

For an algorithm for the efficient uploading of sparse quantum states, see Ref. [55], and for further discussion and details, see Section V A.

### D. Algorithm description

In this subsection, we describe the actual quantum algorithm. A machine learning model is defined partially by a function $\mathcal{L}_\mathcal{A}$, called the *loss function*, as a function of weight vector (variational angle) $\theta \in \mathbb{R}^n$ with components $\theta_\mu$, and the *input training set* $\mathcal{A}$. In instances, we will also write $\boldsymbol{\theta}$ when we emphasize that it is a vector. The weight vector has $n$ components for an $n$-dimensional model. The task is to minimize the function $\mathcal{L}_\mathcal{A}$ by adjusting $\theta$ with $T$ iterations. The most widely utilized algorithm in machine learning is called *(stochastic) gradient descent*. Starting from the initial weight vector $\theta(t = 0)$, we implement the following ordinary differential equation from $t = 0$ to $t = T$,

$$\theta_\mu(t+1) = \theta_\mu(t) - \eta \frac{d\mathcal{L}_\mathcal{A}}{d\theta_\mu}\bigg|_{\theta(t)}. \quad (I.3)$$

Variants of the gradient descent algorithms also include adding random noise $t \mapsto \xi_\mu(t)$ in each step, so-called *stochastic gradient descent algorithms*. One can show that in many cases, at the end of training, $\theta_\mu(t = T)$ could make the loss function $\mathcal{L}_\mathcal{A}(\theta(t = T))$ sufficiently small. In our work, we propose the following quantum algorithm.

---
**Algorithm 1:** Efficient quantum (stochastic) gradient descent algorithm

---
**Input:** An initial weight vector $\theta(0)$, maximal number of iterations $T$, machine learning architecture $\mathcal{L}_\mathcal{A}$ with size $n$.
**Output:** A terminal weight vector $\theta(T)$.

1. Linearize the model $\mathcal{L}_\mathcal{A}$ following Section II with a sparse matrix $M$.
2. Upload the sparse weight vector $\theta(0)$ as a state vector in the quantum device, using the methods of Ref. [55] or more sophisticated architectures like QRAM.
3. Perform the quantum linear system solver described in Section II C once to produce the quantum state vector at the end $t = T$.
4. Perform state tomography described in Section II D to output the classical model parameter $\theta(T)$.

**Result:** $\theta(T)$.

---

As long as the conditions described in Section I E are satisfied, the overall complexity of the efficient quantum (stochastic) gradient descent algorithm scales poly-logarithmically in $n$ and linearly or quadratic in $T$.

### E. Assumptions on the machine learning setup

In this part, we briefly list the assumptions made in the setup.

- *Sparsity of the model.* First, the sparsity of the matrix $F_\ell$ used in the matrix $A$ (see Section II for details) should be a constant when $n$ grows. This is depending on the machine learning model architectures. However, in Section V E, we show that as long as the machine learning model architecture is pruned well, the overall sparsity of QCM is well-controlled.

- *Sparsity of the weight vectors.* The sparsity of the weight vectors is satisfied naturally in sparse training, where the sparsity of the weight vectors stay as a constant and do not scale with the size of the model.

- *Dissipative assumptions.* The model should be sufficiently dissipative, such that either the eigenvalues of *A* are all non-negative, or the number of positive eigenvalues of *A* is much more than the number of negative eigenvalues (see Section II and Section III for detailed theoretical and experimental discussions).

- *Complexity of uploading and downloading quantum states.* The tomographic procedure is efficient due to the assumption of sparse training (see the algorithm described by Section II D).

## II. EFFICIENT QUANTUM ALGORITHMS FOR DISSIPATIVE DIFFERENTIAL EQUATIONS

The idea of Ref. [36] can be directly generalized to differential equations. Roughly speaking, Ref. [36] primarily makes use of Carleman linearization, basically transforming non-linear ODEs into a set of linear ODEs. Then one can discretize ODEs as differential equations. The error in this procedure arises from the following two sources: the linearization error from Carleman linearization, and the discretization error for solving ODEs. If we are interested in the differential equations, the discretization error should not be considered. Other perspectives of the proof will be very similar to the pure ODE case. The results of Ref. [36] can be generalized from the quadratic case to higher-degree cases [56–58]. There are alternative linearization approaches that can be implemented on quantum computers [59–67].

### A. Quantum Carleman linearization and error bounds

In this subsection, we turn to discussing the specific procedure of *quantum Carleman linearization* [36]. As an example, let us consider the $q$-th degree polynomial differential equation given by

$$\delta u = u(t+1) - u(t) = \sum_{\ell=0}^{q} F_\ell u^{\otimes \ell}(t) \tag{II.1}$$

$$= F_q u^{\otimes q}(t) + \ldots + F_2 u^{\otimes 2}(t) + F_1 u(t) + F_0 ,$$

$$u(0) = u_{\text{in}} . \tag{II.2}$$

We have

$$F_0 \in \mathbb{R}^{n \times 1}, \tag{II.3}$$

$$u = (u_1, \ldots, u_n) \in \mathbb{R}^n, F_1 \in \mathbb{R}^{n \times n} , \tag{II.4}$$

$$u^{\otimes 2} = \left(u_1^2, u_1 u_2, \ldots, u_1 u_n, u_2 u_1, \ldots, u_n u_{n-1}, u_n^2\right) \in \mathbb{R}^{n^2}, F_2 \in \mathbb{R}^{n \times n^2} , \tag{II.5}$$

$$u^{\otimes 3} = \left(u_1^3, u_1^2 u_2, \ldots, \right) \in \mathbb{R}^{n^3}, F_3 \in \mathbb{R}^{n \times n^3} , \tag{II.6}$$

$$\ldots$$

$$u^{\otimes q} = \left(u_1^q, u_1^{q-1} u_2, \ldots, \right) \in \mathbb{R}^{n^q}, F_q \in \mathbb{R}^{n \times n^q} . \tag{II.7}$$

Here, we assume that $F_{q,\cdots,2,1}$ does not change over time, while $F_0 = F_0(t)$ might depend on time. We further assume that all $F_q$ are at most $s$-sparse. Moreover, one can define

$$A^j_{j+q-1} := F_q \otimes I^{\otimes j-1} + I \otimes F_q \otimes I^{\otimes j-2} + \cdots + I^{\otimes j-1} \otimes F_q , \tag{II.8}$$

$$\ldots$$

$$A^j_{j+1} := F_2 \otimes I^{\otimes j-1} + I \otimes F_2 \otimes I^{\otimes j-2} + \cdots + I^{\otimes j-1} \otimes F_2 , \tag{II.9}$$

$$A^j_j := F_1 \otimes I^{\otimes j-1} + I \otimes F_1 \otimes I^{\otimes j-2} + \cdots + I^{\otimes j-1} \otimes F_1 , \tag{II.10}$$

$$A^j_{j-1} := F_0 \otimes I^{\otimes j-1} + I \otimes F_0 \otimes I^{\otimes j-2} + \cdots + I^{\otimes j-1} \otimes F_0 . \tag{II.11}$$

We can now formulate the so-called *quantum Carleman linearization*. (There is another contribution due to the discrete nature of the dynamics. We will discuss it in Section V B. The contribution is expected to be ignored if the learning rate is small.)

$$\delta \begin{pmatrix} u \\ u^{\otimes 2} \\ u^{\otimes 3} \\ \vdots \\ u^{\otimes (N-1)} \\ u^{\otimes N} \\ \vdots \end{pmatrix} = \begin{pmatrix} A_1^1 & A_2^1 & A_3^1 & \cdots & A_q^1 & & & \\ A_1^2 & A_2^2 & A_3^2 & \cdots & A_q^2 & A_{q+1}^2 & & \\ & A_2^3 & A_3^3 & A_4^3 & \cdots & & & \\ & & \ddots & \ddots & \ddots & & & \\ & & & A_{N-2}^{N-1} & A_{N-1}^{N-1} & A_N^{N-1} & \cdots & \\ & & & & A_{N-1}^N & A_N^N & \ddots & \\ & & & & & & \ddots & \ddots \end{pmatrix} \begin{pmatrix} u \\ u^{\otimes 2} \\ u^{\otimes 3} \\ \vdots \\ u^{\otimes (N-1)} \\ u^{\otimes N} \\ \vdots \end{pmatrix} + \begin{pmatrix} F_0 \\ 0 \\ 0 \\ \vdots \\ 0 \\ 0 \\ \vdots \end{pmatrix}. \quad (\text{II}.12)$$

Based on these expressions, we can construct the linear system truncated with the parameter $N$ as

$$\delta \hat{y} = A \hat{y} + b, \quad \hat{y}(0) = \hat{y}_{\text{in}}, \quad (\text{II}.13)$$

which is

$$\delta \begin{pmatrix} \hat{y}_1 \\ \hat{y}_2 \\ \hat{y}_3 \\ \vdots \\ \hat{y}_{N-1} \\ \hat{y}_N \end{pmatrix} = \begin{pmatrix} A_1^1 & A_2^1 & \cdots & A_q^1 & 0 & 0 \\ A_1^2 & A_2^2 & \cdots & A_q^2 & A_{q+1}^2 & 0 \\ 0 & A_2^3 & A_3^3 & A_4^3 & 0 & 0 \\ 0 & 0 & \ddots & \ddots & \ddots & 0 \\ 0 & 0 & 0 & A_{N-2}^{N-1} & A_{N-1}^{N-1} & A_N^{N-1} \\ 0 & 0 & 0 & 0 & A_{N-1}^N & A_N^N \end{pmatrix} \begin{pmatrix} \hat{y}_1 \\ \hat{y}_2 \\ \hat{y}_3 \\ \vdots \\ \hat{y}_{N-1} \\ \hat{y}_N \end{pmatrix} + \begin{pmatrix} F_0 \\ 0 \\ 0 \\ \vdots \\ 0 \\ 0 \end{pmatrix}. \quad (\text{II}.14)$$

Here, we specify

$$\hat{y}_j := u^{\otimes j} \in \mathbb{R}^{n^j}, \quad (\text{II}.15)$$

$$\hat{y}_{\text{in}} := \left( u_{\text{in}}, u_{\text{in}}^{\otimes 2}, \ldots, u_{\text{in}}^{\otimes N} \right). \quad (\text{II}.16)$$

Now we define the error of the linearization process as

$$\eta_j(t) := u^{\otimes j}(t) - \hat{y}_j(t). \quad (\text{II}.17)$$

One can show that it will satisfy

$$\delta \eta = A \eta + \hat{b}, \quad \eta(0) = 0, \quad (\text{II}.18)$$

which is

$$\delta \begin{pmatrix} \eta_1 \\ \eta_2 \\ \eta_3 \\ \vdots \\ \eta_{N-1} \\ \eta_N \end{pmatrix} = \begin{pmatrix} A_1^1 & A_2^1 & \cdots & A_q^1 & 0 & 0 \\ A_1^2 & A_2^2 & \cdots & A_q^2 & A_{q+1}^2 & 0 \\ 0 & A_2^3 & A_3^3 & A_4^3 & 0 & 0 \\ 0 & 0 & \ddots & \ddots & \ddots & 0 \\ 0 & 0 & 0 & A_{N-2}^{N-1} & A_{N-1}^{N-1} & A_N^{N-1} \\ 0 & 0 & 0 & 0 & A_{N-1}^N & A_N^N \end{pmatrix} \begin{pmatrix} \eta_1 \\ \eta_2 \\ \eta_3 \\ \vdots \\ \eta_{N-1} \\ \eta_N \end{pmatrix} + \begin{pmatrix} 0 \\ \vdots \\ A_{N+1}^{N-q+2} u^{\otimes (N+1)} \\ \vdots \\ \sum_{\ell=1}^{q-2} A_{N+\ell}^{N-1} u^{\otimes (N+\ell)} \\ \sum_{\ell=1}^{q-1} A_{N+\ell}^{N} u^{\otimes (N+\ell)} \end{pmatrix}. \quad (\text{II}.19)$$

The term $t \mapsto F_0(t)$ could be understood as the noise term in the stochastic gradient descent. The solution of the iterative relation is given by

$$\eta(t+1) = \left( \prod_{i=t}^{0} (1 + A(i)) \right) \eta(0) + \sum_{i=0}^{t} \left( \left( \prod_{j=i}^{i+1} (1 + A(j)) \right) \hat{b}(i) \right). \quad (\text{II}.20)$$

Thus

$$\|\eta(t+1)\| \leq \left\|\left(\prod_{i=t}^{0}(1+A(i))\right)\eta(0)\right\| + \left\|\sum_{i=0}^{t}\left(\left(\prod_{j=t}^{i+1}1+A(j)\right)\hat{b}(i)\right)\right\|. \qquad \text{(II.21)}$$

We call the matrix $A(t)$ here the *quantum Carleman matrix* (QCM). The dimension of the QCM is given by

$$\Delta \equiv n + n^2 + \cdots + n^N = \frac{n^{N+1}-n}{n-1} = O\left(n^N\right). \qquad \text{(II.22)}$$

Assuming that $A(t)$ has $\Delta$ eigenvalues denoted by $a_\mu(t)$, $\mu \in [\Delta]$.

First, we assume that $A(t)$ does not change over time, $A(t) = A$, which corresponds to a time-independent $F_0$. More generally, we use the index notation $\mu, \nu, \ldots \in [\Delta]$. So we have

$$\eta_\mu(t+1) = \sum_{\mu_1,\mu_2}(1+a_{\mu_1})^{t+1}S^{-1}_{\mu,\mu_1}S_{\mu_1,\mu_2}\eta_{\mu_2}(0) + \sum_{\mu_1\mu_2}\left((1+a_{\mu_1})^{t+1}-1\right)a_{\mu_1}^{-1}S^{-1}_{\mu,\mu_1}S_{\mu_1,\mu_2}\hat{b}_{\mu_2}. \qquad \text{(II.23)}$$

Here, the matrix $S$ is diagonalizing the QCM as

$$\sum_{\mu_1}S^{-1}_{\mu,\mu_1}a_{\mu_1}S_{\mu_1,\nu} = A_{\mu,\nu}. \qquad \text{(II.24)}$$

Thus, we have

$$\left(\left(\prod_{i=t}^{0}(1+A)\right)\eta(0)\right)_\mu = \left((1+A)^{t+1}\eta(0)\right)_\mu = \sum_{\mu_1\mu_2}(1+a_{\mu_1})^{t+1}S^{-1}_{\mu,\mu_1}S_{\mu_1,\mu_2}\eta_{\mu_2}(0).$$

$$\left(\sum_{i=0}^{t}\left(\left(\prod_{j=t}^{i+1}1+A\right)\hat{b}(i)\right)\right)_\mu = \left(\sum_{i=0}^{t}\left((1+A)^{t-i}\hat{b}(i)\right)\right)_\mu$$

$$= \sum_{i=0,\mu_1\mu_2}^{t}\left(S^{-1}_{\mu,\mu_1}(1+a_{\mu_1})^{t-i}S_{\mu_1,\mu_2}\hat{b}_{\mu_2}(i)\right). \qquad \text{(II.25)}$$

Making use of the sub-multiplicativity of the operator norm and by bounding the right hand side by a largest element, one can bound the second term by

$$\left\|\sum_{i=0}^{t}\left(\left(\prod_{j=t}^{i+1}1+A\right)\hat{b}(i)\right)\right\| \leq \sum_{i=0}^{t}\|(1+A)\|^{t-i}\|\hat{b}(i)\|. \qquad \text{(II.26)}$$

Moreover, this term can be put into a more refined form, since

$$\sum_{i=0,\mu_1,\mu_2}^{t}\left(S^{-1}_{\mu,\mu_1}(1+a_{\mu_1})^{t-i}S_{\mu_1,\mu_2}\hat{b}_{\mu_2}(i)\right) = \sum_{\mu_1,\mu_2}\left(S^{-1}_{\mu,\mu_1}S_{\mu_1,\mu_2}\sum_{i=0}^{t}(1+a_{\mu_1})^{t-i}\hat{b}_{\mu_2}(i)\right). \qquad \text{(II.27)}$$

Assuming that all $b_\mu$ are upper bounded as

$$\left|\hat{b}_{\mu_2}(i)\right| \leq \bar{b}_{\mu_2} \qquad \text{(II.28)}$$

for all $i$, we find

$$\left|\sum_{i=0,\mu_1\mu_2}^{t}\left(S^{-1}_{\mu,\mu_1}(1+a_{\mu_1})^{t-i}S_{\mu_1,\mu_2}\hat{b}_{\mu_2}(i)\right)\right| \leq \sum_{\mu_1\mu_2}\left(|S^{-1}_{\mu,\mu_1}S_{\mu_1,\mu_2}\bar{b}_{\mu_2}|\left|\sum_{i=0}^{t}(1+a_{\mu_1})^{t-i}\right|\right)$$

$$= \sum_{\mu_1\mu_2}\left(|S^{-1}_{\mu,\mu_1}S_{\mu_1,\mu_2}\bar{b}_{\mu_2}|\left|\frac{(a_{\mu_1}+1)^{t+1}-1}{a_{\mu_1}}\right|\right). \qquad \text{(II.29)}$$

At the same time, we can also write

$$\left|\eta_\mu(t+1)\right| \leq \sum_\nu \left|(1+a_\nu)^{t+1}\right| Q_{\mu,\nu} + C_\mu, \tag{II.30}$$

where

$$Q_{\mu,\nu} = \left|S_{\mu,\nu}^{-1} \sum_{\mu_1} S_{\nu,\mu_1} \eta_{\mu_1}(0)\right| + \left|a_\nu^{-1}\right| \left|S_{\mu,\nu}^{-1} \sum_{\mu_1} S_{\nu,\mu_1} \bar{b}_{\mu_1}\right| = \left|a_\nu^{-1}\right| \left|S_{\mu,\nu}^{-1} \sum_{\mu_1} S_{\nu,\mu_1} \bar{b}_{\mu_1}\right|,$$

$$C_\mu = \left|\sum_{\mu_2,\nu} a_\nu^{-1} S_{\mu,\nu}^{-1} S_{\nu,\mu_2} \bar{b}_{\mu_2}\right|. \tag{II.31}$$

Since the matrix $Q$ and the vector $C$ are time independent, the specific time dependence is completely depending on the eigenvalues of $A$. In the most extreme case, all eigenvalues of $A$ satisfy

$$\left|1+a_\mu\right| < 1, \tag{II.32}$$

then we get a bounded $\eta$, in that

$$\|\eta_\mu(t)\| \leq \left\|\sum_\nu \left|(1+a_\nu)^{t+1}\right| Q_{\mu,\nu}\right\| + \|C_\mu\| \Rightarrow \|\eta(t)\| \leq \|Q\| + \|C\|. \tag{II.33}$$

Actually, we also have

$$\|C_\mu\| \leq \left\|\sum_{\mu_2,\nu} a_\nu^{-1} S_{\mu,\nu}^{-1} S_{\nu,\mu_2} \bar{b}_{\mu_2}\right\| \Rightarrow \|C\| \leq \|A\|^{-1} \|\bar{b}\|, \tag{II.34}$$

$$\|Q_{\mu,\nu}\| \leq \left\|\left|a_\nu^{-1}\right| S_{\mu,\nu}^{-1} \sum_{\mu_1} S_{\nu,\mu_1} \bar{b}_{\mu_1}\right\| \Rightarrow \|Q\| \leq \|A\|^{-1} \|\bar{b}\|. \tag{II.35}$$

Now we have the following theorem.

**Theorem II.1** (Fully dissipative). *If $\|1+A\| < 1$ and $\|\hat{b}(t)\|$ is bounded by $\|\bar{b}\| < \infty$, then we have*

$$\|\eta(T)\| \leq \|Q\| + \|C\| \tag{II.36}$$

*and*

$$\|Q\| \leq \|A\|^{-1} \|\bar{b}\|, \tag{II.37}$$
$$\|C\| \leq \|A\|^{-1} \|\bar{b}\|, \tag{II.38}$$

*where $T$ is the total number of iterations.*

Moreover, we can formulate a probabilistic version. We assume that there are $\Delta_+$ eigenvalues $a_\mu$ satisfying $|1+a_\mu| < 1$. Without loss of generality, we assume that

$$|1+a_\nu| < 1 : 1 \leq \nu \leq \Delta_+, \tag{II.39}$$
$$|1+a_\nu| \geq 1 : \nu > \Delta_+. \tag{II.40}$$

Moreover, we define $\Delta_- := \Delta - \Delta_+$, and we write

$$[\Delta_+] := \{1 \leq \nu \leq \Delta_+\}, \tag{II.41}$$
$$[\Delta_-] := \{\nu > \Delta_+\}. \tag{II.42}$$

Thus, we have

$$\sum_\nu \left|(1+a_\nu)^{t+1}\right| Q_{\mu,\nu} + C_\mu = \sum_{\nu_+ \in [\Delta_+]} \left|(1+a_{\nu_+})^{t+1}\right| Q_{\mu,\nu_+} + \sum_{\nu_- \in [\Delta_-]} \left|(1+a_{\nu_-})^{t+1}\right| Q_{\mu,\nu_-} + C_\mu. \tag{II.43}$$

Then, the exponential convergence will still be approximately valid, if

$$\sum_{\nu_+ \in [\Delta_+]} \left|(1+a_{\nu_+})^{t+1}\right| Q_{\mu,\nu_+} \gg \sum_{\nu_- \in [\Delta_-]} \left|(1+a_{\nu_-})^{t+1}\right| Q_{\mu,\nu_-}. \tag{II.44}$$

This condition means that, weighted by $Q_{\mu,\nu}$, if the contribution of diverging modes in QCM (eigenvalues satisfying $|1+a_\nu| \ge 1$) is significantly smaller than the contribution of the dissipative modes (eigenvalues satisfying $|1+a_\nu| < 1$), then we still get the convergent error $\eta$.

Let us give a simple example at this point. Assuming that we only have two typical $a_\mu$, $a_+$ as a dissipative mode, and $a_-$ as a divergent mode. Furthermore, assuming that $Q_{\mu,\nu}$ are equal for all components. We have

$$\Delta_+ \left|(1+a_+)^{t+1}\right| \gg \Delta_- \left|(1+a_-)^{t+1}\right| \tag{II.45}$$

which will imply

$$\left|\frac{1+a_+}{1+a_-}\right|^{t+1} \gg \frac{\Delta_-}{\Delta_+}. \tag{II.46}$$

Replacing $t+1$ with the total number of iterations $T$, we get a constraint for the time $T$, namely that

$$T \ll \frac{\log \frac{\Delta_-}{\Delta_+}}{\log \left|\frac{1+a_+}{1+a_-}\right|} \tag{II.47}$$

has to hold. This translates into

$$\Delta_+ \gg \frac{\Delta}{\left|\frac{1+a_+}{1+a_-}\right|^T + 1}. \tag{II.48}$$

This calculation gives a precise condition: We have choose the number of dissipative modes to be much larger than the number of diverging modes in the QCM. One can easily generalize the above argument to different modes. for the $\mu$ component, one could derive the condition for the total number of iteration $T_\mu$, as

$$T_\mu \ll \frac{\left(\sum_{\nu_- \in [\Delta_-]} \log |Q_{\mu,\nu_-}| - \sum_{\nu_+ \in [\Delta_+]} \log |Q_{\mu,\nu_-}|\right)}{\left(\sum_{\nu_+ \in [\Delta_+]} \log |(1+a_{\nu_+})| - \sum_{\nu_- \in [\Delta_-]} \log |(1+a_{\nu_-})|\right)}. \tag{II.49}$$

Now we care mostly about $\eta_1$, which is the scalar difference between the original non-linear equation solution and the linearized version, so we can take $\mu = 1$. Finally, in terms of mathematical statement, we can replace $\gg$ as $>$, since the bounds stated can be directly established with $>$. Thus, we arrive at the following claim.

**Theorem II.2** (Almost dissipative). *If $\left\|\hat{b}(t)\right\|$ is bounded by $\left\|\bar{b}\right\| < \infty$, and,*

$$\sum_{\nu_+ \in [\Delta_+]} \left|(1+a_{\nu_+})^T\right| Q_{\mu,\nu_+} > \sum_{\nu_- \in [\Delta_-]} \left|(1+a_{\nu_-})^T\right| Q_{\mu,\nu_-}. \tag{II.50}$$

*Then, we have*

$$\|\eta(T)\| \le 2 \|Q\| + \|C\| \tag{II.51}$$

*and*

$$\|Q\| \le \|A\|^{-1} \left\|\bar{b}\right\|, \tag{II.52}$$
$$\|C\| \le \|A\|^{-1} \left\|\bar{b}\right\|. \tag{II.53}$$

We now turn to discussing the time-dependent cases, which correspond to settings of stochastic gradient descent. Similarly as before, we have the diagonalization

$$\sum_{\mu_1} S^{-1}_{\mu,\mu_1}(i) a_{\mu_1}(i) S_{\mu_1,\nu}(i) = A_{\mu,\nu}(i), \tag{II.54}$$

and we find

$$\left(\left(\prod_{i=t}^{0}(1+A(i))\right)\eta(0)\right)_{\mu} =$$
$$\sum_{\nu}\sum_{\alpha_t,\alpha_{t-1},\ldots,\alpha_1,\alpha_0} f^{\mu,\nu}_{\alpha_t,\alpha_{t-1},\ldots,\alpha_1,\alpha_0}(t;0)\left(1+a_{\alpha_t}(t)\right)\left(1+a_{\alpha_{t-1}}(t-1)\right)\ldots\left(1+a_{\alpha_0}(0)\right)\eta_{\nu}(0) \tag{II.55}$$

where

$$f^{\mu,\nu}_{\alpha_t,\alpha_{t-1},\ldots,\alpha_1,\alpha_0}(t;0) :=$$
$$\sum_{\mu_t,\mu_{t-1},\ldots,\mu_1} S^{-1}_{\mu\alpha_t}(t) S_{\alpha_t,\mu_t}(t) S^{-1}_{\mu_t,\alpha_{t-1}}(t-1) S_{\alpha_{t-1},\mu_{t-1}}(t-1)\ldots S^{-1}_{\mu_2,\alpha_1}(1) S_{\alpha_1,\mu_1}(1) S^{-1}_{\mu_1,\alpha_0}(0) S_{\alpha_0,\nu}(0). \tag{II.56}$$

Furthermore, we find

$$\left(\sum_{i=0}^{t}\left(\prod_{j=t}^{i+1} 1+A(j)\right)\hat{b}(i)\right)_{\mu} =$$
$$\sum_{i=0}^{t}\sum_{\nu}\sum_{\alpha_t,\alpha_{t-1},\ldots,\alpha_{i+2},\alpha_{i+1}} f^{\mu,\nu}_{\alpha_t,\alpha_{t-1},\ldots,\alpha_{i+2},\alpha_{i+1}}(t;i+1)$$
$$\times\left(1+a_{\alpha_t}(t)\right)\left(1+a_{\alpha_{t-1}}(t-1)\right)\ldots\left(1+a_{\alpha_{i+1}}(i+1)\right)\hat{b}_{\nu}(i). \tag{II.57}$$

Thus, the structure is similar as before, if we have sufficiently small numbers of divergent modes, and much more dissipative modes, we will have similar bounds to guarantee that $\eta$ is decaying and gives us the exponentially converging errors.

One might still worry about how realistic this approach is for practically useful machine learning problems. An assumption we could consider is to assume a distribution of the QCM eigenspectra. For simplicity, we assume that $a_\nu$ is independently following the normal distribution $\mathcal{N}(\bar{a},\sigma)$ (with the measure $Da_\nu$). So we get

$$\left(\sum_{\nu} Q_{\mu,\nu}\int Da_\nu(1+a_\nu)^{t+1}\right)+C_\mu$$
$$= (-i)^{t+1} 2^{\frac{t+1}{2}} \sigma^{t+1} U\left(\frac{1}{2}(-t-1),\frac{1}{2},-\frac{(1+\bar{a})^2}{2\sigma^2}\right)\left(\sum_{\nu} Q_{\mu,\nu}\right)+C_\mu. \tag{II.58}$$

Here, $U$ is the $U$-confluent hypergeometric function. If $\bar{a}>0$ (the divergent case), the formula will quickly be dominated by the exponential divergence for large $t$,

$$\left(\sum_{\nu} Q_{\mu,\nu}\int Da_\nu(1+a_\nu)^{t+1}\right)+C_\mu$$
$$\approx \left((1+a)^{t+1}+\frac{t(t+1)}{2}(1+a)^{t-1}\sigma^2+\ldots\right)\left(\sum_{\nu} Q_{\mu,\nu}\right)+C_\mu. \tag{II.59}$$

If $\bar{a}<0$ (the dissipative case), the value of $\sigma$ matters to see if the error is bounded or not. If $\sigma$ is large, we might still get divergence. If $\sigma$ is small enough, the divergence will not show up since it is dominated by the exponential decay $(1+\bar{a})^{t+1}$. For a given value of $\bar{a}$ and $\sigma$, one can plot the time dependence from the confluent hypergeometric function to estimate the error. For instance, one can evaluate

$$\Omega(t,\bar{a},\sigma) := \left|\frac{g(t,\bar{a},\sigma)}{g(0,\bar{a},\sigma)}\right| \tag{II.60}$$

to check the convergence/divergence properties, where

$$g(t, \bar{a}, \sigma) := (-i)^{t+1} 2^{\frac{t+1}{2}} \sigma^{t+1} U\left(\frac{1}{2}(-t-1), \frac{1}{2}, -\frac{(1+\bar{a})^2}{2\sigma^2}\right). \tag{II.61}$$

A particularly interesting example is where $\bar{a} = 1$. In the case where $\sigma < 1$, one can estimate

$$g(t, \bar{a} = 1, \sigma) \approx 1 + \frac{t(t+1)}{2}\sigma^2 + \ldots. \tag{II.62}$$

The expression of $g$ is now dominated by the quadratic term. Thus, the error is quadratic in time $t + 1$, which is much more mild than exponential divergences.

The mathematics is easier when we consider the independent uniform distribution with the range $[\bar{a} - \sigma_a, \bar{a} + \sigma_a]$ (note that now $\sigma_a$ is not exactly the standard deviation). We have

$$\left(\sum_\nu Q_{\mu,\nu} \int Da_\nu (1 + a_\nu)^{t+1}\right) + C_\mu$$
$$\approx \left(\frac{(1 + \bar{a} + \sigma_a)^{t+2} - (1 + \bar{a} - \sigma_a)^{t+2}}{2\sigma_a(t+2)}\right)\left(\sum_\nu Q_{\mu,\nu}\right) + C_\mu. \tag{II.63}$$

Thus, one can get exponential convergence if $-1 < \bar{a} + \sigma_a < 0$. One can use those criteria to numerically check the property of the bounds. Finally, we estimate the bound on $\|\bar{b}\|$. We can express this as

$$\|\hat{b}(t)\| \leq \sum_{\ell'=1}^{q-1} \sum_{\ell=1}^{q-\ell'} \left\|A_{N+\ell}^{N-\ell'+1} u^{\otimes(N+\ell)}\right\|. \tag{II.64}$$

Using

$$\left\|A_{N+\ell}^{N-\ell'+1}\right\| \leq (N - \ell' + 1)\|F_{\ell+\ell'}\|, \tag{II.65}$$

we get

$$\|\hat{b}(t)\| \leq \sum_{\ell'=1}^{q-1} \sum_{\ell=1}^{q-\ell'} (N - \ell' + 1)\|F_{\ell+\ell'}\|\|u\|^{N+\ell} \leq \|u\|^{N+1} \sum_{\ell'=1}^{q-1} \sum_{\ell=1}^{q-\ell'} (N - \ell' + 1)\|F_{\ell+\ell'}\|$$
$$= \|u\|^{N+1} \sum_{\ell=2}^{q} \|F_\ell\|\left((\ell-1)N - \frac{(\ell-1)(\ell-2)}{2}\right)$$
$$\leq N\|u\|^{N+1} \sum_{\ell=2}^{q} (\ell-1)\|F_\ell\|, \tag{II.66}$$

with $N > q$. Moreover, we can assume that

$$\|u(t)\| \leq \|\bar{u}\| < \infty. \tag{II.67}$$

Namely, the condition is that $t \mapsto u(t)$ is bounded in time. By scaling variables, one could without loss of generality, assume that

$$0 < \|\bar{u}\| < 1. \tag{II.68}$$

So we set

$$\|\bar{b}\| := N\|\bar{u}\|^{N+1} \sum_{\ell=2}^{q} (\ell-1)\|F_\ell\| \tag{II.69}$$

and we have

$$\|A\|^{-1} \|\bar{b}\| = N\|A\|^{-1} \|\bar{u}\|^{N+1} \sum_{\ell=2}^{q} (\ell - 1) \|F_\ell\|, \tag{II.70}$$

$$\|Q\| \leq N\|A\|^{-1} \|\bar{u}\|^{N+1} \sum_{\ell=2}^{q} (\ell - 1) \|F_\ell\|, \tag{II.71}$$

$$\|C\| \leq N\|A\|^{-1} \|\bar{u}\|^{N+1} \sum_{\ell=2}^{q} (\ell - 1) \|F_\ell\|. \tag{II.72}$$

So in all cases, we find

$$\|\eta(T)\| \leq 3N\|A\|^{-1} \|\bar{u}\|^{N+1} \sum_{\ell=2}^{q} (\ell - 1) \|F_\ell\|. \tag{II.73}$$

Therefore, if we assume that the linearization error is mostly $\epsilon_{\text{le}}$, we get

$$3N\|A\|^{-1} \|\bar{u}\|^{N+1} \sum_{\ell=2}^{q} (\ell - 1) \|F_\ell\| = \epsilon_{\text{le}}. \tag{II.74}$$

For this reason, there exists a positive integer $N$ such that

$$N\|\bar{u}\|^{N+1} \leq \|\bar{u}\|^{\frac{1}{2}N}. \tag{II.75}$$

So we obtain

$$N \geq \frac{2}{\log(1/\|\bar{u}\|)} \log \frac{3 \sum_{\ell=2}^{q} (\ell - 1) \|F_\ell\|}{\|A\| \epsilon_{\text{le}}}, \tag{II.76}$$

for a sufficiently small $\epsilon_{\text{le}} > 0$.

### B. The QCM spectra

#### 1. Dominance from Hessian matrices

In this part, we will derive how the QCM spectra could be related to the Hessian in the machine learning problems. For simplicity, let us firstly consider

$$F_0 = F_{\ell \geq 2} = 0. \tag{II.77}$$

Now, assuming that for

$$\text{spec}(F_1) = (f_a)_{a=1}^{n}, \tag{II.78}$$

we have

$$\text{spec}(A_j^j) = (f_{a_1} + f_{a_2} + \ldots + f_{a_j})_{a_1, a_2, \ldots a_j = 1}^{n}, \tag{II.79}$$

$$\dim(A_j^j) = n^j. \tag{II.80}$$

Thus we can identify the spectra of the QCM to be

$$\text{spec}(A) = \bigcup_{j=1}^{N} (f_{a_1} + f_{a_2} + \ldots + f_{a_j})_{a_1, a_2, \ldots a_j = 1}^{n}, \tag{II.81}$$

$$\dim(A) = \frac{n^{N+1} - n}{n - 1} = \Delta. \tag{II.82}$$

We can now bring these findings into the context of actual machine learning models. For a machine learning model with trainable parameters $\theta_a$ and the loss function $\mathcal{L}_A$, we have (see Section III for further information, and $\supset$ indicates that we are calculating the first order term in the Taylor expansion)

$$\delta\theta_a = -\eta\frac{\partial \mathcal{L}_A}{\partial \theta_a} \supset -\eta \sum_b \frac{\partial^2 \mathcal{L}_A}{\partial\theta_a \partial\theta_b}\theta_b$$

$$-\eta \frac{\partial^2 \mathcal{L}_A}{\partial\theta_a \partial\theta_b} = (F_1)_{a,b}$$

$$-\eta \times \mathrm{spec}\left(\frac{\partial^2 \mathcal{L}_A}{\partial\theta_a \partial\theta_b}\right) = -\eta \times \mathrm{spec}\,(H) = \mathrm{spec}\,(F_1)\ . \tag{II.83}$$

Thus, more positive Hessian eigenvalues reflect more dissipative modes in QCM.

### 2. Perturbation theory

We here put the findings into the context of perturbation theory. We take the case where

$$\|F_{0,\ell\geq 2}\| \ll \|F_1\|\ . \tag{II.84}$$

In this case, one can use perturbation theory in quantum mechanics to compute the shift of the QCM spectra from the perturbation of $F_{0,\ell\geq 2}$. We assume that

$$F_{0,\ell\geq 2} = g_{0,\ell\geq 2}\hat{F}_{0,\ell\geq 2}, \tag{II.85}$$
$$0 < g_{0,\ell\geq 2} \ll 1. \tag{II.86}$$

We discuss notions of perturbation theory for such cases. For simplicity, we assume that $f_a$s are all different numbers, and there is no extra degeneracy except the permutations,

$$\mathrm{spec}\left(A_j^j\right) \supset \bigcup_{\sigma;a}\left(f_{\sigma(a_1)} + f_{\sigma(a_2)} + \ldots + f_{\sigma(a_j)}\right)^n_{a_1,a_2,\ldots a_j=1}\ . \tag{II.87}$$

Here, we combine the degenerate eigenvalues and their corresponding eigenspaces induced by the permutation $\sigma$ of the list $a_1, a_2, \cdots, a_j$. We give examples about non-degenerate and degenerate perturbations as the following.

- *Non-degenerate perturbation theory:* Consider the eigenvalues

$$\mathrm{spec}\left(A_j^j\right) \supset (j \times f_a)_{a=1}^n\ . \tag{II.88}$$

Those eigenvalues are non-degenerate if those eigenvalues of $A$ have no extra degeneracy and are generically chosen. In this case, the minimal corrections are the second-order terms in perturbation theory $O(g^2)$. We denote the eigenvalue as

$$a_{j,\mu_0} = j \times f_a \tag{II.89}$$

so that the eigenvalue perturbation is given by

$$a_{j,\mu_0} \mapsto a_{j,\mu_0} + g_0^2 \sum_{a_{j-1,\mu}\neq a_{j,\mu_0}} \frac{\left|\langle j,\mu_0|\hat{A}_{j-1}^j|j-1,\mu\rangle\right|^2}{a_{j-1,\mu} - a_{j,\mu_0}}$$

$$+ \sum_{\ell=1}^{q-1} g_{\ell+1}^2 \sum_{a_{j+\ell,\mu}\neq a_{j,\mu_0}} \frac{\left|\langle j,\mu_0|\hat{A}_{j+\ell}^j|j+\ell,\mu\rangle\right|^2}{a_{j+\ell,\mu} - a_{j,\mu_0}}. \tag{II.90}$$

Here,

$$\hat{A}_{j+q-1}^j = \hat{F}_q \otimes I^{\otimes j-1} + I \otimes \hat{F}_q \otimes I^{\otimes j-2} + \cdots + I^{\otimes j-1} \otimes \hat{F}_q\ , \tag{II.91}$$

$$\cdots$$

$$\hat{A}_{j+1}^j = \hat{F}_2 \otimes I^{\otimes j-1} + I \otimes \hat{F}_2 \otimes I^{\otimes j-2} + \cdots + I^{\otimes j-1} \otimes \hat{F}_2\ , \tag{II.92}$$

$$\hat{A}_{j-1}^j = \hat{F}_0 \otimes I^{\otimes j-1} + I \otimes \hat{F}_0 \otimes I^{\otimes j-2} + \cdots + I^{\otimes j-1} \otimes \hat{F}_0\ , \tag{II.93}$$

and the sum is taken over all the eigenvalues of $A^{j-1}_{j-1}$ and $A^{j+\ell}_{j+\ell}$ for $1 \leq \ell \leq q-1$, denoted by $a_{j-1,\mu}$ and $a_{j+\ell,\mu}$. The eigenvectors are given by $|j-1,\mu\rangle$ and $|j+\ell,\mu\rangle$. Similarly, the eigenvector corresponding to the eigenvalue $a_{j,\mu_0}$ is $|j,\mu_0\rangle$. We have

$$\dim\{|j-1,\mu\rangle\} = n^{j-1}, \tag{II.94}$$
$$\dim\{|j,\mu_0\rangle\} = n^{j}, \tag{II.95}$$
$$\dim\{|j+\ell,\mu\rangle\} = n^{j+\ell} \tag{II.96}$$

and $A^j_{j-1} \in \mathbb{R}^{n^j \times n^{j-1}}$, $A^j_{j+\ell} \in \mathbb{R}^{n^j \times n^{j+\ell}}$. Thus, the inner product is well-defined.

- *Degenerate perturbation theory:* We consider the eigenvalue

$$\mathrm{spec}\left(A^j_j\right) \supset \left(a_{j,\sigma} := f_{\sigma(a_1)} + f_{\sigma(a_2)} + \ldots + f_{\sigma(a_j)}\right)_\sigma \tag{II.97}$$

with

$$1 \leq a_1 < a_2 \ldots < a_j \leq n. \tag{II.98}$$

Here, $\sigma$ is a permutation among $a_1, a_2, \cdots, a_j$. Those eigenvalues are degenerate, with respective degeneracy $j!$, since there are $j!$ permutations in total. Thus, for a given $j$, all $a_{j,\sigma}$ are equal, where we could denote as $a_j$. In this case, the degenerate perturbation theory gives a possible eigenvalue splitting with respective to $\sigma$,

$$a_j \mapsto a_j + g_0^2 \sum_{a_{j-1,\mu} \neq a_{j,\sigma}} \frac{\left|\langle j,\sigma|\hat{A}^j_{j-1}|j-1,\mu\rangle\right|^2}{a_{j-1,\mu} - a_{j,\sigma}}$$
$$+ \sum_{\ell=1}^{q-1} g^2_{\ell+1} \sum_{a_{j+\ell,\mu} \neq a_{j,\sigma}} \frac{\left|\langle j,\sigma|\hat{A}^j_{j+\ell}|j+\ell,\mu\rangle\right|^2}{a_{j+\ell,\mu} - a_{j,\sigma}}. \tag{II.99}$$

Here, $|j,\sigma\rangle$ is the eigenvector assigned with the permutation $\sigma$ in the eigenspace with the eigenvalue $a_j$. Thus we get at most $j!$ different $a_{\sigma,j}$s in the second-order degenerate perturbation theory, where we have $j!$ eigenvalue splitting at most.

### C. Quantum linear system solvers

Until know, we have established how to obtain a linearized differential equation. We now aim at approximately solving this equation by making use of a quantum computer. Given the initial condition described by a quantum state, we aim to provide a quantum-state description of the solution given the terminal time $T$. Quantum computers can identify the solution of a $N_h$-dimensional linear system in Hilbert space. Originally proposed by Harrow, Hassidim, and Lloyd [8], quantum linear system algorithms [8, 68–73] have been developed to provide a quantum state encoding the solution with complexity $\mathrm{poly}(\log N_h)$. Building on this body of work, generalizations to solve linear ordinary and partial differential equations [74–78] have been proposed.

**Theorem II.3** (Quantum linear system algorithm [73]). *Let $Mx = b$ be a system of linear equations, where $M$ is an $s_M$-sparse $N_h \times N_h$ matrix with condition number $\kappa := \|M\|\|M^{-1}\|$. Given a sparse matrix oracle for $M$ and an oracle for $|b\rangle$, there exists a quantum algorithm which produces the state vector $|M^{-1}b\rangle$ within error $\epsilon_{\mathrm{HHL}} > 0$ using gate complexity*

$$O\left(s_M \kappa \cdot \mathrm{poly}\log(N_h/\epsilon_{\mathrm{HHL}})\right). \tag{II.100}$$

We consider the linearized differential equation

$$\delta\hat{y} = \hat{y}(t+1) - \hat{y}(t) = A\hat{y}(t) + b, \quad \hat{y}(0) = \hat{y}_{\mathrm{in}}, \tag{II.101}$$

with iterations $t \in [T+1]_0 := \{0, 1, \ldots, T\}$, which we rewrite as one linear system $MX = B$ in the form

$$\begin{pmatrix} I & & & & \\ -(I+A) & I & & & \\ & \ddots & \ddots & & \\ & & -(I+A) & I & \\ & & & -(I+A) & I \end{pmatrix} \begin{pmatrix} \hat{y}(0) \\ \hat{y}(1) \\ \vdots \\ \hat{y}(T-1) \\ \hat{y}(T) \end{pmatrix} = \begin{pmatrix} \hat{y}_{\mathrm{in}} \\ b \\ \vdots \\ b \\ b \end{pmatrix}. \tag{II.102}$$

$M^{-1}$ can actually be written as

$$\begin{pmatrix} I & & & & \\ I+A & I & & & \\ (I+A)^2 & I+A & I & & \\ \cdots & \cdots & \cdots & & \\ (I+A)^T & \cdots & (I+A)^2 & I+A & I \end{pmatrix}, \tag{II.103}$$

such that we find the upper bounds

$$\|M^{-1}\| \leq \|I\| + \|I+A\| + \ldots + \|(I+A)^T\|. \tag{II.104}$$

As discussed in Theorem II.1, if all eigenvalues of $A$ satisfy

$$|1 + a_\mu| < 1, \tag{II.105}$$

or, in other words, $\|I+A\| < 1$ holds true, then

$$\kappa = \|M\|\|M^{-1}\| \leq (1 + \|I+A\|) \left( \sum_{t=0}^{T} \|I+A\|^t \right) = 2(T+1). \tag{II.106}$$

Given a sparse matrix oracle, the cost of encoding an $s$-sparse $N_h \times N_h$ matrix $A$ is found to be $O(sq)$. The upper bound on the condition number shows that $\kappa = 2(T+1)$, which indicates the gate complexity of solving the desired linear system is $O(sqT \cdot \text{poly}(\log(N_h/\epsilon_{\text{HHL}})))$.

Furthermore, we can ask what happens when we have divergent modes? Here, we notice that the HHL algorithm can be implemented only on the well-conditioned eigenspaces by adding auxiliary qubits (see Ref. [8], discussion section), which could be likely applicable to the adiabatic method [73] as well. In the context of the original work [8] when solving $MX = b$, we set

$$|b\rangle := \sum_j \beta_j |u_j\rangle \tag{II.107}$$

where $|u_j\rangle$ are normalized eigenvectors of $M$. The algorithm will give

$$|x\rangle = \left( \sum_j \left| \lambda_j^{-1} \beta_j \right|^2 \right)^{-1/2} \sum_j \lambda_j^{-1} \beta_j |u_j\rangle \tag{II.108}$$

up to the given precision. Instead, one can add auxiliary qubits such that the output is

$$|\tilde{x}\rangle = \left( \sum_{j:|\lambda_j| \geq 1/\kappa_a} \left| \lambda_j^{-1} \beta_j \right|^2 + \sum_{j:|\lambda_j| < 1/\kappa_a} |\beta_j|^2 \right)^{-1/2}$$
$$\times \left( \sum_{j:|\lambda_j| \geq 1/\kappa_a} \lambda_j^{-1} \beta_j |u_j\rangle |\text{well}\rangle + \sum_{j:|\lambda_j| < 1/\kappa_a} \beta_j |u_j\rangle |\text{ill}\rangle \right). \tag{II.109}$$

Here, $\kappa_a > 0$ is an effective positive number and may not necessarily be the condition number of $M$. When $\|M\| = 1$, and $\kappa_a = \kappa$, the algorithm effectively returns to Theorem II.3 regardless of those auxiliary qubits. The auxiliary states $|\text{well}\rangle$ and $|\text{ill}\rangle$ will be post-selected at the end of the algorithm. If we set a general $\kappa_a$, the error on the state vector $|x\rangle$ can be estimated by the fidelity

$$\langle x|\tilde{x}\rangle = \frac{\sum_{j:|\lambda_j| \geq 1/\kappa_a} \left| \lambda_j^{-1} \beta_j \right|^2}{\left( \sum_j \left| \lambda_j^{-1} \beta_j \right|^2 \right)^{1/2} \left( \sum_{j:|\lambda_j| \geq 1/\kappa_a} \left| \lambda_j^{-1} \beta_j \right|^2 + \sum_{j:|\lambda_j| < 1/\kappa_a} |\beta_j|^2 \right)^{1/2}}. \tag{II.110}$$

In an actual machine learning process, the weights and biases might depend highly randomly. A good model to reflect this randomness is to assume that $|b\rangle$ is a Haar-random state vector. Namely, $|b\rangle = U_{\text{Haar}}|0\rangle$ where $U_{\text{Haar}}$ is distributing uniformly in the whole unitary group (see Ref. [79]). We denote $\rho_0 = |0\rangle\langle 0|$ and

$$\Lambda^{-1} = \text{diag}_j(\lambda_j^{-1}), \tag{II.111}$$

$$\bar{\Lambda}^{-1} = \text{diag}_{j:|\lambda_j|\geq 1/\kappa_a}(\lambda_j^{-1}) \oplus \text{diag}_{j:|\lambda_j|<1/\kappa_a}(0), \tag{II.112}$$

where $\Lambda^{-1}$ is the diagonal matrix of all eigenvalue inversions of $M$, and $\bar{\Lambda}^{-1}$ is the *pruned* version of $\Lambda^{-1}$. This is the step of a sparse truncation: Here all eigenvalues satisfying $|\lambda_j| \geq 1/\kappa_a$ are included, while other eigenvalues satisfying $|\lambda_j| < 1/\kappa_a$ are excluded and set to zero, leading to a sparse setting. We have

$$\int dU_{\text{Haar}} \sum_{j:|\lambda_j|\geq 1/\kappa_a} \left|\lambda_j^{-1}\beta_j\right|^2 = \int dU_{\text{Haar}} \langle 0|U_{\text{Haar}}^{-1}\bar{\Lambda}^{-1\dagger}\bar{\Lambda}^{-1}U_{\text{Haar}}|0\rangle$$

$$= \int dU_{\text{Haar}} \text{Tr}(U_{\text{Haar}}^\dagger \bar{\Lambda}^{-1\dagger}\bar{\Lambda}^{-1}U_{\text{Haar}}\rho_0) = \frac{1}{\Delta(T+1)}\text{Tr}\left(\bar{\Lambda}^{-1\dagger}\bar{\Lambda}^{-1}\right). \tag{II.113}$$

Since the expression of $\langle x|\tilde{x}\rangle$ only cares about the inner product that involves a pair of $U_{\text{Haar}}$ and its Hermitian conjugate, based on the *Page theorem* [79, 80], it is equivalently assuming that $|b\rangle$ is a superposition of all computational basis states. Similarly, we have

$$\int dU_{\text{Haar}} \sum_j \left|\lambda_j^{-1}\beta_j\right|^2 = \frac{1}{\Delta(T+1)}\text{Tr}\left(\Lambda^{-1\dagger}\Lambda^{-1}\right), \tag{II.114}$$

$$\int dU_{\text{Haar}} \sum_{j:|\lambda_j|\geq 1/\kappa_a} \left|\lambda_j^{-1}\beta_j\right|^2 + \int dU_{\text{Haar}} \sum_{j:|\lambda_j|<1/\kappa_a} \left|\beta_j\right|^2$$

$$= \frac{1}{\Delta(T+1)}\text{Tr}\left(\bar{\Lambda}^{-1\dagger}\bar{\Lambda}^{-1}\right) + \frac{\#\left(j:|\lambda_j|<1/\kappa_a\right)}{\Delta(T+1)}, \tag{II.115}$$

where $\#\left(j:|\lambda_j|<1/\kappa_a\right)$ refers to the number of eigenvalues that are smaller than the cutoff $1/\kappa_a$.

We can use the above formulas to estimate the inner product (II.110). Note that we average over the numerator, the denominator separately, inside or outside the square root separately. This is the difference between the annealed average and its opposite [79, 81], where the error is likely negligible in the limit of large $(T+1)\Delta$, the dimension of the Hilbert space. We can specify the situation in the machine learning case, where we have $\Delta_+$ dissipative modes with eigenvalues $\lambda_+$. We can simply take $\kappa_a = 2(T+1)$ as the upper bound, such that the divergent modes are ill-conditioned. In this case, we estimate the fidelity as

$$\langle x|\tilde{x}\rangle = \frac{\Delta_+ \sum_{t=0}^T (1+a_+)^t}{((T+1)\Delta)^{1/2}\left(\Delta_+ \sum_{t=0}^T (1+a_+)^t + (T+1)\Delta_-\right)^{1/2}}$$

$$= \frac{\Delta_+ \frac{(a_++1)^{T+1}-1}{a_+}}{((T+1)\Delta)^{1/2}\left(\Delta_+ \frac{(a_++1)^{T+1}-1}{a_+} + (T+1)\Delta_-\right)^{1/2}}. \tag{II.116}$$

In the limit where $a_+ \to 0^-$, the formula is simply

$$\langle x|\tilde{x}\rangle = \frac{\Delta_+}{\Delta}, \tag{II.117}$$

which indicates that only a constant number of errors will be introduced if we set $\kappa_a$ linear in $T$, where the constant is referring to the ratio of dissipative modes. This observation is independent of the number of iterations $T$ or $(T+1)$.

Finally, we claim that filtering out ill-conditioned eigenvalues might cause additional overhead on the algorithm. The naive usage of the auxiliary qubits in Ref. [8] introduces extra $O(\kappa_a^2)$ gates. Further works about this filtering algorithm require rigorous improvements on auxiliary qubits and post-selection using amplitude amplification and adapting the above framework to the adiabatic setting [73] which we leave for future research.

**Theorem II.4** (Truncated quantum linear system algorithm). *Let $Mx = b$ be a system of linear equations, where $M$ is an $s_M$-sparse $N_h \times N_h$ matrix. We define the effective condition number $\kappa_a > 0$ that filters out all eigenvalues of $M$ smaller than $1/\kappa_a$, and define the truncated solution $|\tilde{x}\rangle$ in Eq. (II.109). Given a sparse matrix oracle for $M$ and an oracle for $|b\rangle$, there exists a quantum algorithm which produces the state vector $|\tilde{x}\rangle$ within error $\epsilon_{\text{HHL}} > 0$ using a gate complexity of*

$$O\left(s_M^2 \kappa_a^2 \text{ polylog}(N_h/\epsilon_{\text{HHL}})\right). \tag{II.118}$$

According to the analysis in Eqs. (II.116) and (II.117), it suffices to choose the effective condition number as $\kappa_a = 2(T+1)$ so that the fidelity of $|\tilde{x}\rangle$ to $|x\rangle$ is close to 1. The gate complexity of solving the desired linear system is $O(s^2 q^2 T^2 \, \text{polylog}(N_h/\epsilon_{\text{HHL}}))$. Overall, combining with the error bound in differential equations, we get the gate complexity

$$O\left( sqT \times \text{poly}(\log n, \frac{2}{\log(1/\|\bar{u}\|)} \log \frac{3 \sum_{\ell=2}^{q} (\ell-1) \|F_\ell\|}{\|A\| \epsilon_{\text{le}}}, \frac{1}{\epsilon_{\text{HHL}}}) \right) \tag{II.119}$$

for $\|1 + A\| < 1$, and

$$O\left( s^2 q^2 T^2 \times \text{poly}(\log n, \frac{2}{\log(1/\|\bar{u}\|)} \log \frac{3 \sum_{\ell=2}^{q} (\ell-1) \|F_\ell\|}{\|A\| \epsilon_{\text{le}}}, \frac{1}{\epsilon_{\text{HHL}}}) \right) \tag{II.120}$$

for dissipative QCM.

### D. Tomographic recovery and sparsity

In this subsection, we briefly discuss notions of tomographic recovery and what role they take in sparse training. The requirement of sparsity for the implementation of our quantum algorithms has two components: The sparsity of weight and bias vectors $\theta$ on the one hand and the sparsity of the QCM matrix $A$ on the other hand. The sparsity of the QCM matrix originates from the machine learning architecture and Taylor expansion, while the sparsity of weight matrices comes from the nature of sparse training. Here, we primarily discuss the sparsity of the output weight vector, where we denote by $r$ the number of non-zero entries of a vector with Hilbert space dimension $N$. The vectors to be recovered tomographically in this step are of the form

$$|x\rangle = \sum_{\alpha=1}^{r} v_\alpha |p_\alpha\rangle, \tag{II.121}$$

normalized as

$$\sum_{\alpha=1}^{r} |v_\alpha|^2 = 1, \tag{II.122}$$

where each $|p_\alpha\rangle$ is a computational basis vector of a system of $m = \log_2(n)$ qubits. That is to say, the state is $r$-sparse in the sense that there are at most $r$ non-vanishing coefficients. So in a first step, it has to be determined which computational basis vectors feature and hence which coefficients are non-zero, as the sparsity pattern in this sense that is not known. This problem can be identified as an instance of the *coupon collector's problem* (see, for instance, Ref. [82]). The algorithm in this step goes as follows. One performs projective measurements in the computational Pauli-$Z$ basis and records observed patterns. This procedure is repeated until all $r$ different patterns are observed. The original coupon collector's problem is recovered in this setting exactly in the situation of the state vector $|x\rangle = r^{-1/2} \sum_{\alpha=1}^{r} |p_\alpha\rangle$, as an equal superposition, as then each vector reflecting a different pattern is observed with the same probability. For this problem, the expected number of measurements is known to scale as $O(r \log r)$. For the state vector as specified in Eq. (II.121), the weights are different for each pattern. For this weighted coupon collector's problem, the expected number of steps until all patterns have been observed is also known [82]: The expected number of repetitions is given by

$$\int_0^{+\infty} \left( 1 - \prod_{\alpha=1}^{r} \left(1 - e^{-|v_\alpha|^2 y}\right) \right) dy. \tag{II.123}$$

It takes a moment of thought that for equal probability for each probability one arrives at the minimum expectation value. There is a subtlety that is worth commenting upon that relates to small non-vanishing weights. If weights converge to zero, the expectation values of observing all patterns including those of the small weights clearly diverges. In this situation, one can consider only the significant weights above a given threshold $\delta > 0$, which will lead to an efficient algorithm for an approximation up to weights larger than $\delta$, as a detailed analysis shows. Overall, again, this step can be performed making use of $O(r \log r)$ many measurements.

After this step has been taken of identifying the at most $r$ relevant computational basis vectors, corresponding to a $r$-sparse weight state, we can pursue a variant of state tomography to extract the full pure quantum state including phases. We suggest to

do so by implementing a variant of the *shadow estimation scheme* of Ref. [27] making use of random quantum circuits involving the $n$-qubit Clifford group. For the physical implementation of this, $O(m^2)$ entangling gates are necessary. After implementing random Clifford circuits, one performs computational basis measurements, giving rise to outcomes $o$ reflecting computational basis vectors $|o\rangle$ of the $m$ qubit system. For the input $|x\rangle\langle x|$, the output of this random quantum channel is obtained as

$$\mathcal{M}(|x\rangle\langle x|) = \mathbb{E}\left(U^\dagger |o\rangle\langle o| U\right), \tag{II.124}$$

where $U \in \mathcal{C}_m$ is from the $m$-qubit Clifford group. To compute the inverse as

$$|x\rangle\langle x| = \mathbb{E}\left(\mathcal{M}^{-1}(U^\dagger |o\rangle\langle o| U)\right), \tag{II.125}$$

one has to compute overlaps of computational basis vectors and stabilizer states obtained from applying Clifford circuits to computational basis vectors. Such overlaps can be classically efficiently computed with an effort polynomial in $m$. The inverse map is found to be

$$\mathcal{M}^{-1}(X) = (2^m + 1)X - I. \tag{II.126}$$

For a state vector of the form as in Eq. (II.121), one can take samples of random Clifford circuits applied to $|x\rangle$, and sample from the output distribution obtained from computational basis measurements, creating a classical shadow. From these data, one can infer about the elements $v_\alpha$ for $\alpha = 1, \ldots, r$, requiring an overall effort of $O(m^2 r^3/\epsilon^2)$ in the random measurement procedure, for a recovery with the 1-norm error $\epsilon > 0$, ignoring logarithmic factors. It also involves a polynomial classical effort in $m$, due to computing overlaps of stabilizer states and computational basis states [83]. Alternative methods are also conceivable, e.g., based on compressing circuits and the methods presented in Ref. [84] based on projected least squares.

In the actual machine learning process, the ratio of pruned parameters in sparse training $\delta_{\text{tr}}(n)$ is scaling with the size of the model $n$. In this case, the weight sparsity is given by $r = n(1 - \delta_{\text{tr}}(n))$. For this to be feasible, we expect that $r$ is a constant in $n$. This requires that

$$\delta_{\text{tr}}(n) = 1 - O\left(\frac{1}{n}\right). \tag{II.127}$$

This will guarantee that the cost of our algorithm will have a poly-logarithmic dependence in $n$.

Finally, we wish to mention that in the case of dense input vectors, the tomographic algorithm will read the $r$ vector components with the largest norms, and this could be interpreted as a natural pruning method for further classical processes. In this case, we are assuming a generic dense training approach, and the input dense weight vectors can be uploaded via QRAM. The downloading process could still be efficient if $r$ is not scaling with $n$, and this could be interpreted as a hybrid quantum-classical pruning method.

## III. MACHINE LEARNING AND NUMERICAL ASSESSMENT

In this section, we present substantial numerical evidence regarding the feasibility of accelerating machine learning with our proposed quantum algorithm. First, we formulate the learning dynamics of a *multilayer perceptron* (MLP) machine learning model as an ODE system and make some error scaling statements. Second, we numerically analyze the convergence characteristics of the quantum algorithm for large vision models using their Hessian spectra. This is because computing the original QCM and diagonalize it for large models is numerically intractable. Lastly, we present numerical evidence supporting the use of the Hessian spectra for convergence analysis.

### A. Training dynamics of MLPs as ODE systems

To analytically understand neural networks, we consider the MLP, a simple neural network architecture (see Ref. [85]). The definition of its elements is given by

$$z_i^{(1)}(x) := b_i^{(1)} + \sum_{j=1}^{n_0} W_{i,j}^{(1)} x_j,$$

$$\text{for} \quad i = 1, \ldots, n_1, \tag{III.1}$$

$$z_i^{(\ell+1)}(x) := b_i^{(\ell+1)} + \sum_{j=1}^{n_\ell} W_{i,j}^{(\ell+1)} \sigma\left(z_j^{(\ell)}(x)\right),$$

$$\text{for} \quad i = 1, \ldots, n_{\ell+1}; \quad \ell = 1, \ldots, L-1, \tag{III.2}$$

where $x$ is the input data vector, $W^{(\ell)}$ and $b^{(\ell)}$ are the $\ell$-th layer trainable weights and biases, $z^\ell$ is the $\ell$-th layer preactivation, $n_\ell$ is the $\ell$-th layer width, and $\sigma$ is a non-linear activation function. In vector notation,

$$z^{(\ell+1)}(x) := b^{(\ell+1)} + W^{(\ell+1)}\sigma\left(z^{(\ell)}(x)\right), \tag{III.3}$$

where $\sigma$ is applied element-wise. During training of the neural network, parameters update according to the gradient descent algorithm on the *mean squared error* (MSE) loss

$$\theta_\mu(t+1) = \theta_\mu(t) - \eta \left.\frac{d\mathcal{L}_\mathcal{A}}{d\theta_\mu}\right|_{\theta(t)}, \tag{III.4}$$

$$\mathcal{L}_\mathcal{A} := \frac{1}{2}\sum_{\alpha \in \mathcal{A}}|z^{(L)}(x_\alpha;\theta) - y_\alpha|^2, \tag{III.5}$$

where $\alpha \in \mathcal{A}$ are elements in the training set $\mathcal{A}$, $y$ are the labels (referred to as ground truths), $z^{(L)}$ are the final predictions from the MLP model, $\eta$ is the training rate, and $\theta$ is the vectorized combination of all weights and biases. In the language of ODE, $u \mapsto \theta$ and

$$\sum_{\ell=0}^{q} F_\ell u^{\otimes \ell}(t) \mapsto -\eta\left.\frac{d\mathcal{L}_\mathcal{A}}{d\theta_\mu}\right|_{\theta(t)}. \tag{III.6}$$

Therefore, we need to express $\eta\frac{d\mathcal{L}_\mathcal{A}}{d\theta_\mu}\big|_{\theta(t)}$ as a polynomial in $\theta$. We have

$$\frac{d\mathcal{L}_\mathcal{A}}{d\theta_\mu} = \sum_{\alpha \in \mathcal{A}}\left[z^{(L)}(x_\alpha) \cdot \frac{dz^{(L)}(x_\alpha)}{d\theta_\mu} - y_\alpha \cdot \frac{dz^{(L)}(x_\alpha)}{d\theta_\mu}\right]. \tag{III.7}$$

Since $x, y$ are fixed, $z^{(\ell)}$ are already functions of $\theta$, what remains to be shown is how to express the derivatives as functions of $\theta$ as well. We find

$$\frac{dz_i^{(L)}(x_\alpha)}{db_j^{(\ell)}} = \frac{dz_i^{(L)}(x_\alpha)}{dz_j^{(\ell)}(x_\alpha)}, \quad \frac{dz_i^{(L)}(x_\alpha)}{dW_{jk}^{(\ell)}} = \frac{dz_i^{(L)}(x_\alpha)}{dz_j^{(\ell)}(x_\alpha)}\sigma\left(z_k^{(\ell-1)}(x_\alpha)\right), \tag{III.8}$$

$$\frac{dz_i^{(L)}(x_\alpha)}{dz_j^{(\ell)}(x_\alpha)} = \sum_{k_{\ell+1},k_{\ell+2},\ldots,k_{L-1}}^{n_{\ell+1},n_{\ell+2},\ldots,n_{L-1}} W_{i,j}^{(L)}W_{k_{L-1}j}^{(L-1)}\ldots W_{k_{\ell+2}j}^{(\ell+2)}W_{k_{\ell+1}j}^{(\ell+1)}$$
$$\times \sigma\left(z_j^{(L-1)}(x_\alpha)\right)'\sigma\left(z_j^{(L-2)}(x_\alpha)\right)'\ldots\sigma\left(z_j^{(\ell+1)}(x_\alpha)\right)'\sigma\left(z_j^{(\ell)}(x_\alpha)\right)'. \tag{III.9}$$

We now have increments of $\theta$ in terms of $\theta$ (and training data), and have obtained a non-linear ODE system. The highest order term of the ODE corresponds to the terms derived from $dz^{(L)}/dW_{j,k}^{(1)}$ and $dz^{(L)}/db_j^{(1)}$. Assuming that the non-linearity $\sigma$ is an order $s_p$-polynomial, then $\sigma^{(1)}$ is an order $s_p$-polynomial in $\theta_\mu$, $z^{(2)}$ has order $s_p + 1$, $\sigma^{(2)}$ has order $s_p(s_p + 1)$, and $\sigma^{(\ell)}$ has order $\sum_{m=1}^{\ell} s_p^m = s_p(s_p^\ell - 1)/(s_p - 1)$. From Eq. (III A), we can conclude that $dz_{i;\alpha}^{(L)}/dz_{j;\alpha}^{(1)}$ is of order

$$O\left((L-1) + \sum_{\ell=1}^{L-1}\left(\frac{s_p(s_p^\ell - 1)}{s_p - 1} - 1\right)\right). \tag{III.10}$$

Since $z$ is one order higher than $\sigma^{(L-1)}$, Eq. (III A) leads to the following result.

**Theorem III.1** (Bound to the order). *For an $L$ layer MLP with activation functions of order $s_p$, the ODE $d\theta_\mu/dt = d\mathcal{L}_\mathcal{A}/d\theta_\mu$ has order*

$$O\left(\left(\frac{s_p(s_p^{L-1} - 1)}{s_p - 1} + 1\right) + \left[(L-1) + \sum_{\ell=1}^{L-1}\left(\frac{s_p(s_p^\ell - 1)}{s_p - 1} - 1\right)\right]\right). \tag{III.11}$$

Thus, for a neural network with bounded polynomial activation functions, the degree $q$ of the ODE is finite.

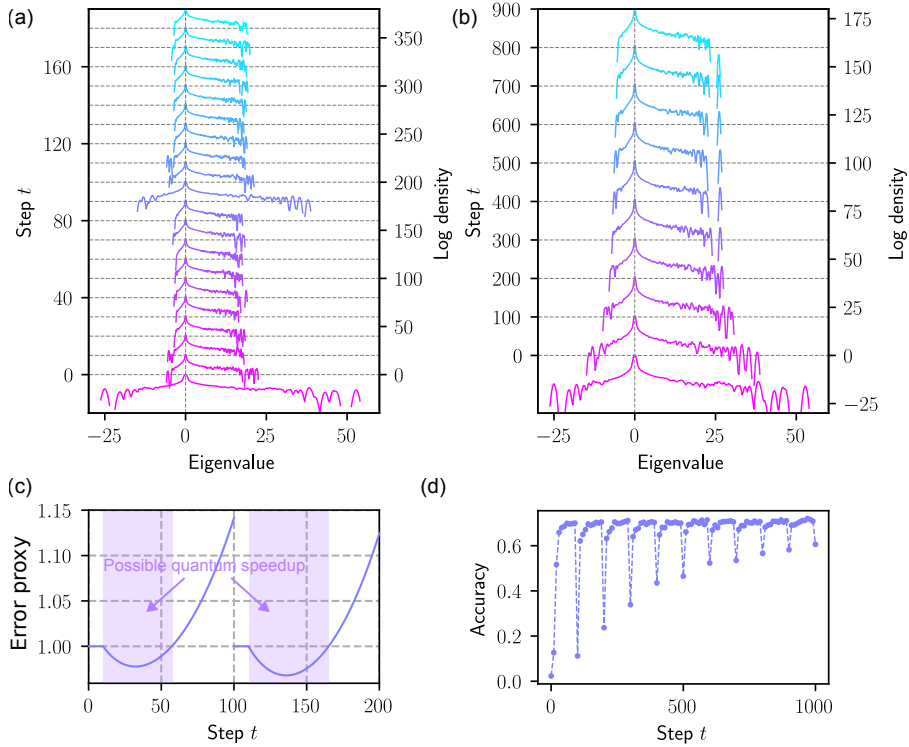

FIG. 2. Numerical results on ResNet as a function of step. Each step corresponds to a step of stochastic gradient descent based on the derivatives of the loss computed from 2048 randomly selected training samples. (a) Hessian spectra of the ResNet during the first 200 steps of training shown for every 10 steps. The neural network is pruned with ratio 0.9 every 100 steps, which corresponds to the sudden broadening of the Hessian spectra. (b) Hessian spectra shown for every 100 steps where spectra broadening take place due to pruning. (c) Estimate of error proxy. The first 10 steps after pruning are trained classically, hence incurring no error. (d) Neural network accuracy during trainig.

### B. Numerical analysis on sparse ResNet for vision

In order to understand the potential of our approach in accelerating training of state-of-the-art large neural networks, we study the training dynamics of the *ResNet* [86] applied to the *CIFAR-100* [87] data set. CIFAR-100 is a set of 60000 images in 100 categories, and the neural network has been trained to classify given images. We first train a dense ResNet, where all parameters are kept as normal, with depth 32 and widen factor of 4. The parameter count for the dense model is 7.451.044. After that, 90% of the parameters are being pruned (set to 0), and a new model is trained. We allow pruned parameters to evolve to non-zero values so that we might obtain a different set of parameters as sparse weights. Further, since the quantum training algorithm deviates from the ideal training trajectory after a number of steps, we download sparse weights (10% of all parameters in this case) every 100 steps from the quantum ODE system and reupload the sparse weights to continue quantum training. We train the ResNet and examine the Hessian spectra at different steps as shown in Fig. 2 (a) and (b), where each step reflects an update using gradient estimated from 2048 randomly selected training samples with a learning rate of 0.01. We observe in Fig. 2 (a) that the Hessian spectra immediately after sparse weights download are broad, but rapidly concentrates and becomes dissipative within 10 steps of training. We also show the Hessian spectra immediately after sparse weights download for the entire training process. The spectra broadening effect of downloads becomes less significant as training progresses. The entire training process contains more dissipative modes than divergent modes as can be seen in the Hessian spectra.

We further quantify the impact of the Hessian spectra on the convergence of the linearized solver. Equation (II.30) gives us a bound on the linearization error. Assuming $Q_{\mu,\nu}$ are the same for all $\mu, \nu$, we test this criterion by examining

$$\text{error proxy} = \frac{1}{\Delta} \sum_{\nu} \left|(1+a_\nu)^t\right|, \tag{III.12}$$

where the $a_\nu$ are the negatives of the Hessian eigenvalues. If all eigenvalues are exactly 1, neither dissipative or divergent, the quantity stays exactly 1. If the divergent contributions outweigh the dissipative contributions, the quantity is greater than 1. We numerically compute the density of Hessian eigenvalues given by $\rho(a)$, and we have $\int_{-\infty}^{\infty} \rho(a)da = 1$. Assuming the Hessian

spectrum stays constant, the error quantity can be estimated as

$$\text{error proxy} = \int_{-\infty}^{\infty} \rho(a)|(1+a)^t|da. \tag{III.13}$$

We estimate the error proxy by sampling the eigenvalue densities at 256 points. Further, since eigenvalues close to 0 are extremely abundant and lead to stationary error proxy, we remove eigenvalues with magnitudes less than 0.4 and renormalize the densities. More explicitly, we define the error proxy as

$$\text{error proxy} := \frac{1}{N_c}\int_{-\infty}^{-0.4} \rho(a)|(1+a)^t|da + \frac{1}{N_c}\int_{0.4}^{\infty} \rho(a)|(1+a)^t|da, \tag{III.14}$$

where

$$\frac{1}{N_c}\int_{-\infty}^{-0.4} \rho(a)da + \frac{1}{N_c}\int_{0.4}^{\infty} \rho(a)da = 1. \tag{III.15}$$

Since the Hessian spectra are broad immediately after download, the error proxy tends to diverge. Therefore, we consider the training scenario where the first 10 steps after download are trained classically without error, and the new dense weights are uploaded to the quantum ODE system with QRAM. As can be seen in Fig. 2 (a), the Hessian spectra after 10 steps following download are already concentrated and mainly dissipative, which results in less rapidly growing linearization error.

Fig. 2 (c) shows the estimated error proxy. The error proxy stays exactly at 1 for 10 steps after download because the model is trained classically and has no linearization error. Afterwards, we observe that the error proxy initially decreases due to the dominance of dissipative modes. However, as time increases, the exponentially growing contribution from divergent modes eventually takes over, leading to an apparently exploding error. We download the weights before the error becomes too unreasonable, and resumes the classical training and weights reuploading scheme. Since the Hessian spectra for larger step numbers have less divergent contributions as indicated in Fig. 2 (b), we can be fairly certain that the error proxy will be even better managed in later stages of training.

Finally, we examine the *generalization ability* of the model following this training scheme. Fig. 2 (d) shows that the test set accuracy drops after download, but the drop becomes less significant as training progresses. Our quantum ODE training scheme, therefore, produces useful sparse machine learning models. In addition, we examine the Hessian of a 103 million parameter ResNet. We start with a pre-trained model and prune 90% of the parameters. Due to the immense computational cost of computing Hessian for a large machine learning model (a relatively large-scale model for computational vision based on our computational resources), we only benchmark the Hessian spectra to provide evidences of dissipation and potential quantum enhancements. Fig. 3 shows the initial Hessian, which clearly shows the dominance of dissipative modes over divergent modes similar to the 7 million parameter model. Since the Hessian improves with training for the 7 million parameter model, we believe this is evidence that the 103 million parameter model will have similarly manageable error growth.

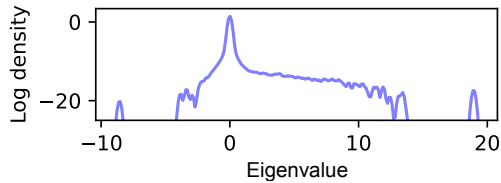

FIG. 3. Hessian of the pruned 103 million parameter model immediately after pruning without any additional training.

### C. Numerical verification of the effect of Hessian on convergence

In this subsection, we provide further evidence that dissipative Hessians actually lead to better convergence. We have simulated systems of quadratic ODEs, where the $F_1$ matrix has a tunable percentage of positive eigenvalues, and the $F_2$ matrix entries are randomly sampled from a uniform distribution $U(-0.015, 0.015)$. To generate random $F_1$ matrices, we first generate positive eigenvalues from $U(0, 1)$ and negative eigenvalues from $U(-0.01, 0)$, form $F_1$ in the diagonal form, and transform its basis using a randomly generated orthogonal matrix. $F_0$ is 0 for all entries.

First, we provide evidence for the suppression of error in time and the QCM order $N$ in the case that all modes are dissipative in Fig. 4(a), which shows the error of a randomly chosen parameter. If all modes are dissipative, an exponential reduction in the

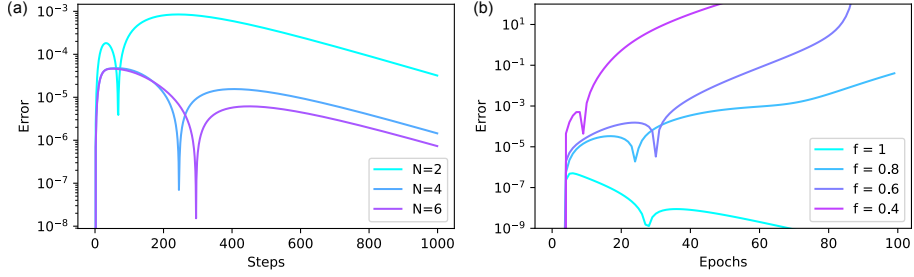

FIG. 4. Numerical evidence for the effect of the Hessian on the approximation error. (a) Error scaling for different QCM orders $N$ for a Hessian with all positive eigenvalues. (b) Error scaling for different fractions $f$ of positive Hessian eigenvalues.

error as time increases is observed due to the exponentially vanishing error bound. Second, for different fractions $f$ of positive Hessian eigenvalues, the error of a randomly chosen parameter for $N = 4$ is shown in Fig. 4(b). For $f < 1$, divergent modes take over and eventually lead to exponentially growing errors in the limit of large times. However, reducing the number of divergent modes significantly suppresses the error, and extends the time period within which the error remains controlled.

## IV. FUTURE IMPROVEMENTS

Our algorithm, as an end-to-end application of the HHL algorithm to the context of quantum machine learning, seems to open up several novel research directions. In this section, we hint at some future improvements biulding on our algorithm that may eventually lead to more effective enhancements of classical machine learning tasks using quantum technologies.

- *Time-dependent version during gradient descent trajectories:* In our work, we have defined the QCMs by Taylor expanding the machine learning models around specific starting points. For models beyond the scale-invariant activation functions, it might, however, be favorable to improve the previous analysis towards a time-dependent picture of generic matrix elements in the QCM matrices. The current strategy we have laid out in Section III is to expand the model around different points of parameters during the gradient descent trajectories. A re-formulation of the existing theory towards fully time-dependent QCM might make our theory more solid and complete.

- *Better criteria for dissipation:* In our current work, we define the contributions of different signs of the eigenvalues of QCM to be the primary criteria of dissipation. These, however, are hard to evaluate in large-scale models. Instead, in practice, we might want to count the contributions from the diagonal elements, or, alternatively, the Hessian eigenvalues as criteriawhen aiming at identifying regimes of possible quantum advantages. The question arises whether there are any better criteria that are both more solid and more computationally efficient.

- *Connections to diffusion models in classical machine learning:* Our work highlights the observation that dissipation is important to obtain such an efficient quantum enhancement. The celebrated diffusion models used in the current LLM [88] are also dissipative systems. It might be interesting to explore their connections.

- *Theoretical improvements on the truncated HHL algorithms:* In Theorem II.4, we propose a truncated version of the HHL algorithm. However, improvements were plausible if one could extend the ideas from Ref. [73] to such a truncated version, which will likely reduce the overhead related to the sparsity and the condition number from $O(s^2 \kappa_a^2)$ down to $O(s\kappa_a)$.

- *Quantum enhancement beyond dissipation:* It is demonstrated in Ref. [36] that there is numerical evidence that the ODE system may not need to feature full dissipation to obtain a quantum speedup. It might be interesting to explore alternative methods to avoid the condition of dissipation and still provide efficient quantum algorithms.

- *Classical counterparts:* In the sparse training process, it might be possible that for certain sparse training schemes with sufficient few non-zero entries in the weight vectors, efficient classical algorithms might also be created to restrict the calculations inside the $r$-dimensional subspaces over the whole weight vector space. It might be interesting also to consider more sophisticated hybrid quantum-classical combinations to ensure maximal performances with low overheads.

- *Large learning rate:* In our theory we assume that the learning rate is small enough and the update of weight vectors is slow. This is a usual setting in machine learning problems. However, the learning rate could be large in principle for specific applications [89, 90]. We give an explicit solution about the Carleman linearization in the discrete case in Section V B. We leave the studies about how the discreteness will affect the performances of machine learning problems in large learning rates for future works.

## V. OTHER COMMENTS

### A. Aaronson's criteria

In this section, we finally briefly comment on our quantum algorithms under the view of Ref. [91]. We regard our quantum algorithm as an end-to-end provable application of the HHL algorithm under certain conditions. Ref. [91] summarizes four perspectives on applying HHL, solving $M|x\rangle = |b\rangle$ and getting $|x\rangle$, into machine learning problems with solid guarantees. We will address how our algorithm takes into consideration about all four conditions:

- *Initial state preparation (resolved by either QRAM or sparse training):* Applying the HHL algorithm requires the efficient implementations of the initial state. In general, QRAM [37] might be required to encode non-sparse classical data into the quantum computer. QRAM as a possible efficient uploading oracle architecture has been extensively studied in Refs. [42, 92–94]. It is practically challenging to realize such a machine in practice [43], although it is realized that polynomially controllable error scaling is possible in certain devices [92]. In our work under the sparse training assumption, the input state is sparse due to the assumption of sparse training, and an efficient algorithm has been constructed without resorting to QRAM. For instance, Ref. [55] proposes an algorithm for loading an $r$-sparse quantum state for use in machine learning models with size $n$, using $O(r^2 \log r \log n)$ classical time, and $O(r \log n)$ quantum circuit depth (which only contains single qubit gates and CNOT gates). Thus, under the sparse training assumption, we are able to avoid using QRAM altogether. On the other hand, in the dense training case where input weight vectors are dense, our analysis is still valid when QRAM is available.

- *The sparsity of the inverted matrix (resolved by machine learning model architectures):* HHL requires the sparsity of $M$ to guarantee the efficiency of the algorithm. In our setup, this issue is related to machine learning architectures. This is closely related to our pruning assumptions. In Section V E, we prove that as long as the machine learning model itself is sufficiently pruned, the matrix $A$ (and naturally $M$) is sufficiently pruned as well. This justifies the sparsity of the HHL matrix from theory to practice in certain cases.

- *Condition number of the inverted matrix (resolved by the dissipative assumptions):* The HHL algorithm also requires a polynomial dependence on the condition number of the matrix $M$ to maintain efficiency. In our setup, the condition number is related to our assumption of dissipation. If our QCM matrix $A$ satisfies $\|1 + A\| < 1$, it is entirely dissipative, and the condition number is linearly bounded by the number of iterations $T$. If not, Theorem II.2 and Theorem II.4 will bound the efficiency of the algorithm at early times of training, as has been shown in Section III.

- *The tomographic overhead of the output state vector (resolved by sparse training):* HHL algorithm provides $|x\rangle$ as a quantum state vector as an output, and tomographic efforts are needed to bring the quantum data back to the classical devices. This issue could be resolved by our assumptions of sparse training, as has been discussed in Section II D. For a $r$-sparse quantum state, the tomographic recovery need $O(m^2 r^3/\epsilon^2)$ additional resource with error $\epsilon > 0$ ignoring logarithmic factors, which is computationally efficient. On the other hand, if we are under the dense training assumption, the tomographic recovery could still work to read $r$ components of the weight vectors with the largest norms, and this is effectively a pruning method toward further training steps in the classical processors.

Overall, our algorithm is considered an end-to-end HHL algorithm applications under our assumptions that provides guarantees satisfying the requirements of the HHL algorithm and offers efficient interfaces between classical and quantum processors.

The potentially exponential quantum enhancement we suggest invites efforts for training general large-scale machine learning models, rather than specific machine learning tasks such as those considered in Refs. [25, 28, 29]. The capability of deep neural network models is far from being well understood, and they may not provide the most efficient resolution for certain tasks due to the over-parameterization. However, the computational overhead in terms of parameterizations $n$ is generally unavoidable and costly in practice. Actually, it seems highly unlikely that any classical algorithm for non-linear dissipative differential equations (and training dynamics) can run in time complexity poly-logarithmic in $n$. If we consider the continuous-time limit of the differential equations as a system of differential equations, and let the dissipation and non-linearity approach zero, then asymptotically we might have a system of linear differential equations with no dissipation. The problem of efficiently simulating non-dissipative linear differential equations is BQP-hard even when the dynamics is entirely unitary, generated by Hamiltonian evolution [95]. In other words, an efficient classical algorithm for non-linear dissipative differential equations would imply a classical algorithms for any problem that can be solved efficiently by a quantum computer, which is considered implausible. A similar argument refers to Section 7 of Ref. [36]. This can be seen as an argument that the de-quantization by classical algorithms along the lines of Ref. [96] of the presented quantum algorithm seems highly implausible, even though to assess this matter could give rise to a fruitful research question.

### B. Discrete contributions from Carleman linearization

#### 1. Elementary deriations

Here we will give a more detail explanation on Eq. (II.12). Say that we have a differential equation,

$$du = f(u),  \tag{V.1}$$

where $d$ is an infinitesimal differential operator. The Taylor expansion tells,

$$f(u) = f(0) + \sum_{\mu} (\partial_\mu f) u_\mu + \sum_{\mu_1 \mu_2} \frac{1}{2} (\partial_{\mu_1 \mu_2} f) u_{\mu_1 \mu_2} + \ldots .  \tag{V.2}$$

There is an alternative way of using this, which is

$$f(u) = \sum_{k=0}^{\Lambda_\infty} \frac{1}{k!} \left( \partial_{\otimes(k)} f \right) u^{\otimes(k)},  \tag{V.3}$$

where

$$\partial_{\otimes(k)} f = \text{vec} \left( \partial_{\mu_1 \mu_2 \ldots \mu_k} f \right),  \tag{V.4}$$

and we use the notation where $\Lambda_\infty = +\infty$. Specifically, when we are dealing with the differential equation

$$du = F_q u^{\otimes(q)}(t) + \ldots + F_2 u^{\otimes(2)}(t) + F_1 u(t) + F_0 .  \tag{V.5}$$

Here, $F_0$ can either be time-dependent or time-independent. Then we can identify

$$\frac{1}{k!} \left( \partial_{\otimes(k)} f \right) = F_k .  \tag{V.6}$$

Using the language of the Kronecker products, we could try to understand Carleman linearization. Let us start with a simple example, given by

$$d \left( u^{\otimes(2)} \right) = d (u \otimes u) = d(u) \otimes u + u \otimes d(u) .  \tag{V.7}$$

One could compute each term as

$$d(u) \otimes u = \sum_{k=0}^{\Lambda_\infty} \left( \text{vec}(F_k) u^{\otimes(k)} \right) \otimes u = \sum_{k=0}^{\Lambda_\infty} \left( \text{vec}(F_k \otimes I) \left( u^{\otimes(k+1)} \right) \right),  \tag{V.8}$$

$$u \otimes d(u) = \sum_{k=0}^{\Lambda_\infty} u \otimes \left( \text{vec}(F_k) u^{\otimes(k)} \right) = \sum_{k=0}^{\Lambda_\infty} \left( \text{vec}(I \otimes F_k) \left( u^{\otimes(k+1)} \right) \right).  \tag{V.9}$$

In general, we get

$$d(u^{\otimes(i)}) = \sum_{k=0}^{\Lambda_\infty - i + 1} G_{i,k} u^{\otimes(k+i-1)},  \tag{V.10}$$

where

$$G_{i,k} = \text{vec}(\sum_{\ell=0}^{i-1} I^{\otimes(\ell)} \otimes F_k \otimes I^{\otimes(i-1-\ell)}) .  \tag{V.11}$$

Now, instead, let us estimate the difference between the discrete difference $\delta u$ and the differential $du$. Starting again with the example we have before, we get

$$\delta \left( u^{\otimes(2)} \right) = \delta (u \otimes u) = u(t+1) \otimes u(t+1) - u(t) \otimes u(t)$$
$$= u(t+1) \otimes (u(t+1) - u(t)) + (u(t+1) - u(t)) \otimes (u(t))$$
$$= u(t+1) \otimes \delta u(t) + \delta u(t) \otimes u(t) .  \tag{V.12}$$

For each term, we have

$$\delta u(t) \otimes u(t) = \sum_{k=0}^{\Lambda_\infty} \left(\text{vec}(F_k) u^{\otimes(k)}(t)\right) \otimes u(t) = \sum_{k=0}^{\Lambda_\infty} \left(\text{vec}(F_k \otimes I) \left(u^{[0,k+1]}(t)\right)\right), \tag{V.13}$$

$$u(t+1) \otimes \delta u(t) = \sum_{k=0}^{\Lambda_\infty} u(t+1) \otimes \left(\text{vec}(F_k) u^{\otimes(k)}\right) = \sum_{k=0}^{\Lambda_\infty} \left(\text{vec}(I \otimes F_k) \left(u^{[1,k]}(t)\right)\right). \tag{V.14}$$

Here, we use the notation

$$u^{[i,j]}(t) := u^{\otimes(i)}(t+1) \otimes u^{\otimes(j)}(t). \tag{V.15}$$

Thus, in general we get

$$\delta\left(u^{\otimes(i)}\right) = \sum_{k=0,\ell=0}^{\Lambda_\infty - i+1, i-1} g_{k,\ell}^i u^{[\ell,k+i-1-\ell]}, \tag{V.16}$$

where

$$g_{k,\ell}^i = \text{vec}\left(I^{\otimes(\ell)} \otimes F_k \otimes I^{\otimes(i-1-\ell)}\right). \tag{V.17}$$

Note that we have the identity

$$G_{i,k} = \sum_{\ell=0}^{i-1} g_{k,\ell}^i. \tag{V.18}$$

For differential equations, we can try to derive the recursion relation for the linearization error. Using the notation in the main text, we denote

$$\eta_j(t) := u^{\otimes(j)}(t) - \hat{y}_j(t). \tag{V.19}$$

With the Taylor expansion truncation at the order $N$, we have

$$\begin{aligned}
d\eta_j(t) &= du^{\otimes(j)}(t) - d\hat{y}_j(t) \\
&= \sum_{k=0}^{\Lambda_\infty - j+1} G_{i,k} u^{\otimes(k+j-1)} - \sum_{k=0}^{N-j+1} G_{i,k} \hat{y}_{k+j-1} \\
&= \sum_{k=0}^{N-j+1} G_{i,k} \eta_{k+j-1} + \sum_{k=N-j+2}^{q} G_{i,k} u^{\otimes(k+j-1)},
\end{aligned} \tag{V.20}$$

which is exactly Eq. (II.19) in the matrix form. Here, we use the fact that $G_{i,k} = 0$ if $k > q$. On the other hand, in the discrete case, we have

$$\begin{aligned}
\delta\eta_j(t) &= \delta u^{\otimes(j)}(t) - \delta\hat{y}_j(t) \\
&= \sum_{k=0,\ell=0}^{\Lambda_\infty - j+1, j-1} g_{k,\ell}^j u^{[\ell,k+j-1-\ell]} - \sum_{k=0}^{N-j+1} G_{i,k} \hat{y}_{k+j-1} \\
&= \sum_{k=0,\ell=0}^{\Lambda_\infty - j+1, j-1} g_{k,\ell}^j u^{[\ell,k+j-1-\ell]} - \sum_{k=0}^{\Lambda_\infty - j+1} G_{i,k} u^{\otimes(k+j-1)} + \sum_{k=0}^{\Lambda_\infty - j+1} G_{i,k} u^{\otimes(k+j-1)} - \sum_{k=0}^{N-j+1} G_{i,k} \hat{y}_{k+j-1} \\
&= \sum_{k=0,\ell=0}^{\Lambda_\infty - j+1, j-1} g_{k,\ell}^j u^{[\ell,k+j-1-\ell]} - \sum_{k=0}^{\Lambda_\infty - j+1} G_{i,k} u^{\otimes(k+j-1)} + \sum_{k=0}^{N-j+1} G_{i,k} \eta_{k+j-1} + \sum_{k=N-j+2}^{q} G_{i,k} u^{\otimes(k+j-1)}.
\end{aligned} \tag{V.21}$$

Thus, we can define the extra piece between $du$ and $\delta u$ as the extra contribution from discretization,

$$\delta\zeta_j(t) \equiv \sum_{k=0,\ell=0}^{\Lambda_\infty - j+1, j-1} g_{k,\ell}^j u^{[\ell,k+j-1-\ell]} - \sum_{k=0}^{\Lambda_\infty - j+1} G_{i,k} u^{\otimes(k+j-1)}. \tag{V.22}$$

In fact, we have

$$\delta\zeta_j(t) = \sum_{k=0,\ell=0}^{\Lambda_\infty-j+1,j-1} g^j_{k,\ell}\left(u^{[\ell,k+j-1-\ell]} - u^{\otimes(k+j-1)}\right)$$

$$= \sum_{k=0,\ell=1}^{\Lambda_\infty-j+1,j-1} g^j_{k,\ell}\delta u^{\otimes(\ell)} \otimes u^{\otimes(k+j-1-\ell)} . \tag{V.23}$$

$\delta\zeta_j(t)$ is $O(\eta^2)$ in learning rate $\eta$, since $g = O(\eta)$ and $\delta u = O(\eta)$. The condition for making $\delta\zeta_j(t)$ ignored is that

$$\left|\sum_{k=0,\ell=1}^{\Lambda_\infty-j+1,j-1} g^j_{k,\ell}\delta u^{\otimes(\ell)} \otimes u^{\otimes(k+j-1-\ell)}\right| \ll \left|\delta u^{\otimes(j)}\right| , \tag{V.24}$$

which is

$$\left|\sum_{k=0,\ell=1,k'=0}^{\Lambda_\infty-j+1,j-1,\Lambda_\infty-\ell+1} g^j_{k,\ell}G_{i,k'}u^{\otimes(k+k'+j-2)}\right| \ll \left|\sum_{k'=0}^{\Lambda_\infty-j+1} G_{i,k'}u^{\otimes(k'+j-1)}\right| , \tag{V.25}$$

in the limit where $\eta$ is small, such that we could do perturbative expansion. There is another more direct condition. We start from

$$\delta u^{\otimes(i)} = \sum_{k_1=0,\ell_1=0}^{\Lambda_\infty-i+1,i-1} g^i_{k_1,\ell_1}u^{\otimes(k_1+i-1)} + \sum_{k_1=0,\ell_1=1}^{\Lambda_\infty-i+1,i-1} g^i_{k_1,\ell_1}\delta u^{\otimes(\ell_1)} \otimes u^{\otimes(k_1+i-1-\ell_1)} , \tag{V.26}$$

and we also have

$$\delta u^{\otimes(\ell_1)} = \sum_{k_2=0,\ell_2=0}^{\Lambda_\infty-\ell_1+1,\ell_1-1} g^{\ell_1}_{k_2,\ell_2}u^{\otimes(k_2+\ell_1-1)} + \sum_{k_2=0,\ell_2=1}^{\Lambda_\infty-\ell_1+1,\ell_1-1} g^{\ell_1}_{k_2,\ell_2}\delta u^{\otimes(\ell_2)} \otimes u^{\otimes(k_2+\ell_1-1-\ell_2)} . \tag{V.27}$$

Using this recursion relation multiple times, we get

$$\delta u^{\otimes(i)} = \sum_{k_1=0,\ell_1=0}^{\Lambda_\infty-i+1,i-1} g^i_{k_1,\ell_1}u^{\otimes(k_1+i-1)} + \sum_{\ell_1=1,\ell_2=0,K=k_1,k_1=0}^{i-1,\ell_1-1,2\Lambda_\infty-i-\ell_1+2,\Lambda_\infty-i+1} g^i_{k_1,\ell_1}g^{\ell_1}_{K-k_1,\ell_2}u^{\otimes(K+i-2)}$$

$$+ \ldots + \sum_{\ell_1=i-1,\ell_2=i-2,\ldots,\ell_{i-1}=1,\ell_i=0}^{i-1,\ell_1-1,\ell_{i-2}-1,\ell_{i-1}-1} \sum_{K=k_1+k_2+\ldots+k_{i-1},k_1=0,k_2=0,\ldots,k_{i-1}=0}^{i\Lambda_\infty-\ell_1-\ell_2-\ell_{i-1},\Lambda_\infty-i+1,\Lambda_\infty-\ell_1+1,\ldots,\Lambda_\infty-\ell_{i-1}+1} g^i_{k_1,\ell_1}g^{\ell_1}_{k_2,\ell_2}g^{\ell_2}_{k_3,\ell_3}\cdots g^{\ell_{i-2}}_{k_{i-1},\ell_{i-1}}g^{\ell_{i-1}}_{K-k_1-\ldots-k_{i-1},\ell_i}u^{\otimes K} .$$

$$\tag{V.28}$$

Note that since $g = O(\eta)$, we schematically can understand the equation as

$$\delta u^{\otimes(i)} = O(\eta) + O(\eta^2) + \ldots + O(\eta^i) . \tag{V.29}$$

This equation can be understood as the exact solution of the Carleman linearization in the discrete case. The linearization error $\eta$ vector, could be written as $\delta\eta = A'\eta + \hat{b}'$, where $A'$ and $\hat{b}'$ will get reduced to $A$ and $\hat{b}$ in Eq. (II.18) when the learning rate $\eta$ is small. In fact, we know that the discretization contribution can be ignored if

$$\left|\sum_{k_1=0,\ell_1=0}^{\Lambda_\infty-i+1,i-1} g^i_{k_1,\ell_1}u^{\otimes(k_1+i-1)}\right| \gg $$

$$\left|\sum_{\ell_1=1,\ell_2=0,K=k_1,k_1=0}^{i-1,\ell_1-1,2\Lambda_\infty-i-\ell_1+2,\Lambda_\infty-i+1} g^i_{k_1,\ell_1}g^{\ell_1}_{K-k_1,\ell_2}u^{\otimes(K+i-2)} + \ldots + \sum_{\ell_1=1,\ell_2=1,\ldots,\ell_i=1,\ell_i=0}^{i-1,\ell_1-1,\ell_{i-2}-1,\ell_{i-1}-1} \sum_{K=k_1+k_2+\ldots+k_{i-1},k_1=0,k_2=0,\ldots,k_{i-1}=0}^{i\Lambda_\infty-\ell_1-\ell_2-\ell_{i-1},\Lambda_\infty-i+1,\Lambda_\infty-\ell_1+1,\ldots,\Lambda_\infty-\ell_{i-1}+1} g^i_{k_1,\ell_1}g^{\ell_1}_{k_2,\ell_2}g^{\ell_2}_{k_3,\ell_3}\cdots g^{\ell_{i-2}}_{k_{i-1},\ell_{i-1}}g^{\ell_{i-1}}_{K-k_1-\ldots-k_{i-1},\ell_i}u^{\otimes K}\right| . \tag{V.30}$$

Factorizing $u$, one could make the inequality purely a condition about the tensor $g$, namely $F_k$. Thus, the condition in Eq. (V.30) is a quantitative measure on how small the learning rate $\eta$ is required to be.

## 2. A more detailed investigation

In this subsection, we discuss the relevant terms in substantial detail. Along the lines, we introduce a tensor network diagrammatic notation. We start from writing Eq. (V.28) in a different way as

$$\delta u^{\otimes(i)} = \sum_{k_1=0,\ell_1=0}^{\Lambda_\infty-i+1,i-1} g^i_{k_1,\ell_1} u^{\otimes(k_1+i-1)} + \sum_{\ell_1=1,\ell_2=0,k_1=0,k_2=0}^{i-1,\ell_1-1,\Lambda_\infty-i+1,\Lambda_\infty-\ell_1+1} g^i_{k_1,\ell_1} g^{\ell_1}_{k_2,\ell_2} u^{\otimes(k_1+k_2+i-2)}$$
$$+ \ldots + \sum_{\ell_1=i-1,\ell_2=i-2,\ldots,\ell_{i-1}=1,\ell_i=0}^{i-1,\ell_1-1,\ldots,\ell_{i-2}-1,\ell_{i-1}-1} \sum_{k_1=0,k_2=0,\ldots,k_i=0}^{\Lambda_\infty-i+1,\Lambda_\infty-\ell_1+1,\ldots,\Lambda_\infty-\ell_{i-1}+1} g^i_{k_1,\ell_1} g^{\ell_1}_{k_2,\ell_2} \cdots g^{\ell_{i-2}}_{k_{i-1},\ell_{i-1}} g^{\ell_{i-1}}_{k_i,\ell_i} u^{\otimes(k_1+k_2+\ldots+k_i)}, \quad (V.31)$$

where

$$g^i_{k,\ell} := I^{\otimes(\ell)} \otimes F_k \otimes I^{\otimes(i-1-\ell)}. \quad (V.32)$$

Fhe first non-trivial term is given by (assuming that $\Lambda_\infty - (i+1) = q$ and in this case $i = 2$)

$$\delta u^{\otimes(2)} = \sum_{\ell_1=0}^{1} \sum_{k_1=0}^{\Lambda_\infty-(2-1)} g^2_{k_1,\ell_1} u^{\otimes(k_1+(2-1))} + \sum_{\ell_1=1,\ell_2=0}^{1,\ell_1-1} \sum_{k_1=0,k_2=0}^{\Lambda_\infty-(2-1),\Lambda_\infty-(\ell_1-1)} g^2_{k_1,\ell_1} g^{\ell_1}_{k_2,\ell_2} u^{\otimes(k_1+k_2)}$$
$$= \sum_{k_1=0}^{q} \left(F_{k_1} \otimes I + I \otimes F_{k_1}\right) u^{\otimes(k_1+1)} + \sum_{k_1=0,k_2=0}^{q,q+1} \left(F_{k_2} \otimes F_{k_1}\right) u^{\otimes(k_1+k_2)}$$
$$= \left(F_0 \otimes I + I \otimes F_0 + F_1 \otimes F_0 + F_0 \otimes F_1\right) u$$
$$+ \left(F_1 \otimes I + I \otimes F_1 + F_1 \otimes F_1 + F_2 \otimes F_0 + F_0 \otimes F_2\right) u^{\otimes(2)}$$
$$+ \left(F_2 \otimes I + I \otimes F_2 + F_1 \otimes F_2 + F_2 \otimes F_1 + F_0 \otimes F_3 + F_3 \otimes F_0\right) u^{\otimes(3)} \quad (V.33)$$
$$+ \ldots$$
$$+ \left(F_{q-1} \otimes I + I \otimes F_{q-1} + F_0 \otimes F_q + F_1 \otimes F_{q-1} + \ldots + F_{q-1} \otimes F_1 + F_q \otimes F_0\right) u^{\otimes(q)}$$
$$+ \left(F_q \otimes I + I \otimes F_q + F_1 \otimes F_q + F_2 \otimes F_{q-1} + \ldots + F_q \otimes F_1 + F_{q+1} \otimes F_0\right) u^{\otimes(q+1)}$$
$$+ \left(F_2 \otimes F_q + F_3 \otimes F_{q-1} + \ldots + F_q \otimes F_2 + F_{q+1} \otimes F_1\right) u^{\otimes(q+2)}$$
$$+ \ldots$$
$$+ \left(F_{q+1} \otimes F_q\right) u^{\otimes(2q+1)}.$$

This may look tedious. We can see the pattern, however, if we explicitly spell out another term, reflecting the $i = 3$ case, as

$$\delta u^{\otimes(3)} = \sum_{\ell_1=0}^{2} \sum_{k_1=0}^{\Lambda_\infty-(3-1)} g^3_{k_1,\ell_1} u^{\otimes(k_1+(3-1))} + \sum_{\ell_1=1,\ell_2=0}^{2,\ell_1-1} \sum_{k_1=0,k_2=0}^{\Lambda_\infty-(3-1),\Lambda_\infty-(\ell_1-1)} g^3_{k_1,\ell_1} g^{\ell_1}_{k_2,\ell_2} u^{\otimes(k_1+k_2+(3-2))}$$
$$+ \sum_{\ell_1=2,\ell_2=1,\ell_3=0}^{2,\ell_1-1,\ell_2-1} \sum_{k_1=0,k_2=0,k_3=0}^{\Lambda_\infty-(3-1),\Lambda_\infty-(\ell_1-1)} g^3_{k_1,\ell_1} g^{\ell_1}_{k_2,\ell_2} g^{\ell_2}_{k_3,\ell_3} u^{\otimes(k_1+k_2+k_3)}$$
$$= \sum_{k_1=0}^{q} \left(F_{k_1} \otimes I^{\otimes(2)} + I \otimes F_{k_1} \otimes I + I^{\otimes(2)} \otimes F_{k_1}\right) u^{\otimes(k_1+2)} \quad (V.34)$$
$$+ \left(\sum_{k_1=0,k_2=0}^{q,q+2} \left(F_{k_2} \otimes F_{k_1} \otimes I\right) + \sum_{k_1=0,k_2=0}^{q,q+1} \left(F_{k_2} \otimes I \otimes F_{k_1} + I \otimes F_{k_2} \otimes F_{k_1}\right)\right) u^{\otimes(k_1+k_2+1)}$$
$$+ \sum_{k_1=0,k_2=0,k_3=0}^{q,q+1,q+2} F_{k_3} \otimes F_{k_2} \otimes F_{k_1} u^{\otimes(k_1+k_2+k_3)}.$$

The first part is found to be

$$\delta u^{\otimes(3)} = \Big(F_1 \otimes F_0 \otimes F_0 + F_0 \otimes F_1 \otimes F_0 + F_0 \otimes F_0 \otimes F_1\Big)u$$
$$+\Big(F_0 \otimes I^{\otimes(2)} + I \otimes F_0 \otimes I + I^{\otimes(2)} \otimes F_0$$
$$+ F_1 \otimes F_0 \otimes I + F_0 \otimes F_1 \otimes I + F_0 \otimes I \otimes F_1 + F_1 \otimes I \otimes F_0 + I \otimes F_0 \otimes F_1 + I \otimes F_1 \otimes F_0$$
$$+ F_1 \otimes F_0 \otimes F_1 + F_0 \otimes F_1 \otimes F_1 + F_1 \otimes F_1 \otimes F_0\Big)u^{\otimes(2)} + \ldots. \quad (V.35)$$

The second term is given by

$$\delta u^{\otimes(2)} = \ldots + \Big(F_{q-2} \otimes I^{\otimes(2)} + I \otimes F_{q-2} \otimes I + I^{\otimes(2)} \otimes F_{q-2}$$
$$+ F_0 \otimes F_{q-1} \otimes I + F_1 \otimes F_{q-2} \otimes I + \ldots + F_{q-2} \otimes F_1 \otimes I + F_{q-1} \otimes F_0 \otimes I$$
$$+ F_0 \otimes I \otimes F_{q-1} + F_1 \otimes I \otimes F_{q-2} + \ldots + F_{q-2} \otimes I \otimes F_1 + F_{q-1} \otimes I \otimes F_0$$
$$+ I \otimes F_0 \otimes F_{q-1} + I \otimes F_1 \otimes F_{q-2} + \ldots + I \otimes F_{q-2} \otimes F_1 + I \otimes F_{q-1} \otimes F_0$$
$$+ \sum_{m_1=0, m_2=0}^{q, m_1} F_{m_2} \otimes F_{m_1-m_2} \otimes F_{q-m_1}\Big)u^{\otimes(q)}$$
$$+ \Big(F_{q-1} \otimes I^{\otimes(2)} + I \otimes F_{q-1} \otimes I + I^{\otimes(2)} \otimes F_{q-1}$$
$$+ F_0 \otimes F_q \otimes I + F_1 \otimes F_{q-1} \otimes I + \ldots + F_{q-1} \otimes F_1 \otimes I + F_q \otimes F_0 \otimes I$$
$$+ F_0 \otimes I \otimes F_q + F_1 \otimes I \otimes F_{q-1} + \ldots + F_{q-1} \otimes I \otimes F_1 + F_q \otimes I \otimes F_0 \quad (V.36)$$
$$+ I \otimes F_0 \otimes F_q + I \otimes F_1 \otimes F_{q-1} + \ldots + I \otimes F_{q-1} \otimes F_1 + I \otimes F_q \otimes F_0$$
$$+ \sum_{m_1=0, m_2=0}^{q, m_1+1} F_{m_2} \otimes F_{1+m_1-m_2} \otimes F_{q-m_1}\Big)u^{\otimes(q+1)}$$
$$+ \Big(F_q \otimes I^{\otimes(2)} + I \otimes F_q \otimes I + I^{\otimes(2)} \otimes F_q$$
$$+ \sum_{m_1=0}^{q} F_{1+m_1} \otimes F_{q-m_1} \otimes I + F_{1+m_1} \otimes I \otimes F_{q-m_1} + I \otimes F_{1+m_1} \otimes F_{q-m_1}$$
$$+ \sum_{m_1=0, m_2=0}^{q-1, m_1+2} F_{m_2} \otimes F_{2+m_1-m_2} \otimes F_{q-m_1} + \sum_{m_1=0}^{q+1} F_{1+m_1} \otimes F_{1+q-m_1} \otimes F_0\Big)u^{\otimes(q+2)} + \ldots.$$

The third part is then

$$\delta u^{\otimes(3)} = \ldots + \Bigg(\sum_{m_1=0}^{q} F_{2+m_1} \otimes F_{q-m_1} \otimes I + \sum_{m_1=0}^{q-1} \big(F_{2+m_1} \otimes I \otimes F_{q-m_1} + I \otimes F_{2+m_1} \otimes F_{q-m_1}\big)$$
$$+ \sum_{m_1=0, m_2=0}^{q-2, m_1+3} F_{m_2} \otimes F_{3+m_1-m_2} \otimes F_{q-m_1} + \sum_{m_1=0}^{q+1} F_{1+m_1} \otimes F_{1+q-m_1} \otimes F_1 + \sum_{m_1=0}^{q} F_{2+m_1} \otimes F_{1+q-m_1} \otimes F_0\Bigg)u^{\otimes(q+3)}$$
$$+ \Bigg(\sum_{m_1=0}^{q} F_{3+m_1} \otimes F_{q-m_1} \otimes I + \sum_{m_1=0}^{q-2} \big(F_{3+m_1} \otimes I \otimes F_{q-m_1} + I \otimes F_{3+m_1} \otimes F_{q-m_1}\big) \quad (V.37)$$
$$+ \sum_{m_1=0, m_2=0}^{q-3, m_1+3} F_{m_2} \otimes F_{4+m_1-m_2} \otimes F_{q-m_1} + \sum_{m_1=0}^{q+1} F_{1+m_1} \otimes F_{1+q-m_1} \otimes F_2 + \sum_{m_1=0}^{q} F_{2+m_1} \otimes F_{1+q-m_1} \otimes F_1$$
$$+ \sum_{m_1=0}^{q} F_{3+m_1} \otimes F_{1+q-m_1} \otimes F_0\Bigg)u^{\otimes(q+4)} + \ldots.$$

The fourth term is found to be

$$\delta u^{\otimes(3)} = \ldots + \Bigg(F_q \otimes F_q \otimes I + F_{q+1} \otimes F_{q-1} \otimes I + F_{q+2} \otimes F_{q-2} \otimes I$$
$$+ \big(F_q \otimes I \otimes F_q + I \otimes F_q \otimes F_q + F_{q+1} \otimes I \otimes F_{q-1} + I \otimes F_{q+1} \otimes F_{q-1}\big)$$
$$+ \sum_{m_1=0}^{q+1} F_{m_1} \otimes F_{q+1-m_1} \otimes F_q + \sum_{m_1=0}^{q+1} F_{1+m_1} \otimes F_{q+1-m_1} \otimes F_{q-1}$$
$$+ \sum_{m_1=0}^{q} F_{2+m_1} \otimes F_{1+q-m_1} \otimes F_{q-2} + \ldots + \sum_{m_1=0}^{2} F_{q+m_1} \otimes F_{1+q-m_1} \otimes F_0 \Bigg) u^{\otimes(2q+1)}$$
$$+ \Bigg(F_{q+1} \otimes F_q \otimes I + F_{q+2} \otimes F_{q-1} \otimes I + \big(F_{q+1} \otimes I \otimes F_q + I \otimes F_{q+1} \otimes F_q\big)$$
$$+ \sum_{m_1=0}^{q+1} F_{1+m_1} \otimes F_{q+1-m_1} \otimes F_q + \sum_{m_1=0}^{q} F_{2+m_1} \otimes F_{q+1-m_1} \otimes F_{q-1}$$
$$+ \sum_{m_1=0}^{q-1} F_{3+m_1} \otimes F_{q+1-m_1} \otimes F_{q-2} + \ldots + \sum_{m_1=0}^{1} F_{1+q+m_1} \otimes F_{1+q-m_1} \otimes F_0 \Bigg) u^{\otimes(2q+2)} \qquad \text{(V.38)}$$
$$+ \Bigg(F_{q+2} \otimes F_q \otimes I$$
$$+ \sum_{m_1=0}^{q} F_{2+m_1} \otimes F_{q+1-m_1} \otimes F_q + \sum_{m_1=0}^{q-1} F_{3+m_1} \otimes F_{q+1-m_1} \otimes F_{q-1}$$
$$+ \sum_{m_1=0}^{q-2} F_{4+m_1} \otimes F_{q+1-m_1} \otimes F_{q-2} + \ldots + F_{2+q} \otimes F_{1+q} \otimes F_0 \Bigg) u^{\otimes(2q+3)}$$
$$+ \Bigg(\sum_{m_1=0}^{q-1} F_{3+m_1} \otimes F_{q+1-m_1} \otimes F_q + \sum_{m_1=0}^{q-2} F_{4+m_1} \otimes F_{q+1-m_1} \otimes F_{q-1}$$
$$+ \sum_{m_1=0}^{q-3} F_{5+m_1} \otimes F_{q+1-m_1} \otimes F_{q-2} + \ldots + F_{2+q} \otimes F_{1+q} \otimes F_1 \Bigg) u^{\otimes(2q+4)} + \ldots .$$

And then, finally, the last term is given by

$$\delta u^{\otimes(3)} = \ldots + \Bigg(\sum_{m_1=0}^{3} F_{q-1+m_1} \otimes F_{q+1-m_1} \otimes F_q + \sum_{m_1=0}^{2} F_{q+m_1} \otimes F_{q+1-m_1} \otimes F_{q-1}$$
$$+ \sum_{m_1=0}^{1} F_{q+1+m_1} \otimes F_{q+1-m_1} \otimes F_{q-2} + F_{q+2} \otimes F_{q+1} \otimes F_{q-3} \Bigg) u^{\otimes(3q)}$$
$$+ \Bigg(\sum_{m_1=0}^{2} F_{q+m_1} \otimes F_{q+1-m_1} \otimes F_q + \sum_{m_1=0}^{1} F_{q+1+m_1} \otimes F_{q+1-m_1} \otimes F_{q-1}$$
$$+ F_{q+2} \otimes F_{q+1} \otimes F_{q-2} \Bigg) u^{\otimes(3q+1)} \qquad \text{(V.39)}$$
$$+ \Bigg(\sum_{m_1=0}^{1} F_{q+1+m_1} \otimes F_{q+1-m_1} \otimes F_q + F_{q+2} \otimes F_{q+1} \otimes F_{q-1} \Bigg) u^{\otimes(3q+2)}$$
$$+ \Bigg(F_{q+2} \otimes F_{q+1} \otimes F_q \Bigg) u^{\otimes(3q+3)} .$$

Using these two *simple* examples as a basis, we can reframe this problem as follows. We introduce a matrix $M^i_K$ such that the relation can be expressed as

$$\delta u^{\otimes(i)} = M^i_K \, u^{\otimes(K)} . \qquad \text{(V.40)}$$

The objective is to determine $M^i_K$. To streamline our approach, we will introduce a bit of notation. Firstly, we represent the order of the correction term as $n$, giving us a representative term in the form

$$I^{\otimes(\ell_n)} \otimes F_{k_n} \otimes I^{\otimes(\ell_{n-1}-1-\ell_n)} \otimes F_{k_{n-1}} \otimes \ldots \otimes F_{k_2} \otimes I^{\otimes(\ell_1-1-\ell_2)} \otimes F_{k_1} \otimes I^{\otimes(i-1-\ell_1)} \,. \tag{V.41}$$

It is apparent that the count of $F_k$ within a term, denoted by $n$, signifies the order of correction for a given expression. Additionally, we establish the upper bounds for $k_\alpha$: $q_1 = \Lambda_\infty - i + 1$, and for $2 \leq \alpha \leq n$, $q_\alpha = \Lambda_\infty - l_{\alpha-1} + 1 = q_1 + i - l_{\alpha-1}$. It is at this point important to note that the number of identity matrices $I$ situated between $F_{k_n}$ is indicated within the upper bound for each $k_\alpha$. This can be visually comprehended through the following illustration: Imagine a lineup of $n$ balls, each representing a $F_{k_\alpha}$.

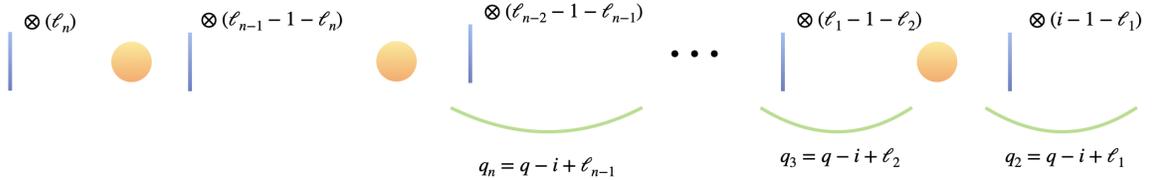

FIG. 5. We visualize this by depicting a blue stick for each identity matrix and a yellow ball to symbolize $F_k$. For a correction term of $n^{th}$ order, there are $i - n$ blue sticks and $n$ yellow balls.

Next, we utilize the $i - n$ blue sticks representing identity matrices to create partitions.

FIG. 6. We divide a sequence of $n$ yellow balls with $i-n$ blue sticks and label the number of sticks between each yellow ball as indicated above. Consequently, the upper bounds $q_\alpha$ can be determined based solely on the number of sticks in between. Please note that $\ell_n$ and $\ell_{n-1} - 1 - \ell_n$ are not involved in calculating the upper bounds, resulting in distinct partitions with identical upper bounds.

The values of $q_\alpha$ are derived from the numbers of blue sticks between the yellow balls. Therefore, when summing over the $\ell$s, it signifies the summation across various partitions with distinct upper bounds. Additionally, although the sum incorporates $\ell_n$, it is not connected to any upper bound. Consequently, different partitions with identical upper bounds need to be included in the summation. Going forward, we omit the identity matrices between $F_{k_\alpha}$ for the sake of brevity in expressing the term

$$F_{k_n} \otimes F_{k_{n-1}} \otimes \ldots \otimes F_{k_2} \otimes F_{k_1} \,. \tag{V.42}$$

$M^i_K$ is the summation of typical terms as described above, ranging from 1 to $n$ where $n \leq i$. The conditions for this summation involve $\tilde{K} := K - i + n = \sum_{\alpha=1}^n k_\alpha$ and, concurrently, adhering to the constraints for each $k_\alpha$, which are defined as $0 \leq k_\alpha \leq q_\alpha$. This can be formally expressed as

$$M^i_K = \sum_{n=1}^{i} \sum_{\ell_1=0,\ell_2=0,\ldots,\ell_n=0}^{i-1,\ell_1-1,\ldots,\ell_{n-1}-1} \widetilde{\sum}_{k_1=0,\ldots,k_n=0}^{q_1,\ldots,q_n} F_{k_n} \otimes F_{k_{n-1}} \otimes \ldots \otimes F_{k_2} \otimes F_{k_1} \,. \tag{V.43}$$

The $\ell$ dependence is embedded within the upper bounds $q_\alpha$, and the symbol $\widetilde{\sum}$ denotes a summation under a condition, where

$\tilde{K} = \sum_{\alpha=1}^{n} k_{\alpha}$. This conditional summation can be expressed recursively. Let us define, for $0 \leq n' \leq n$,

$$\mathcal{F}_{\tilde{K},1}^{(n')} := F_{\tilde{K}}, \quad 0 \leq \tilde{K} \leq q_{n'}, \tag{V.44}$$

$$\mathcal{F}_{\tilde{K},2}^{(1)} := \sum_{m_1=0}^{\min(\tilde{K},q_1)} \mathcal{F}_{\tilde{K}-m_1,1}^{(2)} \otimes F_{m_1}, \tag{V.45}$$

$$\mathcal{F}_{\tilde{K},3}^{(1)} := \sum_{m_1=0}^{\min(\tilde{K},q_1)} \mathcal{F}_{\tilde{K}-m_1,2}^{(2)} \otimes F_{m_1} = \sum_{m_1=0}^{\min(\tilde{K},q_1)} \sum_{m_2=0}^{\min(\tilde{K},q_2)} \mathcal{F}_{\tilde{K}-m_1-m_2,1}^{(3)} \otimes F_{m_2} \otimes F_{m_1}, \tag{V.46}$$

$$\mathcal{F}_{\tilde{K},4}^{(1)} := \sum_{m_1=0}^{\min(\tilde{K},q_1)} \mathcal{F}_{\tilde{K}-m_1,3}^{(2)} \otimes F_{m_1} = \sum_{m_1=0}^{\min(\tilde{K},q_1)} \sum_{m_2=0}^{\min(\tilde{K},q_2)} \mathcal{F}_{\tilde{K}-m_1,2}^{(3)} \otimes F_{m_2} \otimes F_{m_1}, \tag{V.47}$$

$$= \sum_{m_1=0}^{\min(\tilde{K},q_1)} \sum_{m_2=0}^{\min(\tilde{K},q_2)} \sum_{m_3=0}^{\min(\tilde{K},q_1)} \mathcal{F}_{\tilde{K}-m_1-m_2-m_3,1}^{(4)} \otimes F_{m_3} \otimes F_{m_2} \otimes F_{m_1},$$

$$\ldots$$

$$\mathcal{F}_{\tilde{K},n}^{(1)} := \sum_{m_1=0}^{\min(\tilde{K},q_1)} \mathcal{F}_{\tilde{K}-m_1,n-1}^{(2)} \otimes F_{m_1}. \tag{V.48}$$

Therefore, in a general context, we obtain

$$\mathcal{F}_{\tilde{K},n}^{(n')} = \sum_{m_{n'}=0}^{\min(\tilde{K},q_{n'})} \mathcal{F}_{\tilde{K}-m_{n'},n-1}^{(n'+1)} \otimes F_{m_{n'}}. \tag{V.49}$$

Continuing from the previous illustrative explanation, this recursive description can be visualized as follows, with the basic elements defined as shown in the figure.

$$F_{m_{n'}} = \boxed{(1, n')} \hspace{-0.1em}\rule[0.5ex]{2em}{0.4pt}\hspace{-0.1em}^{m_{n'}} \quad (0 \leq m_{n'} \leq q_{n'})$$

$$\mathcal{F}_{\tilde{K},1}^{(n')} = F_{\tilde{K}} = \hspace{-0.1em}\rule[0.5ex]{2em}{0.4pt}\hspace{-0.1em}^{i}\boxed{(1, n')}\hspace{-0.1em}\rule[0.5ex]{2em}{0.4pt}\hspace{-0.1em}^{K} \quad (0 \leq (\tilde{K} = K - (i-1)) \leq q_{n'})$$

FIG. 7. We depict each $F_{m_{n'}}$ as an order 1 tensor with a range set on $m_{n'}$: $0 \leq m_{n'} \leq q_{n'}$. Now, $\mathcal{F}_{\tilde{K},1}^{(n')}$ is introduced as an order 2 tensor, nearly identical to $F_{m_{n'}}$, with the distinction that $\tilde{K}$ relies on both $K$ and $i$, while $F_{m_{n'}}$ is contingent solely on $m_{n'}$.

We are now in the position to formulate the recursion. This involves generating additional purple blobs to contract with the larger one (see Fig. 8). Using this recursive expression, one can systematically generate all typical terms $\mathcal{F}_{\tilde{K},n}^{(1)}$ for any $n$. This can be achieved by repeatedly expressing $\mathcal{F}_{\tilde{K},n}^{(1)}$ in terms of $\mathcal{F}_{\tilde{K},1}^{(n')}$ with $1 \leq n' \leq n$. The maximum value of $K$ corresponds to $n = i$, resulting in

$$\tilde{K} = K = iq_1 + i(i-1)/2. \tag{V.50}$$

At this order of correction and for the same value of $K$, there is always a constant correction term represented by $F_0^{\otimes(iq_1+i(i-1)/2)}$. Consequently, we can now formulate the matrix $M$ as

$$M_K^i = \sum_{n=1}^{i} \sum_{\ell_1=0, \ell_2=0, \ldots, \ell_n=0}^{i-1, \ell_1-1, \ldots, \ell_{n-1}-1} \mathcal{F}_{K-i+n,n}^{(1)}. \tag{V.51}$$

Once again, the initial sum encompasses all orders of correction. The highest order, denoted as $n = i$, represents expressions without any identity matrices. The count of identity matrices between each $F_{k_n}$ influences the upper bounds $q_n$ for $k_n$. Summation over all $\ell$ is equivalent to summing over all potential upper bounds $q$. The introduction of $\tilde{K} = K - i + n$ is to ensure that the identity matrices are counted as 1 in the sum, thereby totalling $K$ to maintain the correct dimension. To summarize, we get

$$\delta u^{\otimes(i)} = \sum_{n=1}^{i} \sum_{\ell_1=0, \ell_2=0, \ldots, \ell_n=0}^{i-1, \ell_1-1, \ldots, \ell_{n-1}-1} \mathcal{F}_{K-i+n,n}^{(1)} u^{\otimes(K)}. \tag{V.52}$$

Reinserting the identity matrices between every term is straightforward, using the upper bound values $q_\alpha$ and $q_1$, enabling the complete expression to be restored.

[Figure 8 diagram showing recursive tensor representation]

FIG. 8. The essence of the recursion is to represent $\mathcal{F}_{\tilde{K},n}^{(n')}$ as an order 2 tensor with a restricted range on $\tilde{K}$. This is achieved by defining it as a sum involving an order 3 tensor $\mathcal{F}_{\tilde{K}-m_{n'},n-1}^{(n'+1)}$ contracted with the order 1 tensor $F_{m_{n'}}$, which has been previously defined. This process is then repeated recursively for $n-1$ iterations. The purple blobs in the representation signify the range confinements associated with each step of this procedure.

### 3. Examples

We will now demonstrate the applicability of this recursive procedure at hand of specific examples. Let us begin by examining the cases where $i = 2$. For $0 \leq K \leq q+1$, the first-order contribution with $\tilde{K} = K - (i-n) = K-1$ is given by

$$\mathcal{F}_{K-1,1}^{(1)} = F_{K-1} \otimes I + I \otimes F_{K-1} . \tag{V.53}$$

Moving on, let us examine the second-order contribution, $F_{k_1} \otimes F_{k_2}$, which is the highest order in this case. This implies that $\tilde{K} = K$. Initially, we have a constant term due to the highest order

$$F_0 \otimes F_0 . \tag{V.54}$$

Now, let us proceed to $1 \leq K \leq q$. We will start by determining the upper bounds. The upper bound for $k_1$ is $q_1 = q$, and for $k_2$ it's $q_2 = q+1$. Using the recursive description, we get

$$\mathcal{F}_{K,2}^{(1)} = \sum_{m_1=0}^{q} \mathcal{F}_{K-m_1,1}^{(2)} \otimes F_{m_1} , \tag{V.55}$$

for which we have

$$\mathcal{F}_{(K-m_1,1)}^{(2)} = F_{K-m_1} , \quad 0 \leq K - m_1 \leq q_2 = q+1 . \tag{V.56}$$

For the range $0 \leq K \leq q+1$, there are no issues with the condition $0 \leq K - m_1 \leq q_2 = q+1$. Thus, for $0 \leq K \leq q+1$, we find

$$\mathcal{F}_{K,2}^{(1)} = \sum_{m_1=0}^{\min(K,q)} F_{K-m_1} \otimes F_{m_1} . \tag{V.57}$$

However, for the range $q+2 \leq K \leq 2q+1$, the upper bound condition can pose issues. Consider the case where $K = q+2$

$$\mathcal{F}_{(q+2-m_1,1)}^{(2)} = F_{q+2-m_1} , \quad 0 \leq q+2-m_1 \leq q+1 . \tag{V.58}$$

As a result, in this case, the constraint on $m_1$ becomes $1 \leq m_1 \leq q$,

$$\mathcal{F}_{q+2,2}^{(1)} = \sum_{m_1=1}^{q} F_{K-m_1} \otimes F_{m_1} . \tag{V.59}$$

Applying the same reasoning, one can deduce the following for the range $q + 2 \leq K \leq 2q + 1$

$$\mathcal{F}^{(1)}_{K,2} = \sum_{m_1=K-(q+1)}^{q} F_{K-m_1} \otimes F_{m_1} \,. \tag{V.60}$$

The last contribution comes from $K = 2q + 1$, given rise to

$$\left(F_{q+1} \otimes F_q\right) u^{\otimes(2q+1)} \,. \tag{V.61}$$

To consolidate all the results, we return to the matrix $M^i_K$, to find

$$\begin{aligned}
\delta u^{\otimes(i)} = M^i_K \, u^{\otimes(K)} &= \left( \sum_{\ell_1=0}^{1} \mathcal{F}^{(1)}_{K-1,1} + \mathcal{F}^{(1)}_{K,2} \right) u^{\otimes(K)} \\
&= \left(F_{K-1} \otimes I + I \otimes F_{K-1}\right) u^{\otimes(K)} + \begin{cases} \sum_{m_1=0}^{\min(K,q)} F_{K-m_1} \otimes F_{m_1} \, u^{\otimes(K)} &, \ 0 \leq K \leq q+1, \\ \sum_{m_1=K-(q+1)}^{q} F_{K-m_1} \otimes F_{m_1} \, u^{\otimes(K)} &, \ q+2 \leq K \leq 2q+1. \end{cases}
\end{aligned} \tag{V.62}$$

This can also be visualized using our diagrams.

$$\mathcal{F}^{(1)}_{\tilde{K},2} = \sum_{m_1=0}^{\min(\tilde{K},q)} \mathcal{F}^{(2)}_{\tilde{K}-m_1,1} \otimes F_{m_1}$$

$(0 \leq \tilde{K} \leq q)$

$(0 \leq \tilde{K} - m_1 \leq q_2)$

$(0 \leq m_1 \leq q_1)$

FIG. 9. Using the recursive approach outlined in Fig. 8, this scenario only necessitates a single iteration, encompassing the two general cases when $n = 2$.

It is at this point evident that this mirrors the exact description we previously presented in Eq. (V.33). Now, let us delve into the $i = 3$ case using the recursive process. The first order entails

$$\mathcal{F}^{(1)}_{\tilde{K},1} = F_{\tilde{K}}, \quad 0 \leq \tilde{K} \leq q \,. \tag{V.63}$$

The situation here aligns with $\tilde{K} = K - (i - n) = K - (3 - 1) = K - 2$. Therefore, the first-order correction spans from $K = 2$ to $K = q + 2$, or more specifically, from $u^{\otimes(2)}$ to $u^{\otimes(q+2)}$. Expressing this in a conventional manner (and expanding the sum over $\ell_1$) gives

$$\mathcal{F}^{(1)}_{K-2,1} u^{\otimes(K)} = \left(F_{K-2} \otimes I^{\otimes(2)} + I \otimes F_{K-2} \otimes I + I^{\otimes(2)} \otimes F_{K-2}\right) u^{\otimes(K)}, \quad 2 \leq K \leq q + 2 \,. \tag{V.64}$$

The second order presents a unique situation ($\tilde{K} = K - (i - n) = K - 1$). It involves three terms divided into two categories. The first category has a single term

$$F_{k_2} \otimes F_{k_1} \otimes I \,. \tag{V.65}$$

For this category, the upper bound is $q_2 = q + 2$, given that $\ell_1 = 1$ while $i = 3$. The second category is

$$F_{k_2} \otimes I \otimes F_{k_1} + I \otimes F_{k_2} \otimes F_{k_1} \,. \tag{V.66}$$

In this case, the upper bound is $q_2 = q + 1$, due to $\ell_1 = 2$ while $i = 3$. There are two terms because we are also summing over $\ell_2$, although it is unrelated to the upper bounds $q_2$ or $q_1$. The second kind is actually trivial, as it follows the same pattern as demonstrated in the previous example,

$$\begin{aligned}
\mathcal{F}^{(1)}_{K-1,2} &= \sum_{m_1=0}^{q} \mathcal{F}^{(2)}_{K-1-m_1,1} \otimes F_{m_1}, \quad 1 \leq K \leq 2q + 2 \text{ and } q_2 = q + 1, \\
&= \begin{cases} \sum_{m_1=0}^{\min(K-1,q)} F_{K-1-m_1} \otimes F_{m_1} \,, & 1 \leq K \leq q + 2, \\ \sum_{m_1=K-1-(q+1)}^{q} F_{K-1-m_1} \otimes F_{m_1} \,, & q + 3 \leq K \leq 2q + 2. \end{cases}
\end{aligned} \tag{V.67}$$

This is analogous to the second order in the $i = 2$ case, with the substitution of $K - 1$ for $K$. The first kind exhibits slight differences but remains similar,

$$\mathcal{F}^{(1)}_{K-1,2} = \sum_{m_1=0}^{q} \mathcal{F}^{(2)}_{K-1-m_1,1} \otimes F_{m_1} \quad , 1 \le K \le 2q+3 \text{ and } q_2 = q+2,$$

$$= \begin{cases} \sum_{m_1=0}^{\min(K-1,q)} F_{K-1-m_1} \otimes F_{m_1}, & 1 \le K \le q+3, \\ \sum_{m_1=K-1-(q+2)}^{q} F_{K-1-m_1} \otimes F_{m_1}, & q+4 \le K \le 2q+3. \end{cases} \quad \text{(V.68)}$$

Summing up all three terms of the first kind, we encompass contributions from $K = 1$ to $K = 2q+2$, while the second kind spans from $K = 1$ to $K = 2q+3$. The third and highest order contribution for $i = 3$ is the third-order contribution. Here, $\tilde{K} = K$, and the upper bounds for $k_1$, $k_2$, and $k_3$ are $q_1 = q$, $q_2 = q+1$, and $q_3 = q+2$,

$$\mathcal{F}^{(1)}_{K,3} = \sum_{m_1=0}^{\min(a,q)} \mathcal{F}^{(2)}_{K-m_1,2} \otimes F_{m_1}. \quad \text{(V.69)}$$

Indeed, the third-order contribution extends from $K = 0$ with a constant term $F_0 \otimes F_0 \otimes F_0$ to $K = 3q+3$ with the term $F_{q+2} \otimes F_{q+1} \otimes F_q \, u^{\otimes(3q+3)}$. Let us take a closer look at this contribution given by

$$\mathcal{F}^{(1)}_{K,3} = \begin{cases} \sum_{m_1=0}^{\min(K,q)} \mathcal{F}^{(2)}_{K-m_1,2} \otimes F_{m_1}, & 0 \le K \le 2q+3, \\ \sum_{m_1=K-2q+3}^{q} \mathcal{F}^{(2)}_{K-m_1,2} \otimes F_{m_1}, & 2q+4 \le K \le 3q+3, \end{cases}$$

$$= \begin{cases} \sum_{m_1=0}^{\min(K,q)} \begin{cases} \sum_{m_2=0}^{\min(K,q+1)} F_{K-m_1-m_2} \otimes F_{m_2} \otimes F_{m_1}, & 0 \le K-m_1 \le q+2, \\ \sum_{m_2=K-m_1-(q+2)}^{q+1} F_{K-m_1-m_2} \otimes F_{m_2} \otimes F_{m_1}, & q+3 \le K-m_1 \le 2q+3, \end{cases} & 0 \le K \le 2q+3, \\ \sum_{m_1=K-(2q+3)}^{q} \sum_{m_2=K-m_1-(q+2)}^{q+1} F_{K-m_1-m_2} \otimes F_{m_2} \otimes F_{m_1} & , 2q+4 \le K \le 3q+3. \end{cases} \quad \text{(V.70)}$$

The recursive nature of the calculation is evident from the fact that the second-order correction is embedded within the third-order correction. This recursive structure can also be visualized using our graphical representation:

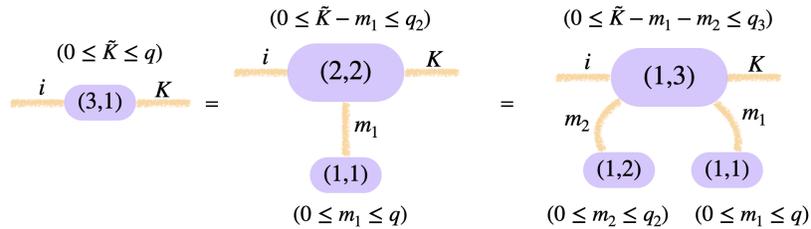

FIG. 10. For $n = 3$, two rounds of recursion are needed. This aligns with the three cases discussed earlier.

The observed pattern becomes clear now: Within the range of $0 \le K \le 3q+3$, each $m_1$ in the appropriate range contributes to a full second-order term. This suggests that the summation is effectively over second-order contributions indexed by $m_1$ for a given $K$. This description is consistent with our initial calculations. This pattern extends to higher values of $i$ as well. For instance, in the case of $i = 4$, we would have fourth-order contributions. In this situation, for a given $K$, we would be summing over bunch of third-order contributions with varying index $m_1$, much like the ones described earlier. The second and third order contributions in the $i = 4$ case would have multiple variations with different upper bounds. The first-order contribution would align with the standard case. This conforms to the recursive procedure detailed above. To present these concepts more

comprehensively, let us explicitly write out a portion of the $M_K^i$ matrix

$$\delta\begin{pmatrix} u \\ u^{\otimes(2)} \\ u^{\otimes(3)} \\ \vdots \\ u^{\otimes(N)} \\ \vdots \end{pmatrix} = \begin{pmatrix} M_1^1 & M_2^1 & \cdots & M_q^1 & & & & & \\ M_1^2 & M_2^2 & \cdots & M_q^2 & \cdots & M_{2q+1}^2 & & & \\ M_1^3 & M_2^3 & \cdots & M_q^3 & \cdots & M_{2q+1}^3 & \cdots & M_{3q+3}^3 & \\ \vdots & \vdots & \vdots & \vdots & \vdots & \vdots & \vdots & \vdots & \ddots \\ M_1^N & M_2^N & \cdots & M_q^N & \cdots & M_{2q+1}^N & \cdots & M_{3q+3}^N & \cdots & M_{Nq+\frac{N(N-1)}{2}}^N & \\ \vdots & \vdots & \vdots & \vdots & \vdots & \vdots & \vdots & \vdots & \vdots & \ddots \end{pmatrix} \begin{pmatrix} u \\ u^{\otimes(2)} \\ \vdots \\ u^{\otimes(q)} \\ \vdots \\ u^{\otimes(2q+1)} \\ \vdots \\ u^{\otimes(3q+3)} \\ \vdots \\ u^{\otimes(Nq+\frac{N(N-1)}{2})} \\ \vdots \end{pmatrix}$$

$$+ \begin{pmatrix} F_0 \\ F_0^{\otimes(2)} \\ \vdots \\ F_0^{\otimes(q)} \\ \vdots \\ F_0^{\otimes(2q+1)} \\ \vdots \\ F_0^{\otimes(3q+3)} \\ \vdots \\ F_0^{\otimes(Nq+\frac{N(N-1)}{2})} \\ \vdots \end{pmatrix}. \tag{V.71}$$

The matrix elements can be derived from the examples provided earlier. The structure of the matrices for different values of $i$ and $K$ follows the described pattern, and the specific elements can be obtained by using the recursive method and considering the upper bounds for each $k_\alpha$,

$$M_1^1 = F_1, \quad M_2^1 = F_2, \quad \ldots \quad M_q^1 = F_q, \tag{V.72}$$

$$M_1^2 = F_0 \otimes I + I \otimes F_0 + F_1 \otimes F_0 + F_0 \otimes F_1, \quad M_2^2 = F_1 \otimes I + I \otimes F_1 + F_2 \otimes F_0 + F_1 \otimes F_1 + F_0 \otimes F_2, \tag{V.73}$$

$$\ldots \quad M_q^2 = F_{q-1} \otimes I + I \otimes F_{q-1} + F_q \otimes F_0 + F_{q-1} \otimes F_1 + \ldots + F_1 \otimes F_{q-1} + F_0 \otimes F_q, \tag{V.74}$$

$$\ldots \quad M_{2q+1}^2 = F_{q+1} \otimes F_q. \tag{V.75}$$

For $M_K^1$, there is only the first-order contribution. For $M_K^2$, both first-order and second-order contributions are present. The first-order contribution extends from $K = 1$ to $K = q + 1$, and the second-order contribution involves specific terms within the

range of $K$ determined by the examples in Eq. (V.62),

$$
\begin{aligned}
M_1^3 &= F_0 \otimes F_0 \otimes I + F_0 \otimes I \otimes F_0 + I \otimes F_0 \otimes F_0 + F_0 \otimes F_0 \otimes F_1 + F_0 \otimes F_1 \otimes F_0 + F_1 \otimes F_0 \otimes F_0 \,, & \text{(V.76)} \\
M_2^3 &= F_0 \otimes I \otimes I + I \otimes F_0 \otimes I + I \otimes I \otimes F_0 & \text{(V.77)} \\
&\quad + I \otimes F_1 \otimes F_0 + I \otimes F_0 \otimes F_1 + F_1 \otimes I \otimes F_0 + F_0 \otimes I \otimes F_1 + F_1 \otimes F_0 \otimes I + F_0 \otimes F_1 \otimes I \\
&\quad + F_2 \otimes F_0 \otimes F_0 + F_1 \otimes F_1 \otimes F_0 + F_0 \otimes F_2 \otimes F_0 + F_1 \otimes F_0 \otimes F_1 + F_0 \otimes F_1 \otimes F_1 + F_0 \otimes F_0 \otimes F_2 \,, \\
&\ldots \\
M_q^3 &= F_{q-2} \otimes I \otimes I + I \otimes F_{q-2} \otimes I + I \otimes I \otimes F_{q-2} + F_{q-1} \otimes F_0 \otimes I + \ldots + F_0 \otimes F_{q-1} \otimes I & \text{(V.78)} \\
&\quad + F_{q-1} \otimes I \otimes F_0 + \ldots + F_0 \otimes I \otimes F_{q-1} + I \otimes F_{q-1} \otimes F_0 + \ldots + I \otimes F_0 \otimes F_{q-1} \\
&\quad + \sum_{m_2=0}^{q} F_{q-m_2} \otimes F_{m_2} \otimes F_0 + \sum_{m_2=0}^{q-1} F_{q-1-m_2} \otimes F_{m_2} \otimes F_1 + \ldots + F_0 \otimes F_0 \otimes F_q \,, \\
&\ldots \\
M_{2q+1}^3 &= F_{q+2} \otimes F_{q-2} \otimes I + F_{q+1} \otimes F_{q-1} \otimes I + F_q \otimes F_q \otimes I + F_{q+1} \otimes I \otimes F_{q-1} + F_q \otimes I \otimes F_q & \text{(V.79)} \\
&\quad + I \otimes F_{q+1} \otimes F_{q-1} + I \otimes F_q \otimes F_q + \sum_{m_2=q-1}^{q} F_{2q+1-m_2} \otimes F_{m_2} \otimes F_0 + \sum_{m_2=q-2}^{q} F_{2q-m_2} \otimes F_{m_2} \otimes F_1 \\
&\quad + \ldots + \sum_{m_2=0}^{q} F_{q+1} \otimes F_{m_2} \otimes F_q, \\
&\ldots \\
M_{3q+2}^3 &= F_{q+2} \otimes F_{q+1} \otimes F_q \,. & \text{(V.80)}
\end{aligned}
$$

The results above can be deduced from Eq. (V.70). It is worth noting that the first-order contribution, as illustrated earlier, is encompassed within the range from $M_2^3$ to $M_{q+2}^3$. Meanwhile, the second-order contribution initiates at $M_1^3$ and extends to either $M_{2q+2}^3$ or $M_{2q+3}^3$ depending on the value of $\ell_1$. This detailed discussion was elaborated upon during the derivation of Eq. (V.70). Following the recursive procedure, we can also write down the corrections from order 1 to $N$ for the $i = N$ case. We leave that as an exercise to the readers.

### C. Alternative methods

The main text discusses the brute-force linearization of the original differential equations. Alternatively, one could possibly make use of some transformations before linearization to simplify the linearization procedure, and even of transformations of a non-dissipative regime into a dissipative one. We discuss the following concrete example from the toy model discussed in Ref. [89] and restrict our analysis in the ODE limit. We consider a data-set with one-dimensional inputs. We write the network function with one hidden layer and linear activations as

$$ f(x) = n^{-1/2} v^T u \,, \tag{V.81} $$

where $n$ is the width, and $u, v \in \mathbb{R}^n$ are the model parameters. We denote the loss function as $L = f^2/2$. The gradient descent equations are

$$ u_{t+1} = u_t - \eta n^{-1/2} f_t v_t \,, \tag{V.82} $$
$$ v_{t+1} = v_t - \eta n^{-1/2} f_t u_t \,, \tag{V.83} $$

for the $t$-th iteration. Taking the limit of $(u_{t+1} - u_t)/\eta \to \dot{u}$ and $(v_{t+1} - v_t)/\eta \to \dot{v}$, we obtain the ODE system

$$ \dot{u} = -n^{-1/2} f v = -n^{-1}(v^T u) v \,, \tag{V.84} $$
$$ \dot{v} = -n^{-1/2} f u = -n^{-1}(v^T u) v \,. \tag{V.85} $$

The gradient descent equations (V.84) can be written in terms of the so-called *neural tangent kernel* (NTK). We denote the primary kernel eigenvalue as

$$ \lambda = n^{-1}(\|u\|_2^2 + \|v\|_2^2) \,. \tag{V.86} $$

Then the gradient descent equations can then be expressed as

$$f_{t+1} = f_t - \eta \lambda_t f_t + n^{-1} \eta^2 f_t^3 , \tag{V.87}$$

$$\lambda_{t+1} = \lambda_t - 4n^{-1} \eta f_t^2 + n^{-1} \eta^2 \lambda_t f_t^2 . \tag{V.88}$$

Taking the limit of $(f_{t+1} - f_t)/\eta \to \dot{f}$ and $(\lambda_{t+1} - \lambda_t)/\eta \to \dot{\lambda}$, we obtain the ODE system

$$\dot{f} = -\lambda f + n^{-1} \eta f^3 , \tag{V.89}$$

$$\dot{\lambda} = -4n^{-1} f^2 + n^{-1} \eta \lambda f^2 . \tag{V.90}$$

We are now in the position to describe the quantum algorithm. Given an initial quantum state vector $u(0), v(0)$ proportional to $u(0), v(0) \in \mathbb{R}^{2n}$, we perform the quantum Carleman linearization algorithm as the main text for the non-linear ODE in Eq. (V.84). One can also regard $f$ as an independent variable to simplfy the non-linearities in Eq. (V.84). We observe that the NTK dynamics as given by Eq. (V.89) is dissipative. Hence, $f$ and $\lambda$ can be computed classically (i.e., regard $f$ as a known time-varying parameter, and thus the quadratic Eq. (V.84) is degenerated to a linear ODE), or quantumly (i.e., simulate the quantum state vector $|f(t)\rangle$ coupled with $|u(t)\rangle$, $|v(t)\rangle$). Another alternative is to simulate $fv$ and $fu$ in Eq. (V.84) in a quantum fashion, such that Eq. (V.84) is always linear for the multi-layer case. An advantage is that the dynamics of $f\dot{f}$ is also dissipative as

$$\frac{d}{dt}(fu) = -\lambda(fu) + n^{-1} \eta f^3 u - n^{-1/2} f^2 v , \tag{V.91}$$

$$\frac{d}{dt}(fv) = -\lambda(fv) + n^{-1} \eta f^3 v - n^{-1/2} f^2 u . \tag{V.92}$$

According to the results of Refs. [36, 56] and the main text, the dissipation in Eq. (V.89), in particular the negative definite kernel $-\lambda$, is beneficial for the convergence of the linearization algorithm. In general, we expect the linearization is efficient whenever $u, v$ is nearly located in the flat local minima of $f$. If that works, there is an exponential enhancement in $n$.

### D. The relationship between our work and Ref. [36]

The foundational algorithm presented in Ref. [36] has significantly influenced our work, and ideas presented there have been further developed here and put to a good use. Our quantum ODE solver has technical elements that indeed echo the concepts from Ref. [36]. This section delves into our unique contributions that generalize and expand the insights presented in that work.

- Building upon the machinery developed in that work, we have expanded its scope to encompass general-order non-linearities, whereas the original publication has been limited to the second order only.

- The most important distinction at the quantum algorithmic level between our work and Ref. [36] is in the introduction of a notion of *discreteness*. We introduce a discrete variant of Carleman linearization and delve into its application in solving stochastic gradient descent equations. This is important for our purposes. In support of this, we offer a comprehensive exposition on discrete Carleman linearization, encompassing a reinterpretation of the Carleman linearization theory, we present a tensor network diagrammatic representation for the discretization error, as well as analytical breakdowns of higher-order amendments, and illustrative instances of lower order expansions.

- In our study, we have taken significant strides in conceptualizing and dissecting innovative pruning algorithms for classical neural networks. This endeavor is particularly geared towards efficiently navigating the unique input-output challenges inherent to quantum devices.

- Moreover, we have discerned that sparsity and dissipation stand out as crucial components for quantum ODE solvers. We have not only identified these elements in general terms, but have also delved into offering a precise definition and theorems relating to dissipation conditions within classical neural networks, specifically focusing on the Hessian matrices and their various derivations.

- Furthermore, we have designed and laid out innovative tomography methods that meld classical shadows with projective measurements. These methods are intriguingly connected to the coupon collector's problem, providing a deeper understanding and a fresh perspective on the topic, as we think.

- Finally, we have conducted numerical experiments involving up to 103 million training parameters, potentially setting a new benchmark for large-scale machine learning models in quantum literature. We have further evaluated the Hessian matrix spectra within these models, ensuring our claims are backed by empirical data and rigorous validation.

### E. Pruning QCM by pruning machine learning models

In our theory, we assume that machine learning model itself is sparse, and pruning conditions are added to ensure the sparsity of the architecture. Here, we show that pruning the model will lead to sufficiently good sparsity condition for the QCM $A$. The matrix we have constructed for Carleman linearization is the matrix $A$, the so-called QCM, given by

$$A = \begin{pmatrix} A_1^1 & A_2^1 & A_3^1 & \cdots & A_q^1 & & & \\ A_1^2 & A_2^2 & A_3^2 & \cdots & A_q^2 & A_{q+1}^2 & & \\ & A_2^3 & A_3^3 & A_4^3 & \cdots & & & \\ & & \ddots & \ddots & \ddots & & & \\ & & & A_{N-2}^{N-1} & A_{N-1}^{N-1} & A_N^{N-1} & \cdots & \\ & & & & A_{N-1}^N & A_N^N & \ddots & \\ & & & & & & \ddots & \ddots \end{pmatrix}. \tag{V.93}$$

To analyze the sparsity of $A$, we have the following theorem.

**Theorem V.1** (Sparsity of $A$). *Consider the differential equations*

$$\delta u_i = f_i(u, u^{\otimes 2}, u^{\otimes 3}, \ldots), \tag{V.94}$$

*for $0 \leq i \leq n$, such that the right hand side $f_i(u, u^{\otimes 2}, u^{\otimes 3}, \ldots)$ has $S_i$ terms. Then the row sparsity of the row corresponding to $u_{i_1} \otimes u_{i_2} \otimes \ldots \otimes u_{i_J}$ in the matrix $A$ is at most $\sum_{j=1}^{J} S_{i_j}$.*

*Proof.* Note that the terms in $u^{\otimes k}$ are linearly independent. For example, if $u \in \mathbb{R}^n$, then

$$u^{\otimes 2} = (u_1 u_1, u_1 u_2, \ldots, u_1 u_n, \ldots, u_{n-1} u_n, u_n u_n). \tag{V.95}$$

Furthermore, each term in $u^{\otimes 2}$ will correspond to an entry in the vector

$$\begin{pmatrix} u \\ u^{\otimes 2} \\ u^{\otimes 3} \\ \vdots \\ u^{\otimes (N-1)} \\ u^{\otimes N} \\ \vdots \end{pmatrix}. \tag{V.96}$$

Similarly, every terms in all $u^{\otimes k}$ ($u^{\otimes k} \in \mathbb{R}^{n^k}$) will be an entry in the above vector Eq. (V.96). In Eq. (V.96), we have just subsumed all these components for simplicity and clarity of notation. Hence, if a differential equation, such as Eq. (V.94), has $S_i$ independent terms in function $f_i(u, u^{\otimes 2}, u^{\otimes 3}, \ldots)$, then there are $S_i$ terms in row $i$ of the matrix $A$ that are non-zero. Now we consider, in general, the row in $A$ that corresponds to $u_{i_1} u_{i_2} \ldots u_{i_J}$. Then we can elaborately write its differential equation as

$$\begin{aligned} \delta \left( u_{i_1} \otimes u_{i_2} \otimes \ldots \otimes u_{i_J} \right) &= \delta u_{i_1} \otimes u_{i_2} \otimes \ldots \otimes u_{i_J} + u_{i_1} \otimes \delta u_{i_2} \otimes \ldots \otimes u_{i_J} \\ &\quad + \ldots + u_{i_1} \otimes u_{i_2} \otimes \ldots \otimes \delta u_{i_J} \\ &= f_{i_1} \otimes u_{i_2} \otimes \ldots \otimes u_{i_J} + u_{i_1} \otimes f_{i_2} \otimes \ldots \otimes u_{i_J} \\ &\quad + \ldots + u_{i_1} \otimes u_{i_2} \otimes \ldots \otimes f_{i_J}. \end{aligned} \tag{V.97}$$

For $0 \leq j \leq J$, the functions $f_{i_j}$ feature $S_j$ terms. If there are no same terms in the summations above in Eq. (V.97), then we achieve the maximum non-zero terms in the selected row, given by $\sum_{j=1}^{J} S_{i_j}$, which also reflects the sparsity of the corresponding row. □

Then the matrix sparsity of $A$ after truncating at $N$ would require to take the maximum of all rows as

$$\text{Sparsity} = \max_{\text{all rows}} \left( \sum_{j=1}^{J} S_{i_j} \right). \tag{V.98}$$

It is less obvious to identify analogous terms in the different functions $f_{i_j}$. This is because the order of $u_i$ in the tensor products matters. As a result, we need to ask terms in functions $f_{i_j}$ to take very specific forms in order to get two identical terms. For example, consider $\delta(u_1 \otimes u_2)$, so

$$\delta(u_1 \otimes u_2) = \delta u_1 \otimes u_2 + u_1 \otimes \delta u_2 = f_1 \otimes u_2 + u_1 \otimes f_2 . \tag{V.99}$$

Hence, for terms in the first and second part in Eq. (V.99) to be the same up to a factor irrelevant to $u$, we need to require the term in $f_1$ start with $u_1$, the term in $f_2$ to end with $u_2$, and all the intermediate terms in the two have to be exactly identical. To give a simple example, if $f_1 = C_1 u_1 \otimes u_3 \otimes u_4$ and $f_2 = C_2 u_3 \otimes u_4 \otimes u_2$, then plugging this into Eq. (V.99) gives

$$\delta(u_1 \otimes u_2) = C_1\, u_1 \otimes u_3 \otimes u_4 \otimes u_2 + C_2\, u_1 \otimes u_3 \otimes u_4 \otimes u_2 = (C_1 + C_2)\, u_1 \otimes u_3 \otimes u_4 \otimes u_2 . \tag{V.100}$$

There are $S_1 = 1$ and $S_2 = 1$ terms in $f_1$ and $f_2$. So the row sparsity for $u_1 u_2$ is $S_1 + S_2 - 1 = 1$, instead of just $S_1 + S_2 = 2$. Now we truncate $A$ at $N$, such that the dimension of $A$ is given by

$$\Delta = n + n^2 + \ldots + n^N = \frac{n^{N+1} - n}{n - 1} = O\left(n^N\right) . \tag{V.101}$$

We define $S_{i_j}^{(k)}$ as the number of terms in $f_{i_j}$ that is contained in $F_k$. Then $S_{i_j} = \sum_{k=0}^{q} S_{i_j}^{(k)}$. For a row $J$ in $A$ such that $1 \leq J \leq N - q$, the sparsity row $J$ is given by

$$\sum_{k=0}^{q} \sum_{j=1}^{J} S_{i_j}^{(k)} . \tag{V.102}$$

For $N - q \leq J \leq N$, the sparsity is found to be

$$\sum_{k=0}^{q-(J-(N-q))} \sum_{j=1}^{J} S_{i_j}^{(k)} . \tag{V.103}$$

Hence, the sparsity of $A$ truncated at $N$ is

$$\max_{1 \leq J \leq N} \begin{cases} \sum_{k=0}^{q} \sum_{j=1}^{J} S_{i_j}^{(k)}, & 1 \leq J \leq N - q, \\ \sum_{k=0}^{q-(J-(N-q))} \sum_{j=1}^{J} S_{i_j}^{(k)}, & N - q \leq J \leq N . \end{cases} \tag{V.104}$$

Overall, the sparsity is a function of $N$ and $S$, since we maximize over $J$, so it is a fixed and bounded number. The order of the ODE is also determined by the problem set up, so $q$ is also a fixed number. Thus, overall, as long as we prune the original machine learning model sufficiently well (reflecting a small $S$), then the overall sparsity will at most be equipped with a factor $N$ and $q$, which will not affect the overall result. In our real example in the main text, the model itself has been already significantly pruned to 1%. However, it might be interesting to look at the exact numbers and track the real-time evolution of sparsity of quantum Carleman matrices in machine learning. Some of those issues will be discussed in future research.


}

\end{document}